\documentclass[sigplan,10pt]{acmart}

\usepackage{xcolor}
\usepackage{wrapfig}
\usepackage{subcaption}
\usepackage{listings}
\usepackage{cleveref}
\usepackage{multirow}
\usepackage{enumitem}

\crefname{section}{\S}{\S\S}
\crefname{lstlisting}{algorithm}{algorithms}
\Crefname{lstlisting}{Algorithm}{Algorithms}
\setlist{nolistsep}

\definecolor{codegreen}{rgb}{0,0.6,0}
\definecolor{codegray}{rgb}{0.5,0.5,0.5}
\definecolor{codepurple}{rgb}{0.58,0,0.82}
\definecolor{backcolour}{rgb}{0.95,0.95,0.92}
\definecolor{anti-flashwhite}{rgb}{0.95, 0.95, 0.96}
\definecolor{highlight}{rgb}{1.0, 0.0, 0.0} % -> use this for red text
\newcommand{\tmname}[1]{NV-HALT}

\usepackage{soul} %needed for hl

\renewcommand{\paragraph}[1]{\noindent\textbf{#1.}}
\lstdefinestyle{mystyle}{
  mathescape=true,
  backgroundcolor=\color{anti-flashwhite},
  commentstyle=\color{codegray},
  numberstyle=\tiny\color{codegray},
  stringstyle=\color{codepurple},
  basicstyle=\ttfamily\footnotesize,
  breakatwhitespace=false,         
  breaklines=true,
  numberblanklines=false,
  captionpos=b,                    
  keepspaces=false,                 
  numbers=left,                    
  numbersep=5pt,                  
  showspaces=false,                
  showstringspaces=false,
  escapeinside={<@}{@>},
  showtabs=false,                  
  tabsize=1,
  keywordstyle=\color{codegreen},
  otherkeywords={thread_local, for,each,true,false, class, struct, def, goto, uint, List*}, morekeywords={type,subtype,break,continue,if,else,elif,end,loop,while,do,done,exit, when,then,return,read,and,or,not,boolean,procedure,invoke,iteration,until,wait}
}
\usepackage{tikz}
\usetikzlibrary{backgrounds}
\pgfdeclarelayer{edgelayer}
\pgfdeclarelayer{nodelayer}
\pgfsetlayers{background, edgelayer,nodelayer}
\newcommand{\edit}[1]{#1}
\newcommand{\smalledit}[1]{#1}

\title{Persistent HyTM via Fast Path Fine-Grained Locking} 
\author{Gaetano Coccimiglio}
\email{gccoccim@uwaterloo.ca}
\affiliation{%
  \institution{University of Waterloo}
  \streetaddress{200 University Ave W}
  \city{Waterloo}
  \state{Ontario}
  \country{Canada}
  \postcode{N2L 3G1}
}

\author{Trevor Brown}
\email{trevor.brown@uwaterloo.ca}
\affiliation{%
  \institution{University of Waterloo}
  \streetaddress{200 University Ave W}
  \city{Waterloo}
  \state{Ontario}
  \country{Canada}
  \postcode{N2L 3G1}
}

\author{Srivatsan Ravi}
\email{srivatsr@usc.edu}
\affiliation{%
  \institution{University of Southern California}
  \streetaddress{4676 Admiralty Way}
  \city{Los Angeles}
  \state{California}
  \country{USA}
  \postcode{90292}
}

\acmConference[Arxiv]{Arxiv}{2025}{}
\acmYear{2025}
\acmISBN{} 
\acmDOI{} 
\startPage{1}
\setcopyright{none}

 %TODO 

\begin{document}

\begin{abstract}
Utilizing hardware transactional memory (HTM) in conjunction with non-volatile memory (NVM) to achieve persistence is quite difficult and somewhat awkward due to the fact that the primitives utilized to write data to NVM will abort HTM transactions. 
We present several persistent hybrid transactional memory (HyTM) that, perhaps counterintuitively, utilize an HTM fast path primarily to read or acquire fine-grained locks which protect data items. 
Our implementations guarantee durable linearizable transactions and the STM path satisfies either weak progressiveness or strong progressiveness.
We discuss the design choices related to the differing progress guarantees and we examine how these design choices impact performance.
We evaluate our persistent HyTM implementations using various microbenchmarks. 
\smalledit{Despite the challenges and apparent awkwardness of using current implementations of HTM to achieve persistence, our implementations achieve up to 10x improved performance compared to the existing state of the art persistent STMs and up to 2.6x improved performance compared to the existing state of the art persistent HyTMs.}
\end{abstract}

\maketitle

\section{Introduction}
The recent commercial availability of byte-addressable Non-Volatile Memory (NVM) has sparked increased efforts towards designing persistent concurrent data structures capable of recovering from system crashes.
These efforts have proven to be challenging since effectively utilizing NVM introduces new issues on-top of the already difficult task of ensuring correctness and efficiency in a completely crash free setting.
% For many existing processors, NVM exists only as a main memory while processor caches, registers and DRAM remain volatile. 
We consider system crashes caused by power failures. 
In the first few generations of processors with support for NVRAM, the NVRAM itself is non-volatile and retains its contents in the event of a power failure, but the caches and registers (and, of course, DRAM) remain volatile.
% 
% If a crash occurs, only data in NVM can be recovered.
Data in the caches must be explicitly \textit{flushed} to NVM in order to be persisted.
In doing so, one must consider issues such as when should data be written back to NVM, write back ordering and at what frequency to issue write backs.
Data in the cache can also be automatically flushed in the background at any time by the system without the programmer's knowledge.
To guarantee that the data is persisted in a timely manner, one nominally adds flushes at key places in the program.
The most recent Intel processors are equipped with \textit{eADR}.
% Recent Intel processors are equipped with \textit{eADR}.
This makes the processor cache effectively non-volatile as it is automatically flushed to NVM in the event of a power failure \cite{eadr}.
eADR removes the need for explicit flushes.
However, eADR does not remove the complexity of designing correct persistent data structures.
The programmer still needs to carefully order writes to any persistent data or risk corrupting the state of the data structure in NVM (we discuss this further in \cref{para-eadr}).
In this work, we simultaneously solve two problems: correctly persisting a program's data and synchronizing its threads, relieving the programmer from the burden of having to meticulously design synchronization like fine-grained locking protocols.

Much of the existing research regarding persistent concurrent data structures has focused on hand-crafted approaches often involving persisting an existing volatile data structure \cite{david2018log, zuriel2019efficient, friedman2020nvtraverse, friedman2021mirror, wei2021flit, QueueFriedman18}. 
Unfortunately, designing efficient and correct hand-crafted persistent data structures is often difficult, time consuming and error prone.
Much of the challenge stems from the fact that it is already difficult to achieve correct volatile concurrent data structures.

Synchronization between threads can be achieved through the use of \textit{transactional memory, (TM)}.
TMs are synchronization mechanisms that allows users to execute sequences of memory accesses as atomic \textit{transactions}.
A transaction either \textit{commits} and appears as a single indivisible step, or \textit{aborts} and has no visible effect.
Similarly, a \textit{persistent TM (PTM)} provides the same functionality but also ensures that the effects of transactions are written back to NVM.
We say that a TM \textit{instruments} memory accesses if it performs additional accesses to metadata for each memory access.

Originally \textit{software TMs (STMs)} were the only available option for TM. 
STMs typically require read/write instrumentation.
Some processors now have \textit{hardware TM (HTM)}, offering instructions that allow users to mark blocks of code %memory accesses
as transactional where the processor is responsible for ensuring the atomicity of these \textit{hardware transactions} \cite{rtm, le2015transactional}.
Current implementations of HTM can abort unconditionally, requiring a software based fallback code path.
TMs that combine a \textit{hardware path} executed on HTM and a \textit{software path} executed in software are known as \textit{hybrid TMs (HyTMs)}.

In the volatile setting, HyTMs can typically avoid instrumenting memory accesses on the hardware path which generally leads to better performance compared to STMs.
\smalledit{By extension, one might think that for a \textit{persistent HyTM}, it is obvious that the use of HTM would similarly lead to better performance.}
However, \textit{flush} instructions, which are required for persistence, force hardware transactions to abort!
\smalledit{This means that if we want to persist the effects of a hardware transaction, we must flush \textit{after} the hardware transaction has completed.}
Moreover, we need additional synchronization if we want to ensure that the transaction appears to be persisted at the same point that it commits, 
As a result, extra care is needed to ensure that the flush instructions and synchronization required to persist transactions do not cause poor performance.
% (or even render the hardware path useless). %instrumentation is required to persist the effects of a hardware transaction.
%This adds considerable overhead to the hardware path.

In this work, we present a family of persistent HyTMs that we call \tmname~ (\underline{N}on-\underline{V}olatile \underline{H}ardware \underline{A}ssisted \underline{L}ocking \underline{T}ransactions).
% that combine the volatile synchronization from the HyTM of Brown and Ravi \cite{brown2017cost} with the persistence mechanism of Ramalhete et al. \cite{ramalhete2021efficient}.
As the name suggests, hardware transactions in our persistent HyTM are mainly used to read and acquire fine-grained locks.
These locks are used to prevent threads from observing transactional stores that have not yet been written back to NVM.
Locking in this manner allows us to ensure a transaction appears to be persistent atomically at the same time as it commits.
% One might hope to leave synchronization entirely to hardware
% While this may be counterintuitive, 
While one might hope to leave synchronization entirely to hardware, we show that this type of instrumentation performs well on real hardware.
Unlike the existing state of the art, \tmname~ utilizes a non-trivial fallback path allowing for concurrency between hardware and software path transactions.
We design and implement several versions of \tmname~ offering different progress guarantees and levels of performance in practice.
We evaluate and compare our persistent HyTM against a recent state of the art persistent STM as well as a state of the art persistent HyTM.
Our evaluation shows that \tmname~ achieves up to 10x improved performance compared to the state of the art persistent STM and up to 2.6x improved performance compared to the state of the art persistent HyTM.

\paragraph{Contributions}
We provide the following specific contributions:
\begin{itemize}[leftmargin=*]
    \item We present several versions of \tmname~, a novel persistent HyTM that features a distinct usage of fine-grained locks to ensure that transaction commits and persistence appear (together) to be atomic.
    \tmname~ allows for greater concurrency and offers improved progressiveness compared to the prior state of the art persistent HyTM, SPHT \cite{castro2021spht}.
    Unlike prior work, \tmname~ has a non-trivial fallback path that allows for concurrency between hardware and software transactions.
    \tmname~ does not require additional background threads.    
    In \tmname~ only transactions with intersecting accesses can abort one another which is an improvement compared to SPHT where transactions that access unrelated data can still abort each other \cref{sec:design}.
    \item We implement \tmname~ on a real system that supports hardware transactions and is equipped with NVM \cref{sec:impl}.
    \item As part of this implementation we provide a custom allocator and memory reclamation mechanism.
    Our approach to memory management avoids memory leaks unlike SPHT.
    Moreover, our allocator avoids introducing additional aborts unlike the existing state of the art PSTM \cref{sec:memmanagment}.
    \item We experimentally evaluate \tmname~ against the state of the art PSTM and state of the art persistent HyTM.
    Our evaluation demonstrates a clear improvement compared to the existing state of the art.
    In many of our benchmarks, \tmname~ achieves more than 2x more throughput compared to existing PTMs and up to 3.5x better energy efficiency \cref{sec:eval}.  
    \item %Minor contribution:
    We introduce the notion of O(1)-abortable transactions to address the fact that there does not exist any formalism to describe how many aborts can be induced by a hardware fast path in a HyTM \cref{sec:abortable}. Our definition is a more realistic model of the behavior of real hybrid TM algorithms.     
\end{itemize}
%
% \section{Background}
% In this section we will review some relevant background information.
% 
\section{Background and Related Work}
% In this section we will review some relevant background information and discuss related work.

% \subsection{Model}

We consider an asynchronous shared memory system in which $N$ processes communicate via shared \emph{objects} which offer \textit{operations}.
A thread can \textit{invoke} an operation on an object, which can (optionally) change the state of the object and return a \textit{response}. 
The system can crash at any time in which case all $N$ processes crash simultaneously.
After a crash, a recovery procedure is invoked.
New operations cannot be invoked until the recovery procedure returns.

\edit{Current implementations of NVM and HTM exist only on Intel systems which use a total store ordering (TSO) memory model.
Thus, we assume a TSO memory model. 
This default assumption of sequential consistency aligns with the default behavior for atomics in C++. 
Implementers can rely on this default behavior, or fence as appropriate for their architecture.}

\subsection{Persistent Memory}
Persistent memory is a part of main memory that retains its contents after a system crash.
The DRAM along with processor registers and caches are not persistent (unless you have eADR in which case the cache is effectively non-volatile). 
% We consider systems in which persistent memory exists only as main memory where writes first take effect in the volatile cache memory.
In order to guarantee that the effects of a write are persisted, the programmer must explicitly \textit{flush} data from the cache to NVM.
On Intel platforms the prescribed way of flushing is via the \texttt{clflushopt} instruction.
The \texttt{clflushopt} instruction initiates a flush of an entire cache line (nominally 64 bytes) to NVM and invalidates the cache line in all caches. 
It returns before the flush has been completed, allowing many flushes to be initiated in parallel and pipelined.
The \texttt{clwb} instruction is similar, but it does not necessarily invalidate the cache line, which may make it more efficient \cite{clwb}. 
However, many processors with NVM support (including the ones in our experimental system) implement \texttt{clwb} identically to \texttt{clflushopt}. 
In systems with such processors, \texttt{clwb} is no more or less efficient than \texttt{clflushopt}.
The programmer can utilize fence instructions (\texttt{sfence} on Intel) to block until all previously initiated flushes have completed.

The processor may also arbitrarily flush data to NVM at any time, even if no explicit flush instructors have been issued.
We refer to this as \textit{background flush}.
Background flushes typically occur as a result of the cache coherence protocol.
These background flushes can cause data to be written back to NVM earlier than the programmer expects.
Depending on the order of writes, this can leave the contents of NVM in an inconsistent state.
To understand how such problems might arise consider a simple example of updating a singly linked list. 
One might add a key to a list by allocating a new node, locking the node containing the predecessor key and then performing the following writes: 
1) update the new node's next pointer to point to the old successor then
2) update the predecessor's next pointer to point the new node.
Suppose that we do not perform any explicit flushes but a background flush causes the effects of the second update to be written back to NVM.
If we crash after this background flush completes the first update will not be reflected in NVM and as a result of this, we will not be able to recover the latter part of the list.

\paragraph{eADR the Silver Bullet}
The need to perform explicit flushes complicates the design of persistent algorithms.
eADR removes the need for flush instructions by making the cache effectively non-volatile.
Processors equipped with eADR will ensure that if a write has occurred in the cache, then it will eventually be written back to NVM.
With eADR, the issues caused by background flushes as in our linked list example will no longer be a problem.
% One might assume that that with a persistent cache we can easily design persistent algorithms.
Unfortunately, a volatile cache is not the only concern when designing persistent algorithms.

\paragraph{eADR is not a Silver Bullet}
\label{para-eadr}
eADR does not solve all of our NVM problems, it only removes the need for explicit flushes and it does not remove the need for careful write ordering which is the main challenge of utilizing NVM.
\edit{Moreover, fences are still required to maintain write ordering correctness \cite{eadr_fence}}.

Recall the linked list example.
Suppose that we swap the order of the updates to be the following:
\smalledit{1) update the predecessor's next pointer to point to the new node then}
2) update the new node's next pointer to point to the old successor.
In a volatile only setting, the order of the writes does not matter since we have already locked the predecessor and thus there is no problem with this ordering. 
However, in the persistent setting, if a crash occurs before the last update, we will not be able to recover part of the list.
This is true even if we have we have eADR. 
In fact, with this new order of writes, eADR has the same effect as the problematic background flush that we discussed before.
This example demonstrates how easy it is to corrupt the state of a data structure after a crash if one does not carefully order writes to persistent data, even if we have eADR.
Moreover, this demonstrates that conventional designs that work fine with volatile memory may not work when using NVM.
Ultimately, while eADR helps by removing the need for flushes, it does not save us from the challenges of designing correct and efficient persistent algorithms.
It is also worth noting that there are still many systems with earlier generation hardware that are not equipped with eADR and we want our algorithms to function on these systems.

\edit{To summarize, with eADR, flushing is made trivial, but persistent algorithm design is not.
This is similar to how fencing is made trivial by memory\_order\_seq\_cst in C++, but concurrent algorithm design remains non-trivial.
Our algorithm confers all of the same benefits with eADR, as without.}
In the following sections we specify flush and fence locations in our algorithm that suffice for systems without eADR, and on systems with eADR they can be safely omitted (as we have ordered writes appropriately to ensure data is not lost).
\\

\paragraph{Correctness in NVM}
Durable linearizability \cite{IzraelevitzMS16} is one of the most common correctness properties for persistent algorithms.
It defines how to apply linearizability to models where crashes are possible.
Intuitively, linearizability states that each operation appears to take effect atomically (as a single indivisible step) at some time during the operation. 
That time is called the linearization point (LP) for that operation.
Moreover, each operation must return the same value as it would if all of the operations were run sequentially in the order of their LPs.
Durable linearizability requires that any operation that completed before a crash will be reflected in the state of the object after recovery by requiring executions, excluding crashes, to be linearizable.

\subsection{Transactional Memory}
A transaction is a sequence of transactional accesses, (reads and writes), performed on a set of \textit{transactional addresses}.
A TM implementation provides operations to %provides a set of processes with deterministic algorithms that implement
start a transaction, read and write transactional addresses, commit a transaction, and voluntarily abort a transaction.
These operations are implemented using a set of \textit{base objects}.
If two transactions are concurrent and one or both of the transactions writes to an address that the other has already accessed then we say that these transactions have a \textit{data conflict} (or simply that the transactions conflict).
A transaction that aborts due to a conflict will typically be retried until it either succeeds and commits, or until it invokes a voluntary abort operation. %, or it will abort due to conflicts and retry until it commits.
A HyTM implementation additionally specifies whether a transaction is executed in hardware or software.
%More specifically, accesses in hardware operate on cached state while accesses in software operate directly on shared memory. 

\paragraph{Hardware TM}
We consider a system equipped with HTM with the same semantics as Intel's RTM \cite{rtm}.
Hardware transactions maintain a \textit{tracking set} comprised of a \textit{read set} and a \textit{write set}.
Conflicts are tracked via the cache coherence protocol.
A conflict occurs whenever a thread writes to an address that is in the tracking set of an ongoing transaction.
If two transactions conflict, \textit{at least} one will abort, but hardware transactions can also abort spuriously (for any reason).
%Hardware transactions can abort spuriously (for any reason), and if two transactions conflict, \textit{at least} one will abort. %and are guaranteed to abort due to \textit{conflicts}. % and \textit{capacity aborts}.
Note that a non-transactional access can also abort a transaction.

While we model all involuntary aborts other than conflict aborts as spurious, a common cause of HTM non-conflict aborts is related to the bounded capacity of HTM read and write sets.
For example, with Intel RTM, the write set of a hardware transaction is limited to the L1 cache.
(Capacity aborts can occur when as few as 9 addresses are accessed on a typical Intel CPU with an 8-way associative L1-cache.)

% The write set of hardware transactions is limited to the size of the L1 cache.
% Capacity aborts occur whenever the write set of the hardware transaction exceeds the maximum capacity.
% Any access that causes an eviction of a cache line belonging to the tracking set of a hardware transaction will cause the transaction to abort.

To initiate a hardware transaction we utilize the \texttt{xbegin} instruction.
Calling the \texttt{xend} instruction attempts to commit the hardware transaction.
We can also explicitly (voluntarily) abort a hardware transaction with \texttt{xabort}.
When a hardware transaction is aborted, control returns to the call of \texttt{xbegin} which will then immediately return a failure code signifying that the last hardware transaction aborted.
Since hardware transactions can abort for any reason a fallback path for execution via software transactions is required to make any kind of progress guarantee.

One cause of spurious aborts is the use of flush instructions.
On any hardware that we are aware of, flush instructions force hardware transactions to abort (although the specification of these instructions does not guarantee that they will abort a hardware transaction).
\edit{Similarly, it is not entirely clear how eADR and HTM (specifically Intel RTM) are intended to function together.
While hardware transactions appear to take effect atomically, each cache line modified during a hardware transaction must be transitioned to a state in which other threads can access it.
The actual implementation details of Intel's RTM are proprietary.
We suspect that caches lines modified during an RTM transaction are held in exclusive mode and are moved to shared mode once the transaction commits.
Regardless of the specifics, for eADR to be of any use in this case, the entire write set of the hardware transaction would need to be somehow marked as completed via a single atomic write and we are not aware of any evidence to suggest this is the case.
Anything that performs multiple distinct updates, say to individual cache lines, would allow for a write back of partially completed hardware transactions.}
So, once again, eADR cannot save us.
We must have a software based fallback path and therefore, we still have problems related to write ordering even with a persistent cache.

\paragraph{Correctness in TM}
We consider the TM-correctness property of \emph{opacity}~\cite{tm-book}.
Intuitively, opacity requires that the set of all concurrent transactions (including aborted ones) be equivalent to a sequential one in which every read returns the value of the latest written write.
For PTMs, we say that a TM offers durable transactions, if it guarantees that the effects of all committed transactions will be reflected in the state of the transactional addresses following recovery. 

\paragraph{Progress in STM}
For STMs, the de facto \textit{progress} properties are (weak) progressiveness and strong progressiveness.
Intuitively, progressiveness states that if a transaction is aborted then there exists a concurrent conflicting transaction. 
Strong progressiveness states that if a set of concurrent transactions conflict, then at least one of them is not aborted.

In principle, progressiveness can be defined for hybrid TMs as well, but since current HTM implementations can abort transactions for any reason (and thus offer no progress guarantees), one cannot actually achieve progressiveness in HyTMs in practice.
Ignoring spurious aborts allows us to reason in terms of these existing progress properties, however, this is unrealistic---it is surprisingly easy to write software in which all transactions will spuriously abort, forever.

\subsubsection{A New Progress Guarantee for Hybrid TM} %C-Abortable Progressiveness for Hybrid TMs}
\label{sec:abortable}
We suggest the following alternative: Think of a  %reasoning about %To reason about progress in a more realistic way, we can think of a
transaction as a sequence of \textit{attempts} culminating in a \textit{commit} or \textit{voluntary abort}.
One can then imagine a transaction having %In this way, a transaction will have
a bounded number of attempts that run on the hardware path, possibly followed by some number of attempts on the software path.
%Under this definition of transactions, we 
We say that a TM is \textit{C-abortable} weak (resp., strong) progressive if, for all executions of the TM, each transaction in an execution can abort unconditionally (for any reason) at most $C$ times, after which any subsequent aborts by that transaction must satisfy the requirements of weak (resp., strong) progressiveness.
When $C$ is a constant we can also say that the TM is O(1)-abortable weak (resp., strong) progressive.

This generalizes these classical definition of progressiveness to deal with the reality of current HTM systems, without relying on an unrealistic assumption that spurious aborts can simply be ignored.
Note that 0-abortable progressiveness is simply progressiveness (and similarly for strong progressiveness).
As an example, a hybrid TM that first attempts a transaction up to 10 times on the hardware path before falling back to a progressive software path could be shown to be 10-abortable progressive.

\subsection{Related Work}
% \section{Related Work}
% In this section we present an overview of some relevant related work. 

\subsubsection{Volatile STMs}
Volatile STMs were originally proposed by Herlihy and Moss \cite{herlihy1993transactional}.
Since then there has been extensive research on the topic of STMs \cite{dice2006transactional, felber2008dynamic, fraser2007concurrent, memoryprinciples, ennals2006software, shavit1995software}.
There has also been attempts to integrate volatile TM into programming languages including Haskell, C++ and Java.
% Below we describe two examples of volatile STMs.
%TL2 is one of the most well known STMs. %Below we describe the well known STM, TL2.

\paragraph{TL2}
Transactional Locking II (TL2) \cite{dice2006transactional} is one of the most well known volatile STMs. %TL2 has two main selling points: it does not rely on garbage collection, and thus can be implemented in unmanaged languages like C/C++.
% (as opposed to garbage collected systems). % and does not require any specialized managed runtime environments. 
%
%TL2 is a word based STM.
%In the basic TL2 algorithm, a global version number is read at the start of each transaction, and incremented at commit time if the transaction performs any writes. %It relies on a global version number.
%The original algorithm required that the global version number be read by every transaction and incremented at commit time by every write transaction. 
%Many different algorithms have since been proposed for when to increment the global version number \cite{lev2009anatomy}.
In the basic TL2 algorithm, each transaction first obtains a version by reading a global version number.
This global version number is incremented by each transaction that performs writes. %to obtain a version that is read at the beginning of each transaction, and incremented at the end of every writing transaction.
Every memory address is protected by a versioned lock that contains the version of the last transaction that wrote to it. %, and a global version number if .
%In the basic TL2 algorithm, a global version number is read at the start of each transaction, and incremented at commit time if the transaction performed any writes.
%Every memory address is protected by a versioned lock that contains the value of the global version.
%Transactions first read and store the global version number.
Each transaction uses its version to \textit{validate} the values it reads, ensuring that it only sees values written before it began. % During a transaction, the value read used to validate every memory address that is read by the transaction.
% TL2 tracks both read and write sets.
% Read and written addresses are explicitly tracked in read and write sets implemented using linked lists. % (with optional Bloom filters). %During a transaction, TL2 tracks the memory addresses that are read/written by maintaining read and write sets.

The public implementation by Dice et al. uses \textit{buffered writes} and deferred \textit{commit time locking}, meaning the write set is locked only when the transaction attempts to commit. 
This is in contrast to \textit{in-place writes} and \textit{encounter time locking}.
After acquiring the write set locks, the global version is incremented and the read set re-validated.
This validation is skipped if, after the increment, the global version is only 1 greater than than the transaction's version.
If the write set is locked in some total order (for example, by increasing memory address), then TL2 guarantees strong progressiveness.
Numerous variants of TL2 have since appeared, suggesting different ways to update the global version number \cite{lev2009anatomy, ramalhete2024scaling}. %, and identifying special cases in which one can avoid incrementing the global version number, or avoid unnecessary aborts~\cite{?? new pedro paper ??}.

% \subsubsection{TinySTM}
% Like TL2, TinySTM is a word based STM that relies on a global clock and versioned locks to protect memory addresses accessed via the STM \cite{felber2008dynamic}.
% Compared to the original TL2, TinySTM differs in that it utilizes encounter time locking as opposed to commit time locking.
% TinySTM also provides the ability for write transactions to utilize write-through accesses or write-back accesses.
% With write-through accesses a write transaction immediately updates an address after locking it whereas with write-back accesses the writes are deferred to commit time. 
% The access strategy changes the way in which transactions need to maintain and interact with write logs.
% It should be noted that since the introduction of TL2, the implementation from Dice et al. has since been updated to provide an \textit{eager} mode which can be enabled to utilize encounter time locking instead of deferred commit time locking.

\subsubsection{Persistent STMs}
There are many existing persistent STMs (PSTMs) \cite{ramalhete2021efficient, correia2020persistent, volos2011mnemosyne, correia2018romulus, gu2019pisces, krauter2021persistent, assa2023tl4x, pmdk, coburn2011nv}.
These algorithms typically rely on techniques such as logging (both undo and redo logging), data replication and custom memory management mechanisms/allocators.
We provide the details of the state of the art PSTM, Trinity.
% Some PSTMs such as Mnemosyne or NV-heaps focus more persistent memory management by providing systems that expose persistent memory to user programs in a more efficient or easy to use manner on-top of which they provide mechanisms for executing transactions. 
% In the case of Mnemosyne this is achieved via compiler support for transactions.
% 
% 
% 
% Below we describe the details of some PSTMs focused on providing persistent transactions. 
% Below we describe the details of a state of the art, PSTM, Trinity. 

% \paragraph{CX-PTM and Redo-PTM}
% Correia et al. presented CX-PTM and Redo-PTM \cite{correia2020persistent}.
% These are both wait-free PTMs.
% Guaranteeing wait-freedom is difficult and adds considerable overhead.
% As a result, these PSTMs perform noticeably worse compared to PSTMs that do not guarantee wait-freedom.
% Since we are not interesting in guaranteeing wait-freedom we provide only a very brief discussion of these algorithms.

% Both CX-PTM and Redo-PTM are based on Herlihy's combining consensus algorithm.
% CX-PTM is adapted from the CX universal construction from the same authors \cite{correia2020wait}.
% CX-PTM utilizes utilizes 2$n$ replicas of the underlying data structure (where $n$ is the number of threas), a pointer to the most up-to-date replica and a volatile log of update operations.
% All replicas must be persisted.
% CX-PTM relies on a turn queue to establish the order on which the update are applied to each replica.
% Redo-PTM is an improvement over CX-PTM as it attempts to avoid re-executing the same update operations and it requires only $n+1$ replicas.

\paragraph{Trinity}
Ramalhete et al. presented Trinity, a technique that is combined with TL2 or flat combining \cite{hendler2010flat} to produce a PSTM  \cite{ramalhete2021efficient}. 
It is effectively undo-logging where the logs are colocated with user data in persistent memory.
It exploits the fact that x86 processors guarantee that writes to the same cache line will never be persisted out of order \cite{cohen2017efficient}.
Trinity uses a shadow data approach where a copy of user data exists in both volatile and persistent memory.

In Trinity, every word of memory that will be accessed transactionally is augmented with % is annotated to contain the user data word which the authors refer to as \textit{main},
an adjacent replica word which the authors call \textit{back} and an adjacent sequence number.
Collectively, these fields must fit in a single cache line.
The metadata is required only for recovery, so rather than having a one-to-one copy of all data in volatile and persistent memory, volatile memory need only contain user data while persistent memory must contain the user data along with the metadata.
% We describe Trinity ignoring this optimization.

Writes to any word must first update \textit{back}, then the sequence number and finally the user word, after which the cache line containing these fields is flushed.
At the end of a transaction a fence is performed after which the global sequence is incremented and persisted (requiring another flush and fence).
After a crash one must search through persistent memory to find any annotated words with sequence numbers that differ from the global sequence number and revert each such word to its old value stored in \textit{back}.

\subsubsection{Volatile Hybrid TMs}
There are many volatile hybrid TMs 
\cite{riegel2011optimizing, brown2017template, damron2006hybrid, kumar2006hybrid, dalessandro2011hybrid, matveev2015reduced, nussbaum2007phtm, brown2017cost, rajwar2001speculative}.
HyTMs often utilize a \textit{two path} approach wherein hardware transactions represent the fast path and software transactions represent the fallback slow path.
% A well known example of this two path style HyTM is transactional lock elision (TLE) wherein software transactions immediately claim a global lock preventing concurrency between hardware and software transactions (and other software transactions). 
Some HyTMs such as hybrid NOrec and reduced hardware NOrec do not allow concurrency between hardware and software transactions \cite{dalessandro2011hybrid, matveev2015reduced}.
Others, such as the work of Damron et al. and that of Kumar et al. allow for concurrency between hardware and software transactions \cite{damron2006hybrid, kumar2006hybrid}.
Brown and Ravi present two examples of two path HyTMs with different levels of concurrency between hardware and software transactions \cite{brown2017cost}. 

\subsubsection{Persistent HyTMs}
There is relatively little existing work on persistent HyTMs.
Many proposed algorithms do not work with existing hardware and instead are based on proposals for new hardware or extensions to existing HTM \cite{avni2015hardware, brown2016phytm, joshi2018dhtm, jeong2020unbounded}.
There are even fewer persistent HyTMs that are supported by existing hardware \cite{gencc2020crafty, castro2021spht, castro2019hardware}.
% We provide details for some persistent HyTMs with examples of algorithms based on proposals for new hardware as well as algorithms that function with current hardware.
We describe SPHT, a state of the art persistent HyTM.

% \paragraph{PHyTM}
% Avni and Brown presented Persistent Hybrid Transactional Memory (PHyTM) \cite{brown2016phytm}.
% This is a three path HyTM that relies on redo-logging to achieve persistence.
% PHyTM also relies on an extension of Intel's HTM implementation that allows for the ability to flush a single bit to persistent memory as part of the commit of a hardware transaction.
% This extension was originally proposed by Avni et al. in \cite{avni2015hardware}.
% This feature has not been implemented in current hardware and as a result PHyTM is not functional on any existing hardware.

% \paragraph{DHTM}
% Joshi et al. presented Durable Hardware Transactional Memory (DHTM).
% DHTM is not functional on any existing hardware.
% DHTM proposes integrating redo logging into a restricted transactional memory (RTM)-like HTM \cite{joshi2018dhtm}.
% DHTM also proposes the use of logging to allow the write set of hardware transactions to exceed the L1 cache.
% Experimental analysis of DHTM was conducted using a simulator.

\paragraph{SPHT}
Scalable Persistent Hardware Transactions (SPHT), which is based on NV-HTM \cite{castro2019hardware}, is a redo-logging based HyTM that runs on current hardware \cite{castro2021spht}.
%SPHT is based on NV-HTM \cite{castro2019hardware} which utilizes redo logging.
The software fallback path immediately claims a global lock and all hardware transactions read this lock and abort if it is held.
SPHT logs operations in NVM and orders these logs using timestamps.
When the logs are full they must be replayed in NVM, both to ensure that NVM is up to date, and to allow the logs to be emptied (avoiding the need for unbounded logs).
%Transaction logs are ordered using timestamps.

% SPHT also includes two optimizations for linking transactions in the log, namely forward linking and backward linking.
% These optimizations improve performance of replaying the log.

SPHT has a global \textit{persistent marker} that stores the timestamp of the most recently persisted transaction.
Each thread has its own persistent log, its own timestamp and its own persistent marker.
The thread's timestamp is marked to indicate whether its last hardware transaction has been persisted.
When a thread updates the global persistent marker it also updates its local persistent marker to match its timestamp.

Before beginning a new hardware transaction, the thread updates its timestamp and marks it as not persistent.
During a hardware transaction writes are logged.
At the end of the transaction the thread gets an updated timestamp.
After the hardware transaction ends, the thread's log is written to NVM.
The log must be ordered relative to concurrent transactions.
There is a non-trivial algorithm for negotiating the ordering, involving threads collecting timestamp vectors and blocking until timestamps are marked.
Notably, this means that transactions can be blocked by other concurrent transactions even if they access disjoint data.

\subsubsection{Miscellaneous}
SpecPMT [54] can be used to retrofit a program to ensure crash consistency, assuming the program \textit{already} correctly synchronizes its threads \cite{ye2023specpmt}.
SpecPMT is effectively an optimized form of undo logging.
SpecPMT is neither a STM nor a hybrid TM since it does not offer any method of synchronizing between threads to guarantee atomicity, isolation or consistency (see 4.3.3 in \cite{ye2023specpmt}).
These types of algorithms are only tangentially related to persistent hybrid TMs since TMs solve a much harder problem.
Algorithms like SpecPMT that only guarantee crash consistency cannot be meaningfully compared against TMs.
In fact, SpecPMT would crash on any program in our benchmarks since the benchmarks rely on TM for synchronization.
\section{\tmname~ Design}
\label{sec:design}

In this section we present the details of our two-path persistent HyTM, \tmname~.
We begin with a simple overview of \tmname~.
We then describe the software path utilized by \tmname~ after which we discuss our hardware path and our approach to persisting transactions that execute in hardware.
% Next we present our approach to persisting transactions focusing on our novel technique for persisting hardware transaction.
% Finally we discuss the correctness and progress guarantees of \tmname~.

\begin{figure}[t]
\centering
\begin{lstlisting}[frame=single, language=python]
tid; sRdSet; sWrdSet; pVerNum # Thread local data

def TxRead(addr):
 if found = sWrSet.find(addr): return found.val
 sRdSet.insert(addr, getLock(addr))    
 if !validate(sRdSet): abort()
 return val

def TxWrite(addr, val):
 sWrSet.insert(addr, getLock(addr), val)

def TxCommit():
 if sWrSet.size == 0: return    
 if !acquireLocks(sWrSet): abort()
 if !validate(sRdSet): abort()
 for (addr, val) in sWrSet:<@\label{line:commit1}@>
   <@\color{magenta}{old = *addr;}@> <@\color{magenta}{pA = vmemAddrToPmem(addr)}@>
   <@\color{magenta}{pA.old = old;}@> <@\color{magenta}{pA.pver = \{tid, pVerNum\};}@> <@\color{magenta}{pA.new = val;}@> 
   <@\color{magenta}{flush(\&pA)}@>
   *addr = val
 <@\color{magenta}{pVerNum++;}@> <@\color{magenta}{flush(\&pVerNum);}@>
 releaseLocks(sWrSet)<@\label{line:commit2}@>
\end{lstlisting}
\vspace{-5mm}
\caption{\tmname~ software path. $\color{magenta}{\blacksquare}$ code for persistence.}
\Description{Pseudocode for \tmname~ software path. Code in $\color{magenta}{magenta}$ is needed for persistence.}
\label{code-gl-software}
\end{figure}

\subsection{High Level Overview}
\tmname~ is an O(1)-abortable progressive persistent HyTM that provides durable transactions and guarantees opacity.
\tmname~ aims to overcome the challenges related to persisting transactions that execute in hardware while avoiding the shortcomings of the existing state of the art, SPHT. 
Unlike SPHT which sacrifices concurrency of disjoint transactions to enable persistent hardware transactions, our design allows disjoint transactions to execute concurrently.
Moreover, in a crash-free execution, SPHT has added overhead related to ordering and replaying persistent logs; 
\tmname~ avoids this overhead.
SPHT also relies on a global lock for its software fallback path which completely eliminates concurrency by blocking other transactions whenever the lock is held.
\tmname~ offers a more practical approach utilizing a non-trivial software fallback path that allows transactions on the software path to execute concurrently with transactions on the hardware path.
This is especially important for update heavy workloads which we discuss further in \Cref{sec:eval}.
With these goals and considerations in mind, we now discuss the high level details of \tmname~.

\tmname~ is a word based TM with a hardware fast path and a software fallback path.
\tmname~ attempts transactions a fixed number of times in hardware before falling back to a progressive software path, meaning \tmname~ is O(1)-abortable progressive.
We utilize versioned locks to protect transactional addresses. 
These fine grained locks serve a dual purpose in \tmname~.
Broadly speaking, the fine grained locks are used to guarantee consistency by ensuring threads synchronize on the locks before modifying or reading a transactional address.
% We discuss this in more detail below.
The fine grained locks are also used to enable durable transactions.
Our persistence mechanism exploits the fact that a transactional address can be non-durable only while its corresponding lock is held.
% \tmname~ requires that a thread hold the lock corresponding to any transactional address that it needs to modify and a thread must validate the version of the lock corresponding to any address that it needs to read.

% One would likely find it unsurprising that this is true for all transactions on the software path. 
% However, it is perhaps counterintuitive that this is also true for transactions on the hardware path.

\begin{figure*}[!t]
    \centering
    \includegraphics[width=0.6\linewidth]{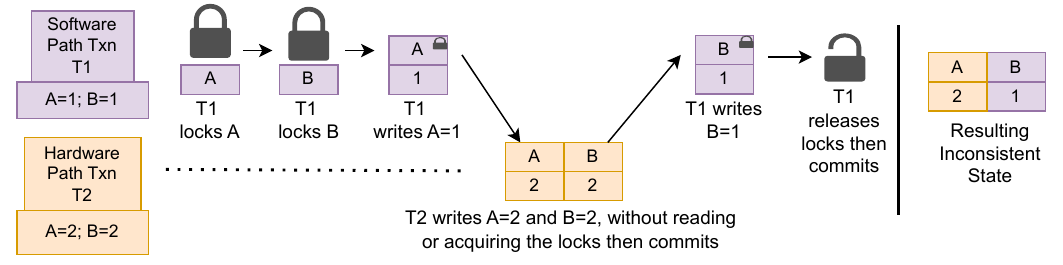}
    \vspace{-4mm}
    \caption{Execution of a HyTM with a lock based software path and hardware path that ignores metadata leading to an inconsistent state. 
    This execution violates opacity because the resulting state could not be produced by any sequential execution.}
    \Description{Execution of a HyTM with a lock based software path and hardware path that ignores metadata leading to an inconsistent state. 
    This execution violates opacity because the resulting state could not be produced by any sequential execution.}
    \label{fig:htm-metadata}    
\end{figure*}

\begin{figure*}[t]
    \centering
    \includegraphics[width=0.9\linewidth]{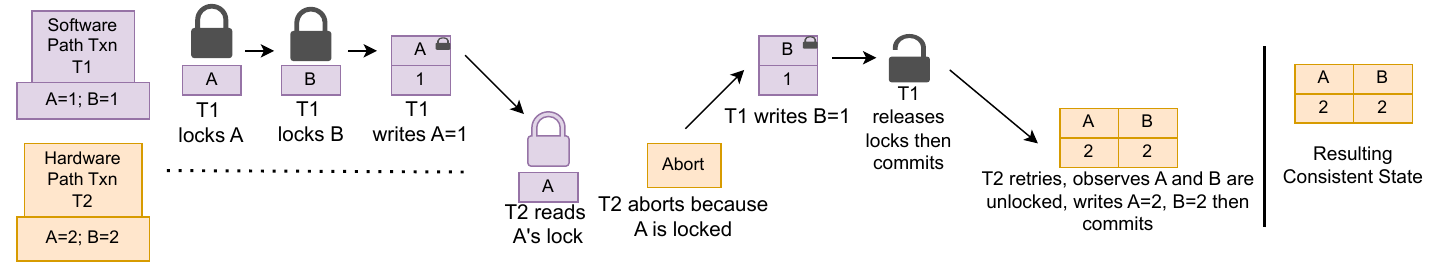}
    \vspace{-4mm}
    \caption{Example execution of a corrected version of the HyTM shown in \Cref{fig:htm-metadata} where, in the volatile setting, instrumenting the hardware path to read shared metadata is sufficient to guarantee opacity.}
    \Description{Example execution of a corrected version of the HyTM shown in \Cref{fig:htm-metadata} where, in the volatile setting, instrumenting the hardware path to read shared metadata is sufficient to guarantee opacity.}
    \label{subfig:htm-volatile-opacity-only}
\end{figure*}

\subsection{Software Path Transactions}
The software path of \tmname~ instruments both reads and writes.
The pseudocode for the software path of \tmname~ is shown in \Cref{code-gl-software}.
Throughout this pseudocode, we omit the implementations of various simple helper functions when the implementation can be inferred from the name.

During a software path transaction, threads track read and write sets which, for each access, log the address and its associated lock version.
\tmname~ utilizes deferred (commit time) locking for writes on the software path meaning that the actual writes are buffered until commit time.
Prior to commit time while accumulating the write set, the thread will check to ensure that the lock corresponding to the address that it plans to write is unlocked, aborting otherwise.

Transactions also maintain a read set.
The read set is revalidated on each read, and the transaction aborts if validation fails.
Validation can fail if an address in the read set is currently locked (by a different thread) or if its associated lock version has changed since the thread first encountered it.

At commit time, threads on the software path attempt to acquire their write set locks.
Each entry in the write set will contain the encounter time version of the lock, that is, the version that the thread observed when it first encountered the associated address during the transaction.
A thread acquires a lock by attempting a \textit{compare and swap (CAS)} on the lock's version to swap it to one greater than the encounter time version.
% When acquiring a lock, the thread first needs to verify that the lock's version matches the version it first encountered during the transaction. 
% If the version has not changed, then the thread can acquire the lock by incrementing its version.
Once all of the write set locks have been acquired, the thread revalidates its read set. 
If successful, the thread performs the writes, then releases the write set locks by incrementing the lock versions a second time.

\paragraph{Persisting Software Transactions}
The effects of all committed transactions need to be persisted.
A transactional address modified by a software path transaction will be non-durable for some period of time after it is written to and before it is written back to persistent memory.
The difficulty in persisting software transactions is primarily related to the necessary synchronization in volatile memory.
More specifically, we need to ensure that threads are blocked when attempting to access non-durable data. 
With correct synchronization, the process of persisting the data is straightforward.

Since writes on the software path are buffered until the thread has claimed all of its write set locks, a transactional address will become non-durable at some point while the address is locked (and the transaction is guaranteed to commit).
If we ensure that the all addresses in the write set are written back to persistent memory before we release the write set locks then we can guarantee that other threads will always observe durable data (so long as they respect the address locks). 
This approach only adds overhead to the commit phase of software transactions that write.
% With this approach, from the perspective of other concurrent transactions, the effects of any one software path transaction will appear to be made durable atomically in the same way that the accesses of transaction itself appear to execute atomically.

While we hold the write set locks, we utilize the undo logging technique of Trinity to persist the write set.
Note that the \textit{back} and a version number fields used by are only required for recovery and do not need to be stored in volatile memory.
With this approach each thread has its own (thread local) sequence number.
We will refer to this as the thread's \textit{persistent version number}.
Persisting the write set follows the same steps as in Trinity.
We reiterate these steps below.

When applying the write set logs of a transaction we iterate over each entry in the write set and do the following: 
first, we get the persistent memory address corresponding to the entry.
This gives us the address of the cache line that we need to modify and write back to persistent memory.
Next, we update this cache line by writing the old value, a tuple combining the thread's persistent version number and thread ID, and the new value.
We need to combine the thread ID and the thread's persistent version number since multiple threads might have the same version.
We then flush the cache line after which we write the new value in volatile memory.
After persisting the write set, we increment the thread's persistent version number and flush it to NVM.

\subsection{Persisting Hardware Transactions}
\smalledit{Before we describe the hardware path of \tmname~, it is helpful to understand the minimum synchronization and instrumentation needed by the hardware path in order to correctly allow for persisting hardware transactions.}
The amount of required hardware path instrumentation differs between the volatile and persistent setting.

In the volatile setting, it is known that for a (C-abortable) progressive HyTM guaranteeing opacity requires the hardware path to access shared metadata \cite{brown2017cost}.
\Cref{fig:htm-metadata} shows an example execution of a HyTM with a software path which relies on metadata in the form of fine grained locks as in the software path of \tmname~.
The hardware path in this example HyTM does not access any metadata.
As a result of this, the given execution results in an inconsistent state violating opacity.
If we want a progressive HyTM that guarantees opacity then transactional accesses on the hardware path must be instrumented in some manner.

\begin{figure*}[t]
    \centering
    \includegraphics[width=0.9\linewidth]{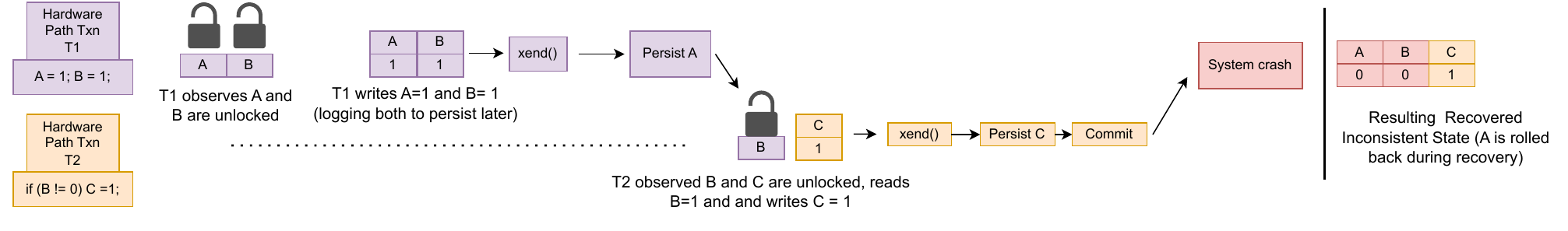}
    \vspace{-4mm}
    \caption{Example execution of the same HyTM as \Cref{subfig:htm-volatile-opacity-only} demonstrating that instrumenting the hardware path to read shared metadata is not sufficient to guarantee opacity in the persistent setting.}
    \Description{Example execution of the same HyTM as \Cref{subfig:htm-volatile-opacity-only} demonstrating that instrumenting the hardware path to read shared metadata is not sufficient to guarantee opacity in the persistent setting.}
    \label{subfig:htm-no-durability}
\end{figure*}

The obvious next question is how much instrumentation is necessary?
The execution in \Cref{fig:htm-metadata} specifically demonstrates a need for instrumenting the hardware path to read shared metadata.
\Cref{subfig:htm-volatile-opacity-only} depicts the corrected example HyTM which guarantees opacity by forcing the hardware path to read the locks.
In the volatile setting, there is no need for the hardware path to write to metadata.
This is because a hardware path transaction that writes to some address $X$ will be aborted if any other concurrent transaction (hardware or software) also writes to $X$.
For any hardware path transaction, in any execution, the hardware path transaction will commit iff any conflicting writes occur strictly before the hardware transaction begins or strictly after it commits.
Either way it is easy to construct an equivalent sequential execution meaning opacity will not be violated.
Note that so far we have ignored the possibility of crashes.

In the persistent setting, we require more hardware path instrumentation.
There are two challenges related to persisting the effects of transactions that execute in hardware.
\smalledit{The first is the fact that current implementations of flush instructions force hardware transactions to abort.}
As a result of this, the effects of a hardware transaction can only be written back to persistent memory after the hardware transaction has completed (i.e. after it successfully returned from xend()).
This inherently requires hardware transactions to keep track of a write set, otherwise we would not know which addresses need to be written back to persistent memory.
However, this represents instrumentation to write to private metadata. 

Correctly persisting hardware path transactions is not as simple as tracking a write set. 
\Cref{subfig:htm-no-durability} shows an example execution of a HyTM demonstrating that, in the persistent setting, instrumenting the hardware path to read shared metadata is not sufficient to guarantee opacity after a crash.
This execution specifically depicts a problem that arises when hardware path transactions do not utilize any mechanism to synchronize persisting modified addresses after the hardware transaction has completed. 
Avoiding this requires the hardware path to write to shared metadata.
With this in mind we can now discuss the hardware path of \tmname~ which uses hardware assisted locking.

\begin{figure}[t]
\centering
\begin{lstlisting}[frame=single, language=python]
def TxRead(addr):
 <@\color{cyan}{lock = getLock(addr)}@><@\label{glock}@>
 <@\color{cyan}{if isLocked(lock) and lock.owner != tid:}@> <@\color{cyan}{xabort()}@>
 return *addr

def TxWrite(addr, val):
 <@\color{cyan}{if !htmAcquireLock(getLock(addr): xabort()}@> 
 <@\color{cyan}{hWrSet.insert(addr, *addr)}@>
 *addr = val

def TxCommit():
 xend()
 <@\color{magenta}{for (addr, old) in hWrSet:}@>  
  <@\color{magenta}{pA = vmemAddrToPmem(addr)}@>
  <@\color{magenta}{pA.old = old;}@> <@\color{magenta}{pA.pver = \{tid, pVerNum\};}@> <@\color{magenta}{pA.new = *addr;}@>
  <@\color{magenta}{flush(\&pA);}@>
 <@\color{magenta}{pVerNum++}@> <@\color{magenta}{flush(\&pVerNum);}@> <@\color{magenta}{releaseLocks(hWrSet)}@>
\end{lstlisting}
\vspace{-5mm}
\caption{\tmname~ hardware path. $\color{cyan}{\blacksquare}$ code for opacity and persistence. $\color{magenta}{\blacksquare}$ code for persistence only.}
\Description{Pseudocode for \tmname~ software path.}
\label{code-gl-hardware}
\end{figure}

\subsection{Hardware Assisted Locking}
The hardware path of \tmname~ is shown in \Cref{code-gl-hardware}.
As we have already mentioned addresses cannot be flushed inside of a hardware transaction.
To this end, \tmname~ instrument writes to keep track of all addresses modified by the hardware transaction using a thread-local append-only log.
After the hardware transaction has completed, we can persist all of the addresses in the log since, the transaction will no longer be executing in hardware so we are free to flush as needed.

\begin{figure*}[!t]
    \centering
    \includegraphics[width=0.95\linewidth]{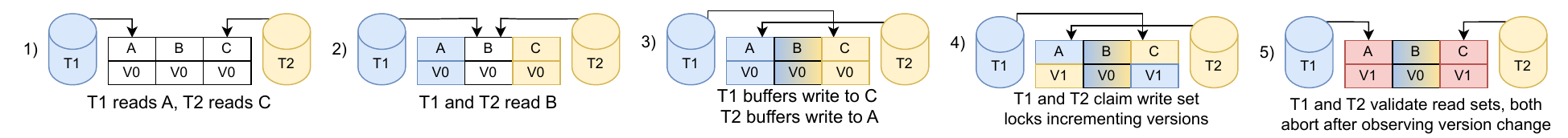}
    \vspace{-4mm}
    \caption{The need for strong progressiveness: A partial execution of the O(1)-abortable (weakly) progressive \tmname~ where two transactions will abort forever.}
    \Description{The need for strong progressiveness: A partial execution of the O(1)-abortable (weakly) progressive \tmname~ where two transactions will abort forever.}
    \label{fig:strong-prog}
\end{figure*}

As with the software path, the difficulty in persisting the modified addresses is the synchronization needed to ensure that other threads do not access non-durable data and to ensure that the state of persistent memory remains consistent.
Addresses modified by a hardware transaction will be non-durable immediately after the hardware transaction completes.
Moreover, there is a theoretically infinite amount of time between the hardware transaction completing and the next instruction performed by the thread.
This means that addresses modified by a hardware transaction need to be protected after the hardware transaction has completed.
This also means that whatever mechanism we utilize to prevent concurrent threads from accessing data modified by our hardware transaction must be activated during the hardware transaction.
On the software path, we utilize the fine grained locks to protect addresses while they are non-durable.
This raises the question, why not apply a similar approach on the hardware path?
This leads to our approach to persisting hardware transactions: hardware assisted locking.

\begin{figure}[t]
\centering
\begin{lstlisting}[frame=single, language=python]
global gClock ; thread_local rClock  #new fields

def htmAcquireLock(lk):         #H/W Path Changes
 if isLocked(lk): return false
 lk.sLockVer++; lk.hLockVer++;<@\label{line:hlock}@>
 return true
 
def TxStart(): rClock = gClock  #S/W Path Changes

def TxCommit():
 if sWrSet.size == 0: return
 sort(sWrSet)
 if not acquireLocks(sWrSet): abort()
 if CAS(gClock, rClock, rClock+1):
  if foundHtxConflict(sRdSet): abort()
 elif not validate(sRdSet): abort()
 #... same as <@\ref{line:commit1}@>-<@\ref{line:commit2}@> in <@\Cref{code-gl-software}@>  
\end{lstlisting}
\vspace{-5mm}
\caption{Pseudocode for the O(1)-abortable strongly progressive version of \tmname~.}
\Description{Pseudocode for the O(1)-abortable strongly progressive version of \tmname~.}
\label{code-sp}
\vspace{-4mm}
\end{figure}

Our hardware path further instruments writes to acquire the associated fine grained lock. 
Before acquiring a lock, the transaction must confirm that it is unlocked or locked by the current thread.
% Since we are in a hardware transaction, acquiring a lock can be done via a write.
Importantly, during the hardware transaction, the thread will only acquire locks.
This means that the addresses will remain locked even after the hardware transaction completes.
After a hardware transaction completes, with the addresses locked, we proceed as in the software path by persisting the write set, updating the thread's persistent version number and releasing the locks in that order.

The hardware path of \tmname~ also instruments reads to check that the associated lock is unlocked (or locked by the current thread).
This is necessary both to ensure that hardware path transactions and software path transactions can execute concurrently without violating opacity and to ensure that hardware transactions do not observe non-durable data.
One would likely find this counterintutive since, in a volatile only setting, we would prefer to avoid instrumenting reads on the hardware path.
However, in a non-volatile setting, we have limited other options.
Allowing hardware transactions to claim locks and hold them after the hardware transaction completes necessitates instrumenting hardware path reads.
This is true even if we disallow concurrency with software path transactions (for example by using a global lock fallback path like SPHT).
One could instead choose an approach like SPHT.
This introduces an interesting trade-off.
One can trade instrumenting hardware path reads and a non-trivial software path (\tmname~) for uninstrumented reads, allowing disjoint transactions to block each other and a trivial global locking fallback path (SPHT). 
% With even a moderate amount of transactional writes the latter (SPHT) introduces significant overhead.
We argue that the former (hardware assisted locking) is preferable.

\subsection{Recovery}
Since we persist individual addresses via Trinity's persistence mechanism, our recovery procedure follows the same approach as in Trinity.
This involves traversing persistent memory and reverting to their old value any addresses that have a version number greater than the version number of the thread that last modified the address.

\subsection{Correctness and Progress}
\label{sec:correctness}

\begin{theorem}
\tmname~ offers durable transactions, guarantees opacity and O(1)-abortable weak progressiveness.    
\end{theorem}

Aborts on the software path of \tmname~ occur when a transaction fails to claim its write set locks or fails to validate its read set.
Both of these scenarios can occur iff some other concurrent transaction has claimed a lock meaning the transactions conflict.
In these scenarios there is no guarantee that one of the conflicting transactions will commit so the software path is progressive.
Since \tmname~ attempts transactions a fixed number of times in hardware, the means \tmname~ guarantees O(1)-abortable weak progressiveness.

\tmname~ guarantees opacity since it is prevents leaking any information from aborted transactions.
No transaction can observe the effects of a concurrent transaction that aborts since all modifications are made only once the transaction has claimed its write set locks and is guaranteed to commit.

As in Trinity, following a crash an address that was last updated by a committed transaction will have a version number that is less than the persistent version number of the owning thread (all other addresses will be reverted to their old values during recovery).

\paragraph{Guaranteeing C-Abortable Strong Progressiveness}
It is possible for a progressive TM to experience live locking.
Consider a simple array being accessed by two threads as shown in \Cref{fig:strong-prog}.
Here, we have one transaction, $T_1$, reads all elements in ascending index order then updates the last element while another concurrent transaction, $T_2$, reads all elements in descending index order and updates the first element.
A C-abortable progressive TM like \tmname~ would consistently abort both transactions since both $T_1$ and $T_2$ will consistently fail to validate their read sets after acquiring all of their write set locks.
Similarly, if both $T_1$ and $T_2$ wanted to update every element neither would be able to acquire their write set locks since they will eventually need a lock that the other has claimed.
A C-abortable strongly progressive TM avoids these issues.

Guaranteeing O(1)-abortable strong progressiveness requires our software path to be strongly progressive.
The latter scenario where a C-abortable progressive TM can live lock can be avoided by acquiring write set locks in a fixed order similar to TL2.
% This requires sorting the addresses in the write set.
In the former scenario, the two transactions are both about to claim their write set locks but both abort.
% Note that if we acquire the locks in a fixed order, a concurrent software path transaction that has not acquired all of its write set locks cannot possibly conflict.
To avoid this, we can use a global \textit{clock}.
Transactions start by reading the clock.
The clock is incremented after acquiring the write set locks.
Read set validation is not necessary if, after the increment, the clock is exactly one greater than the value read at the start of the transaction, (i.e. no concurrent transactions have performed any writes).
% (other than acquiring locks which can be released without issue).

In our case, the use of a global clock on the software path is not too concerning, however, this would be a major bottleneck if we were to increment a shared global on the hardware path.
To avoid this, we instead change the definition of the fine-grained locks to include two version numbers.
The first, which we will call \texttt{sLock}, remains the same.
It is incremented whenever a transaction (software or hardware) acquires or releases the lock.
The added version number, which we will call \texttt{hLock}, is incremented only by hardware transactions.
Software transaction can use the \texttt{hLock} version to identify conflicts with concurrent hardware transactions.

To summarize, we utilize a global clock on the software path where, at the beginning of the transaction the thread reads the clock and, after acquiring its write set locks, attempts a CAS to advance the clock by one greater than the value it read at the start.
If the CAS is successful, the transaction must check for conflicts with concurrent hardware transactions by iterating over its read set to confirm that the \texttt{hLock} versions have not changed.
If the CAS fails then the transaction performs standard read set validation, checking that the \texttt{sLock} versions have not changed.
This gives us a O(1)-abortable strongly progressive version of our persistent HyTM.
We refer to this version as \tmname~-SP (\underline{S}trong\underline{P}rogressive).
We show the pseudocode for the changes compared to \tmname~ in \Cref{code-sp}.
For the hardware path, the only difference is the addition of line \ref{line:hlock}.

\section{Implementation Details}
\label{sec:impl}
In this section we discuss some relevant technical details of \tmname~.
We begin by discussing how we allocate and manage memory within transactions.
% We then discuss our usage of fine-grained locks.

\subsection{Memory Allocation in Transactions}
%Memory allocation within a transaction requires a safe concurrent memory allocator.
Memory management within a TM is an important and challenging task. 
The allocation and freeing of memory needs to be directly tied to transaction commits and aborts.
Otherwise a transaction that aborts could leak memory or free memory that is still in use.

One could choose to ignore the problem simply by never freeing memory.
Allocations within transactions that abort would leak memory and any committed deallocations are effectively no-ops which also leak memory.
Obviously this is not ideal, but it is not too uncommon to defer solving the problem of safe memory reclamation.
In fact, SPHT leaks memory since it never frees memory.
This is particularly problematic when using NVM since memory leaks are essentially persistent forever.
If we wanted to utilize this algorithm in practice we would need to solve the problem of correct memory reclamation.

There are two approaches to actually managing memory within transactions.
One method is to utilize an allocator where the internal state is accessed via transactions (meaning the TM is used to implement the allocator).
This is the approach taken by Trinity.
With durable transactions, this strategy allows one to recover the allocator state following a crash.
However, this approach increases the write set size of transactions and can also cause additional data conflicts.
For HyTM algorithms like \tmname~, this can lead to increased transaction aborts, especially on the hardware path.

\paragraph{Quantifying the Cost of Trinity's Allocator}
\label{para:alloc-problem}
We attempt to quantify the additional cost imposed by accessing allocator metadata through transactions in Trinity.
One might think that the best way to quantify the cost would be to compare Trinity with another version in which we utilize a different allocator that does not rely on transactions; however, using a different allocator would lead to changes in the memory layout, which can drastically affect performance. 
Instead, we constructed some controlled workloads and used the authors publically available implementation of Trinity combined with TL2 (TrinityVRTL2 in \cite{ramalhete2021efficient}) to extract some easily quantifiable metrics that summarize the cost of Trinity's approach to memory management.

The workloads that we utilize perform the following common actions:
First, threads execute transactions to populate an array of pointers to 16-byte objects.
Each thread will access a randomly generated and disjoint subset of the array.
When accessing an element in the array, a thread executes a single transaction which will perform 3 transactional stores.
One store writes the pointer into the array and two stores update the data fields of the object.
We refer to these types of transactions as \textit{type A transactions} (for \underline{A}lloc).
Once the array has been populated, the threads access a new randomly generated and unique subset of the array.
At each access a thread will again perform 3 transactional stores.
Two stores update the data fields of the object and one store removes the object from the array by replacing the pointer with \texttt{null}.
We refer to these types of transactions as \textit{type F transactions} (for \underline{F}ree).
We utilize an array of 1 million objects, meaning that we will execute 2 million transactions.
It is important to note that since each thread accesses a unique subset of the array, this workload has no data conflicts outside of those that might arise in the allocator.
This means that there we will be 0 aborts excluding those that are caused by accesses made by the allocator (and those that can arise due to collisions in the TL2 lock table).

We use this template to construct 3 different workloads that evaluate Trinity's approach to memory management.
In the first workload, we use Trinity's allocator to perform allocations and deallocations.
In type A transactions, each thread will allocate a new object to insert into the array.
In type F transactions, each thread will free the object that they remove from the array.
This workload captures the full cost of Trinity's approach to memory management.
In the second workload, we utilize Trinity's allocator to perform allocations but we do not perform any freeing.
Here type A transactions will again require threads to allocate a new object but type F transactions are the same as in the template.
In the last workload, we preallocate all objects before the measurement period begins (these allocations still use Trinity's allocator but they do not contribute to our data collection).
In this case, both transaction types are unmodified compared to the template (meaning there are no allocations or frees in any measured transactions).
Note that in all 3 cases, outside of allocator functions, the number of transactional accesses is equivalent.

\begin{table*}[t]
\resizebox{\textwidth}{!}{%
\begin{tabular}{|l|c|crrr|crrr|}
\hline
\multicolumn{1}{|c|}{\multirow{3}{*}{\textbf{\begin{tabular}[c]{@{}c@{}}Trinity Allocator \\ Functions Accessed via \\ Transactions\end{tabular}}}} &
  \multirow{3}{*}{\textbf{\begin{tabular}[c]{@{}c@{}}Average\\ Aborts\end{tabular}}} &
  \multicolumn{4}{c|}{\textbf{Average Data Set Sizes}} &
  \multicolumn{4}{c|}{\textbf{Max Data Set Sizes}} \\ \cline{3-10} 
\multicolumn{1}{|c|}{} &
   &
  \multicolumn{2}{c|}{\textbf{Overall}} &
  \multicolumn{2}{c|}{\textbf{At Abort}} &
  \multicolumn{2}{c|}{\textbf{Overall}} &
  \multicolumn{2}{c|}{\textbf{AtAbort}} \\ \cline{3-10} 
\multicolumn{1}{|c|}{} &
   &
  \multicolumn{1}{c|}{\textbf{Read Set}} &
  \multicolumn{1}{c|}{\textbf{Write Set}} &
  \multicolumn{1}{c|}{\textbf{Read Set}} &
  \multicolumn{1}{c|}{\textbf{Write Set}} &
  \multicolumn{1}{c|}{\textbf{Read Set}} &
  \multicolumn{1}{c|}{\textbf{Write Set}} &
  \multicolumn{1}{c|}{\textbf{Read Set}} &
  \multicolumn{1}{c|}{\textbf{WriteSet}} \\ \hline
\textbf{Both Alloc and Free} &
  \multicolumn{1}{r|}{186338.4} &
  \multicolumn{1}{r|}{4.86} &
  \multicolumn{1}{r|}{6.02} &
  \multicolumn{1}{r|}{7.80} &
  5.68 &
  \multicolumn{1}{r|}{15} &
  \multicolumn{1}{r|}{13} &
  \multicolumn{1}{r|}{13} &
  8 \\ \hline
\textbf{Alloc only} &
  \multicolumn{1}{r|}{10504.5} &
  \multicolumn{1}{r|}{2.50} &
  \multicolumn{1}{r|}{4.48} &
  \multicolumn{1}{r|}{2.86} &
  0.45 &
  \multicolumn{1}{r|}{7} &
  \multicolumn{1}{r|}{9} &
  \multicolumn{1}{r|}{5} &
  6 \\ \hline
\textbf{None} &
  \multicolumn{1}{r|}{2153.7} &
  \multicolumn{1}{r|}{2.50} &
  \multicolumn{1}{r|}{2.99} &
  \multicolumn{1}{r|}{0.55} &
  0.75 &
  \multicolumn{1}{r|}{1} &
  \multicolumn{1}{r|}{3} &
  \multicolumn{1}{r|}{1} &
  3 \\ \hline
\end{tabular}%
}
\caption{Evaluation of the cost of accessing allocator metadata via transactions in Trinity.
Each row shows the results from running one of the synthetic array modification workloads described in \cref{para:alloc-problem} using 64 threads. All workloads execute 2 million transactions. Excluding allocator accesses, there are 0 data conflicts and each transaction performs 3 transactional stores.}
\label{table:allocator}
\end{table*}

\Cref{table:allocator} shows the results of this experiment for the machine described in \cref{sec:eval} using 64 threads. 
In particular, we show the averages of the number of aborts, the average read and write set sizes overall commits and aborts, as well as the maximum read and write set size observed at an abort.
One alarming difference is the significantly larger number of aborts when accessing allocator metadata via transactions. 
\smalledit{We observe nearly a 100x increase in aborts when we utilize the full allocator functionality.}
This result is especially important since these aborts would never happen if the allocator were not accessed transactionally.
We can also see that the average and maximum read and write set sizes are much larger for this workload.
One might expect the third workload where we do not utilize Trinity's allocator to have exactly 0 aborts. 
All aborts in this case are a result of collisions in the TL2 lock tables.
These results demonstrates that there is a significant cost associated with Trinity's memory management strategy.

\paragraph{\tmname~ Memory Management}
\label{sec:memmanagment}
\smalledit{In \tmname~, we utilize a custom memory management system where synchronization is achieved without the use of transactional accesses.}
Our algorithm ensures that memory allocated during a transaction is freed if the transaction ultimately aborts and defers freeing memory at least until the transaction that tried to free has committed.
To achieve this, we utilize a custom memory allocator based on mimalloc \cite{leijen2019mimalloc} combined with epoch-based reclamation to ensure safe memory reclamation.
When a user tries to allocate memory in a transaction, the allocation occurs immediately, but we keep track of these allocated objects so that we can reclaim them if the transaction aborts.
Deallocations on the other hand are handed via epoch based reclamation.
At the start of each transaction, the thread will read a global epoch counter and post its current epoch to a global announcement array.
When a transaction commits, threads can read the global epoch to determine if it is safe to reclaim retired objects and threads can read the announcement array to determine if it is safe to advance the global epoch counter.
This strategy allows us to avoid the performance concerns that exist with Trinity's allocator while still guaranteeing that we do not leak memory.

As is the case with many algorithms that utilize NVM, our allocator initially maps a large contiguous address range in virtual memory, and then distributes it to various threads, on demand.
The use of a contiguous address range allows for a direct mapping between volatile and persistent memory addresses.
Unlike Trinity, the internal state of our allocator is not persisted.
This means that we need to reconstruct the allocator state from scratch during recovery. 
To make this possible, the user must provide an iterator that the allocator can utilize to determine which parts of memory are in use.
Trinity's approach optimizes for faster recovery whereas our approach optimizes for improved performance during regular crash free execution.
We argue that this is a preferable approach since system crashes model power failures, which are typically assumed to occur infrequently. 

\subsection{Fine-Grained Locks}
\tmname~ relies on fine-grained locks.
Similar to TL2, we utilize a fixed size lock table.
This means multiple addresses might map to the same lock, but it ensures that the memory layout of user data is unaffected.
In some cases it is favorable to colocate the locks with user data.
In this case each address has a unique lock.
Depending on the layout of user data, this also allows prefetching locks when caching the associated user data.
Our \tmname~-CL implementation utilizes colocated locks.
We achieve this via the use of custom types.
For languages where custom types are not supported other strategies such as compiler extensions could be used to colocate the locks.
% We evaluate the benefit of colocated locks in \Cref{sec:eval}.

% 
\section{Evaluation}
\label{sec:eval}

\begin{figure*}[t!]
    \begin{subfigure}{0.02\linewidth}        
        \raisebox{0.3\height}{\rotatebox{90}{1 million keys}}
    \end{subfigure}
    \begin{subfigure}{0.97\linewidth}
        \begin{subfigure}{0.24\linewidth}
            \centering
            0\% Read-only        
            \includegraphics[width=1\linewidth]{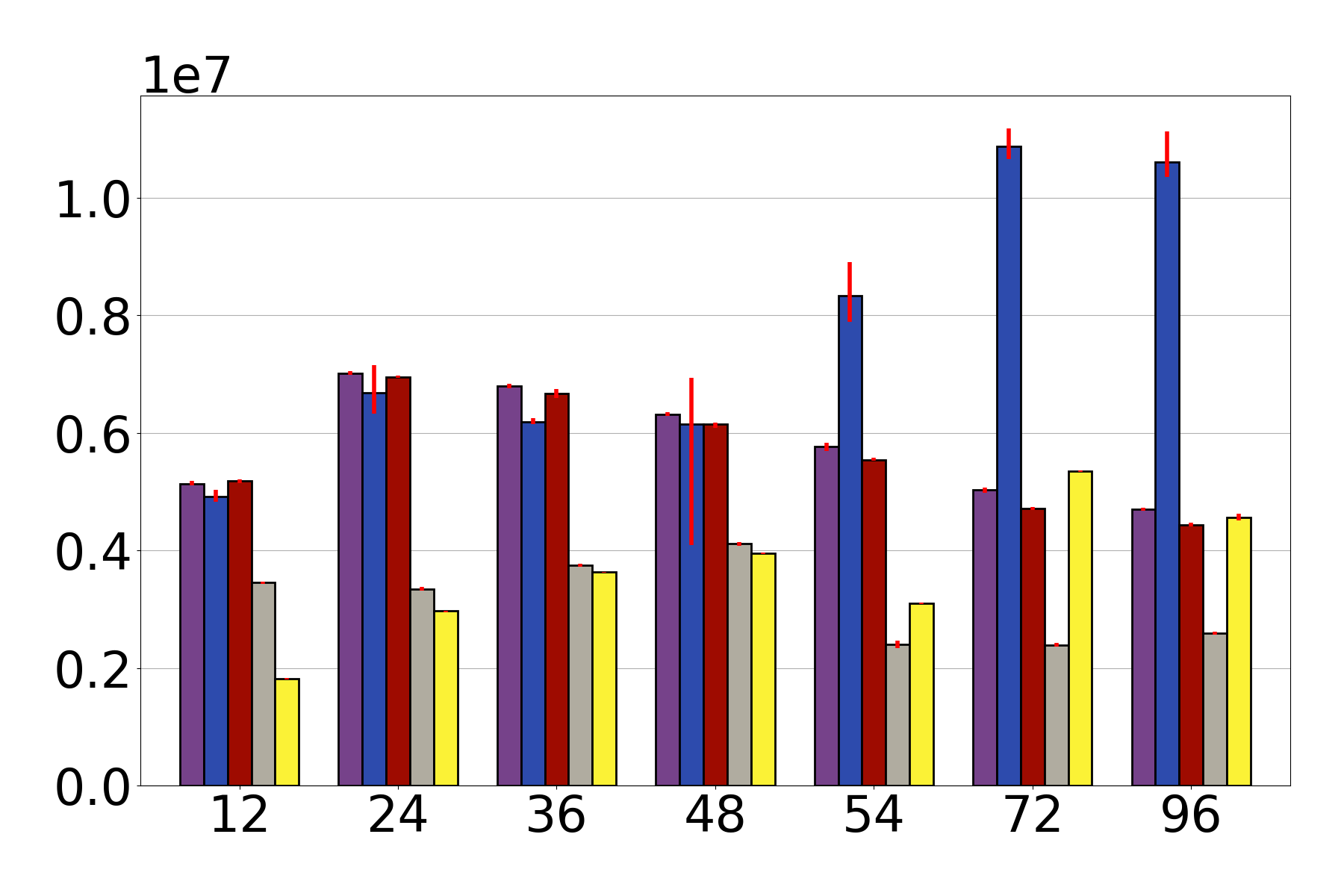}
        \end{subfigure} 
        % \vspace{-4mm}
        \begin{subfigure}{0.24\linewidth}
            \centering        
            50\% Read-only
            \includegraphics[width=1\linewidth]{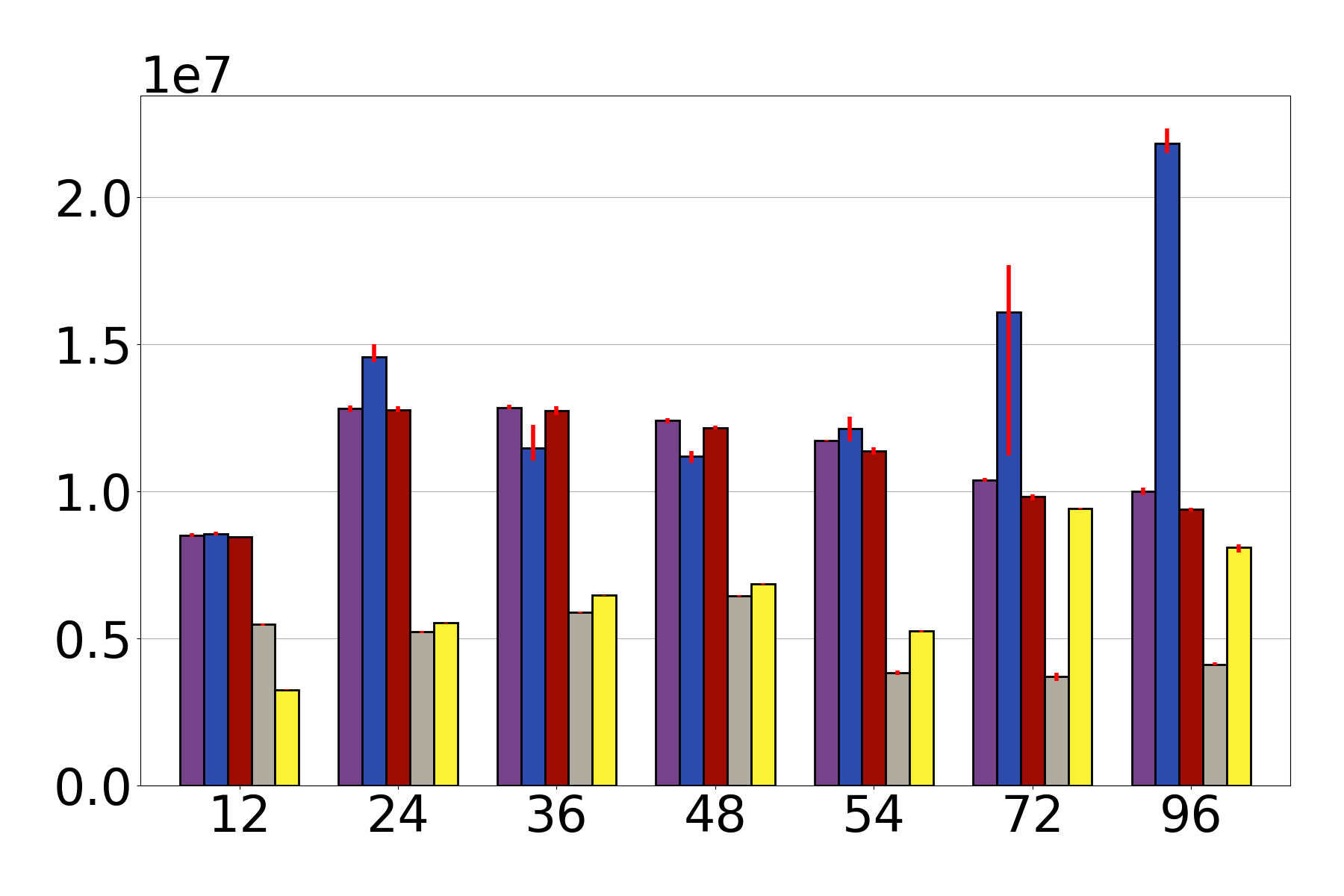}
        \end{subfigure}
        \begin{subfigure}{0.24\linewidth}
            \centering    
            90\% Read-only
            \includegraphics[width=1\linewidth]{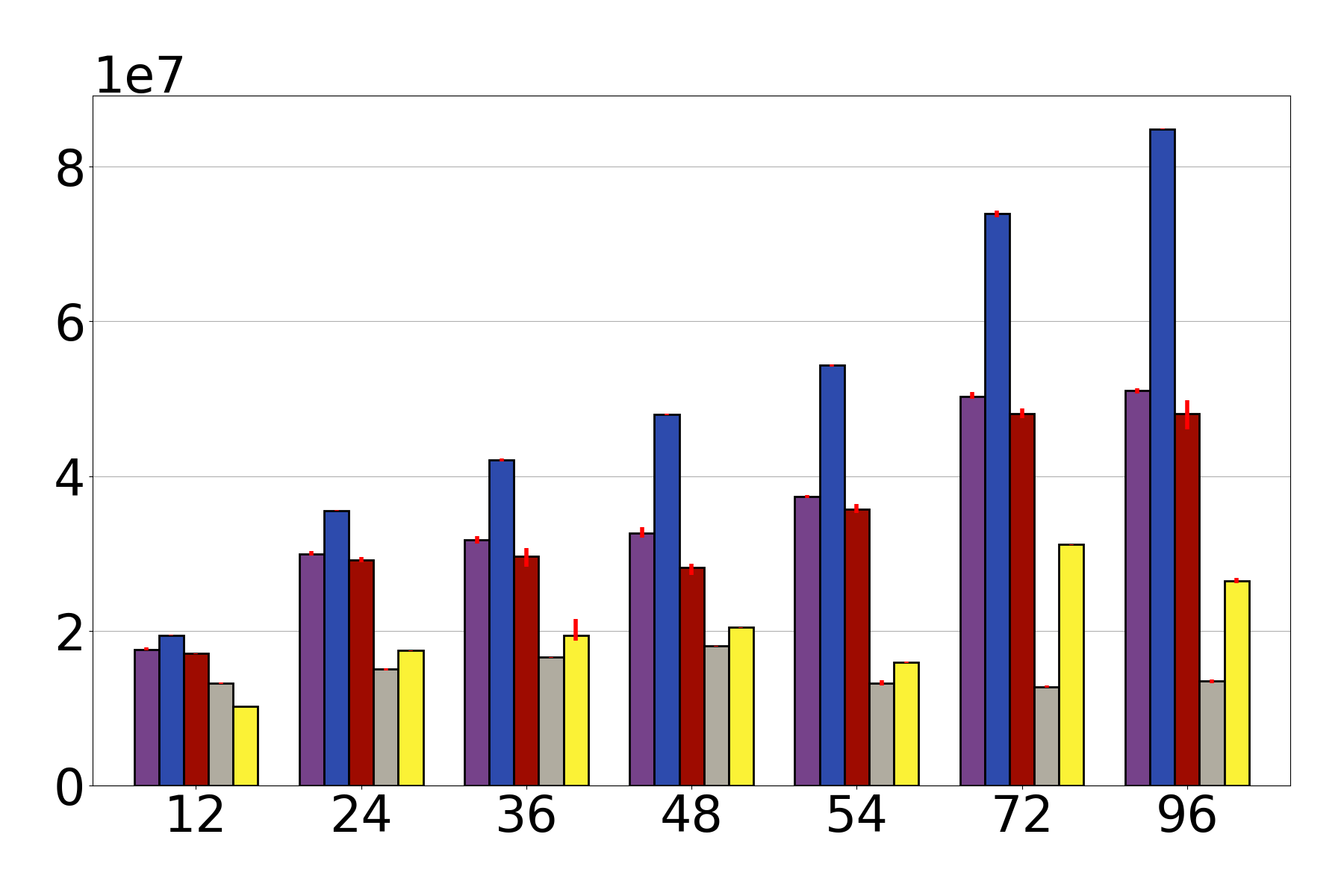}
        \end{subfigure}        
        \begin{subfigure}{0.24\linewidth}
            \centering 
            99\% Read-only
            \includegraphics[width=1\linewidth]{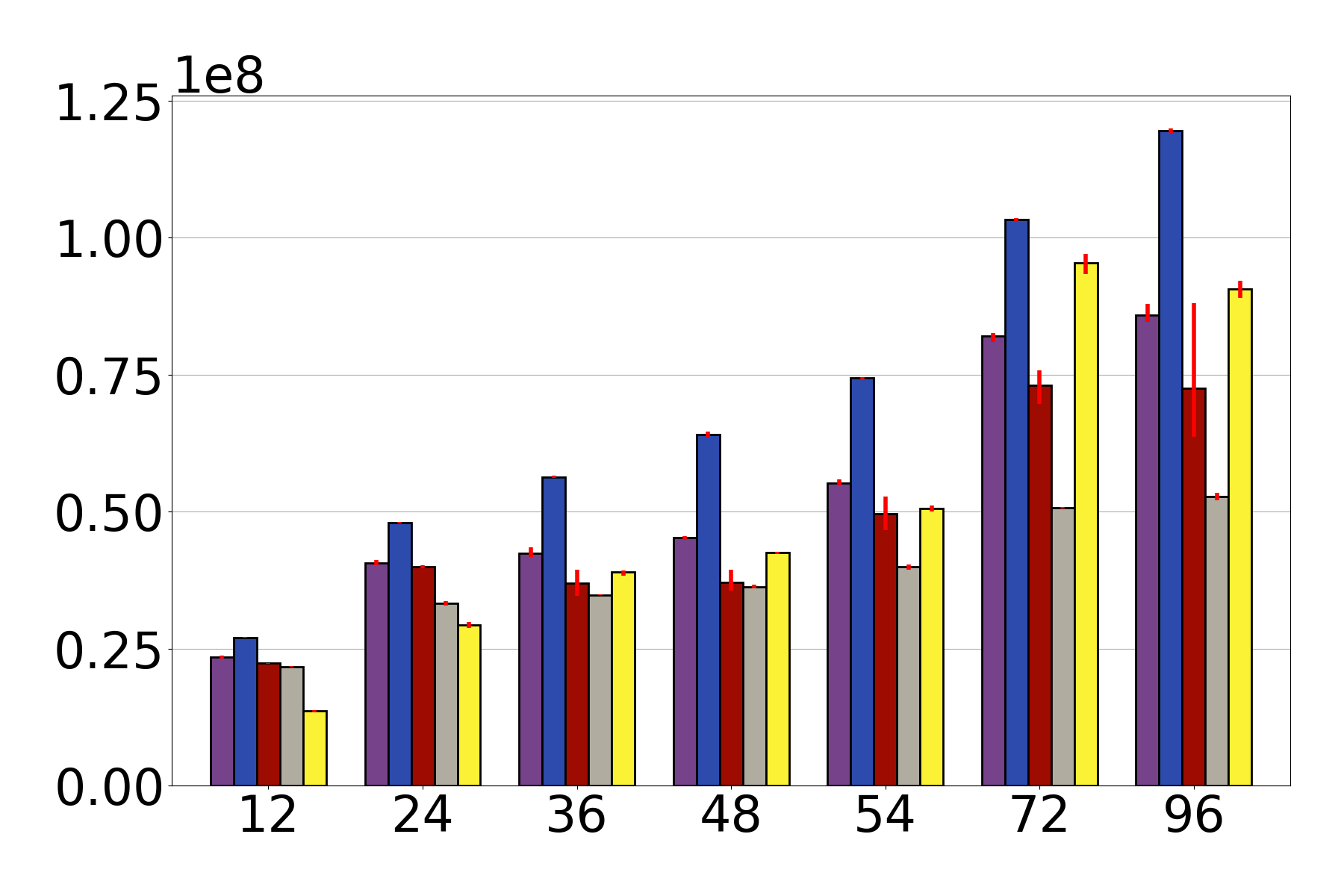} 
        \end{subfigure}     
    \end{subfigure}
    \begin{subfigure}{0.02\linewidth}
        \raisebox{-.5\height}{\rotatebox{90}{\edit{5 million keys}}}
    \end{subfigure}
    \begin{subfigure}{0.97\linewidth}
        \begin{subfigure}{0.24\linewidth}
            \centering
            \raisebox{-.5\height}{\includegraphics[width=1\linewidth]{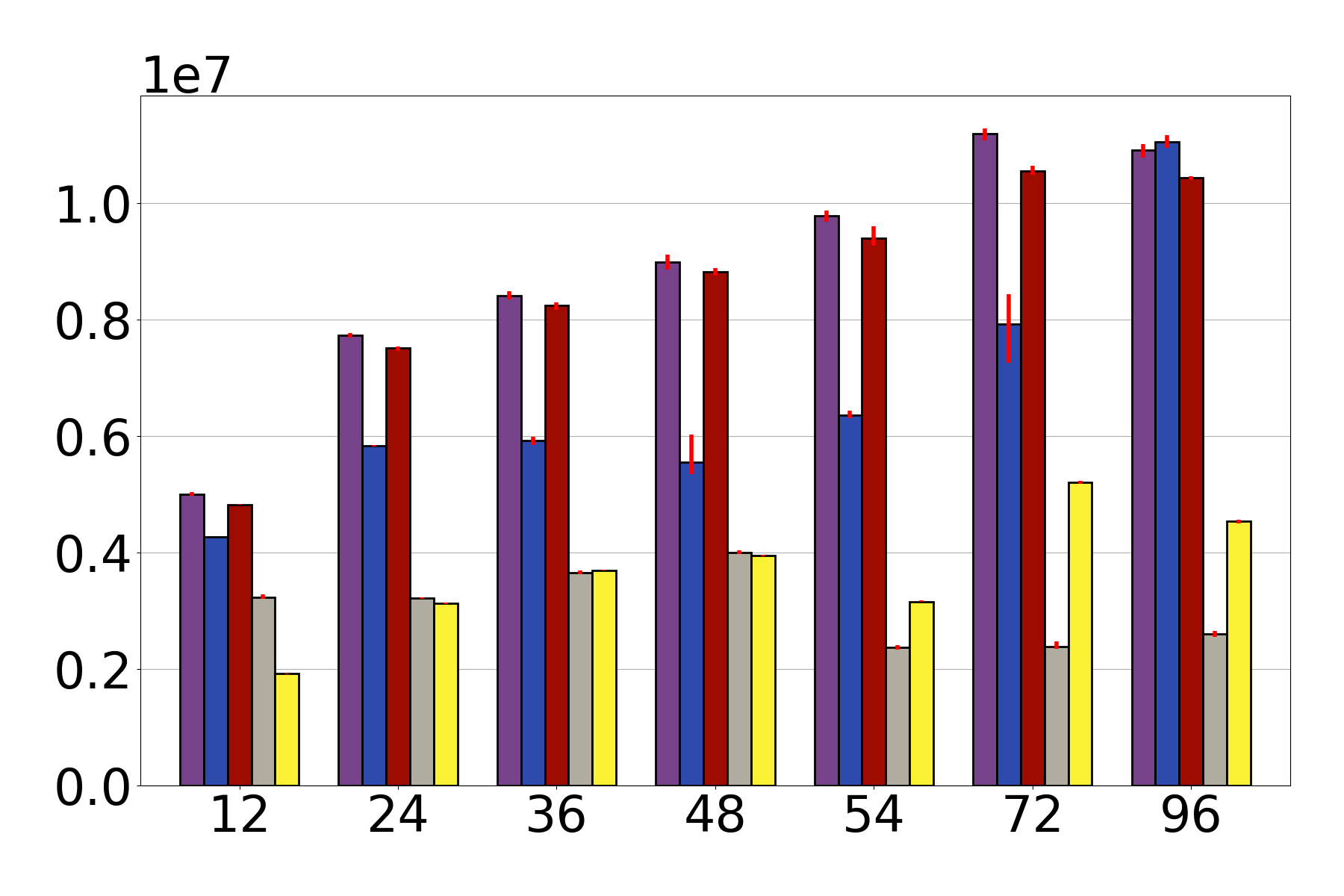}}
        \end{subfigure}
        \begin{subfigure}{0.24\linewidth}
            \centering        
            \raisebox{-.5\height}{\includegraphics[width=1\linewidth]{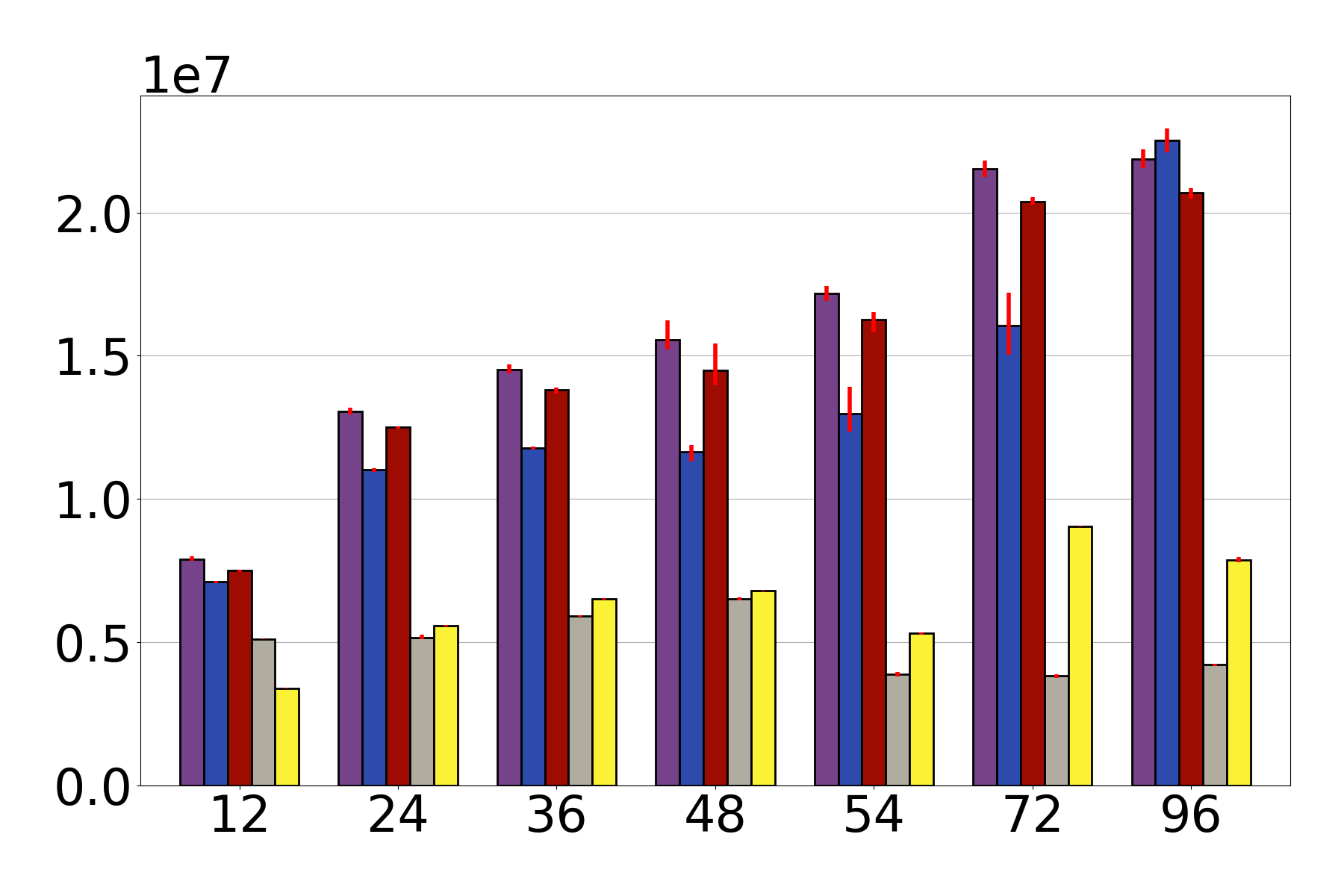}}
        \end{subfigure}
        \begin{subfigure}{0.24\linewidth}
            \centering    
            \raisebox{-.5\height}{\includegraphics[width=1\linewidth]{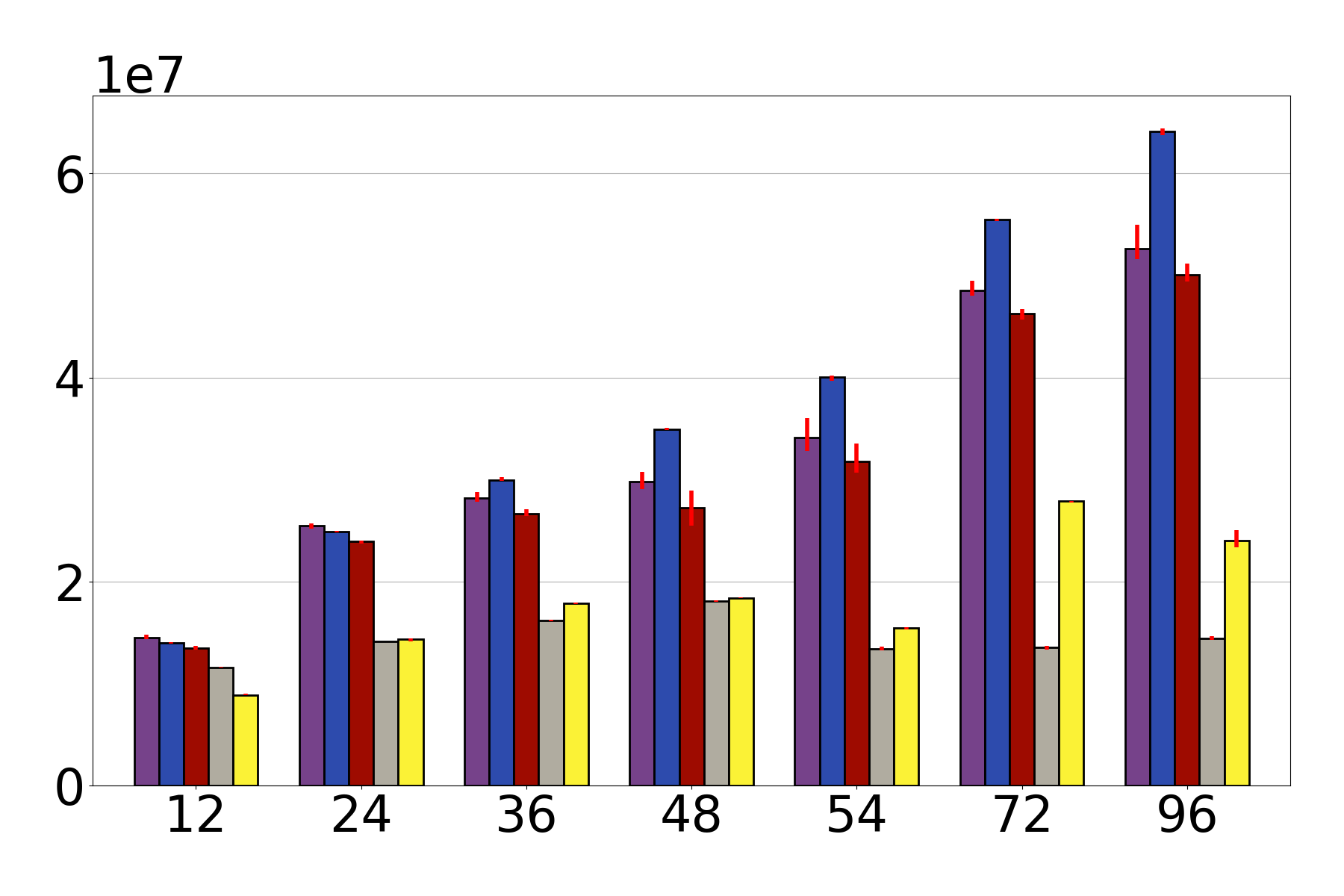}}
        \end{subfigure}        
        \begin{subfigure}{0.24\linewidth}
            \centering     
            \raisebox{-.5\height}{\includegraphics[width=1\linewidth]{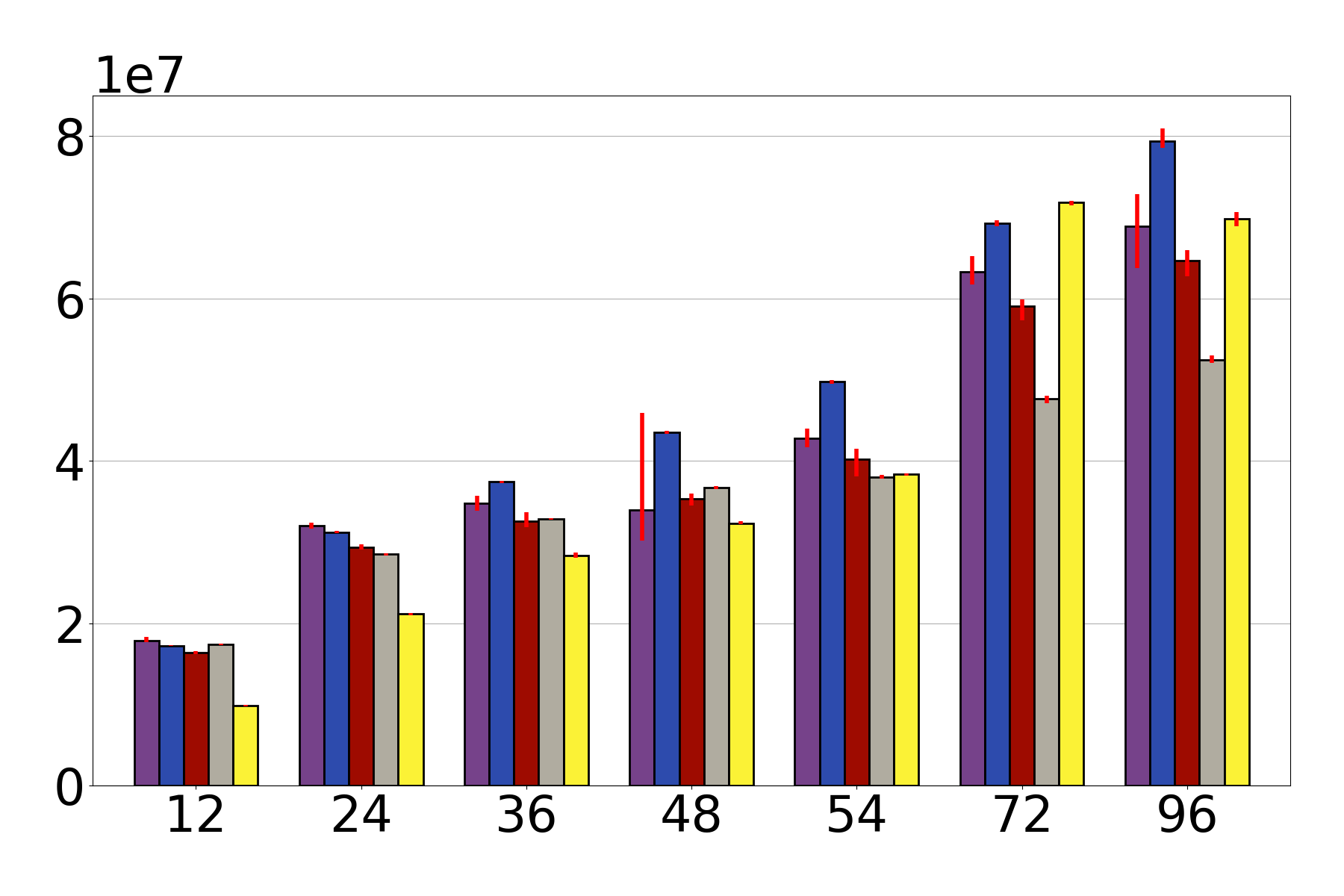}}
        \end{subfigure} 
    \end{subfigure}
    \begin{subfigure}{1.0\linewidth}
        \centering
        \includegraphics[width=0.6\linewidth]{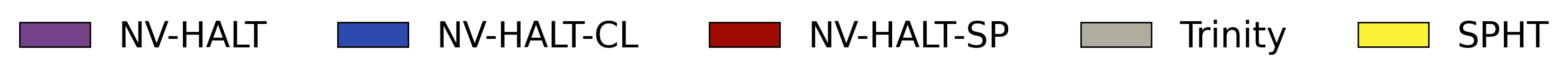}
    \end{subfigure}     
    \vspace{-8mm}
    \caption{\centering Throughput for (a,b)-tree. Y-axis is ops/sec. X-axis is number of threads. Keys are accessed according to a uniform distribution.
    }
    \Description{Throughput for (a,b)-tree. Y-axis is ops/sec. X-axis is number of threads. Keys are accessed according to a uniform distribution}
    \label{fig:throughput-abtree}
\end{figure*}

\begin{figure*}[t!]
    \begin{subfigure}{0.02\linewidth}        
        \raisebox{0.3\height}{\rotatebox{90}{1 million keys}}
    \end{subfigure}
    \begin{subfigure}{0.97\linewidth}
        \begin{subfigure}{0.24\linewidth}
            \centering
            0\% Read-only        
            \includegraphics[width=1\linewidth]{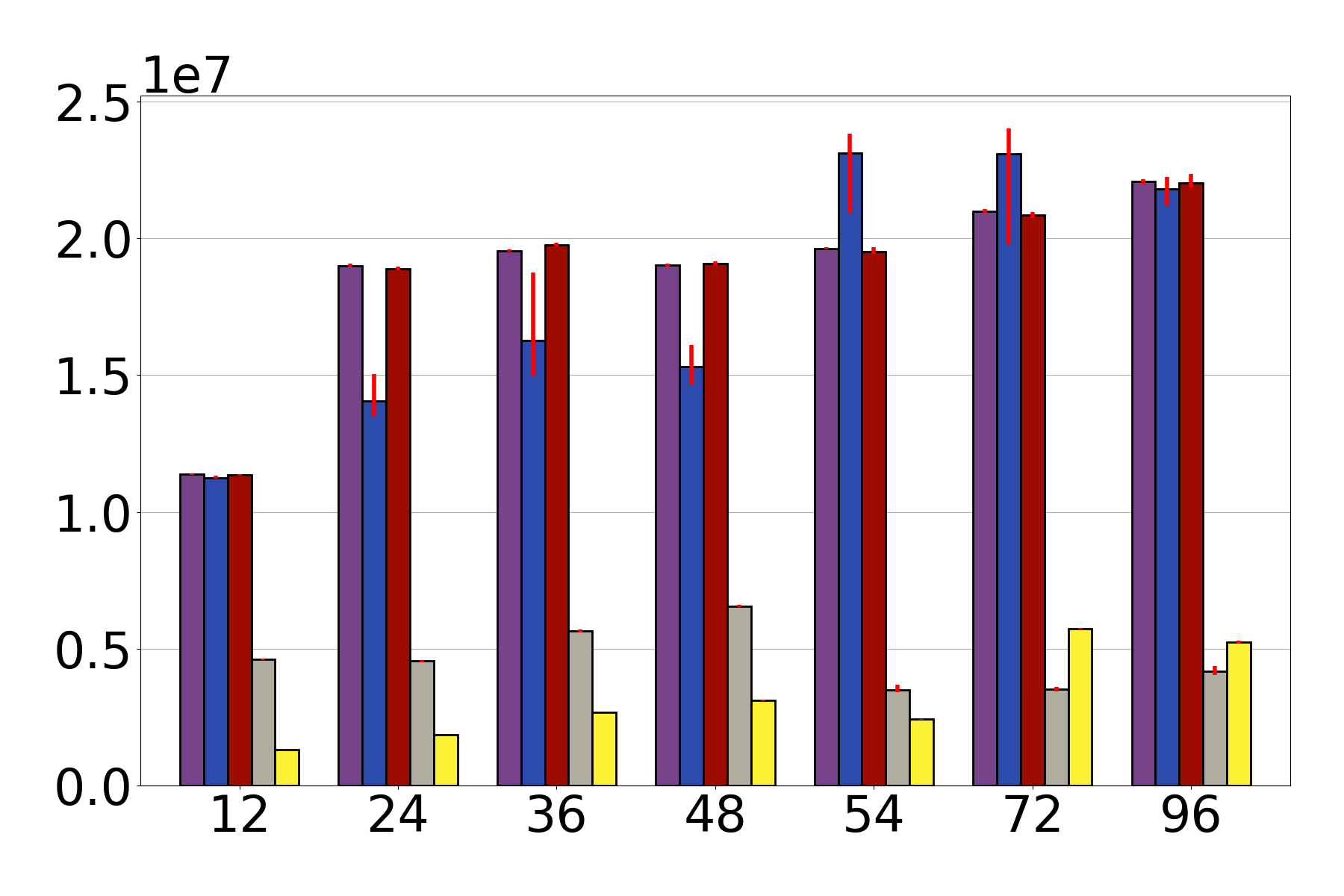}
        \end{subfigure} 
        % \vspace{-4mm}
        \begin{subfigure}{0.24\linewidth}
            \centering        
            50\% Read-only
            \includegraphics[width=1\linewidth]{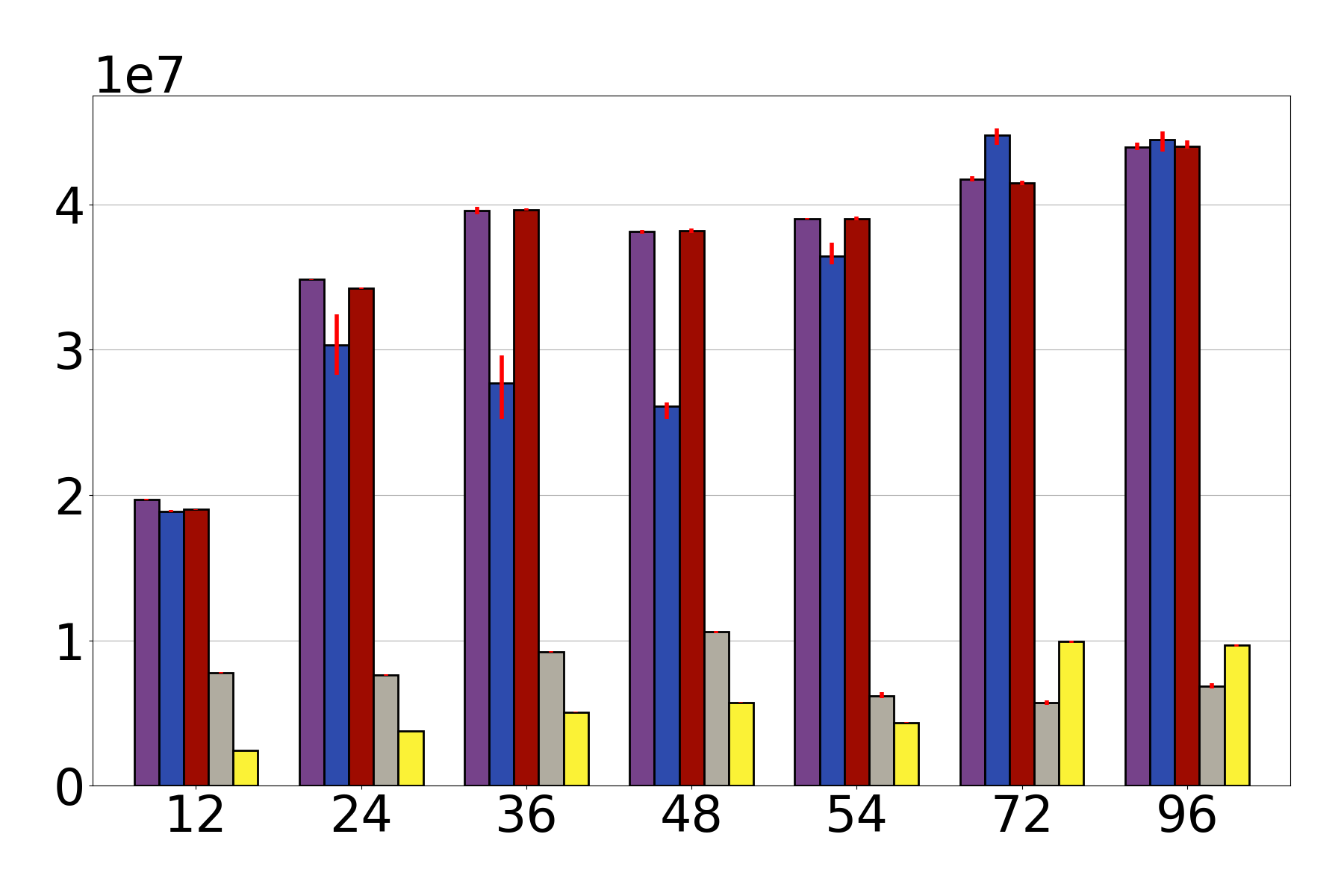}
        \end{subfigure}
        \begin{subfigure}{0.24\linewidth}
            \centering    
            90\% Read-only
            \includegraphics[width=1\linewidth]{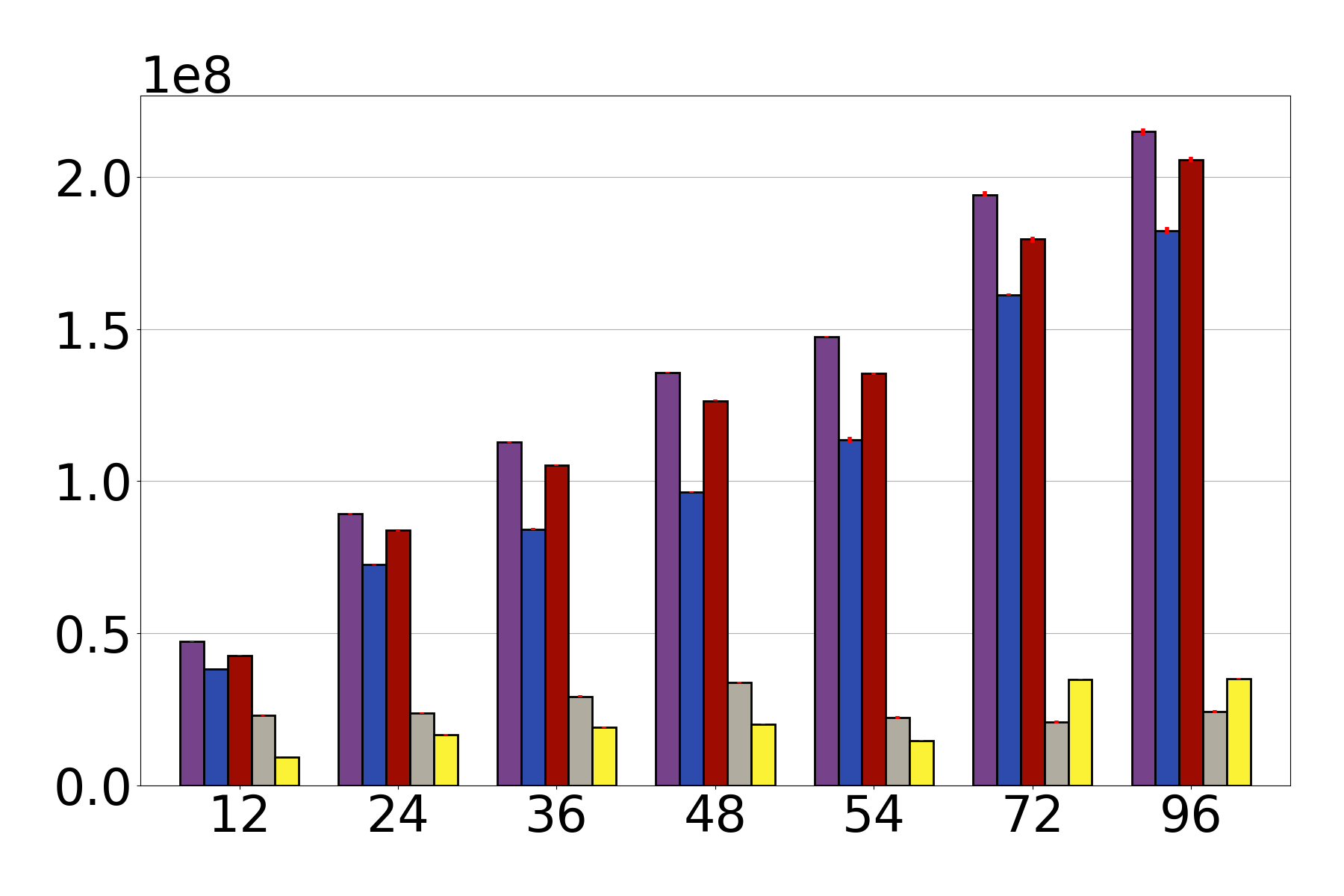}
        \end{subfigure}        
        \begin{subfigure}{0.24\linewidth}
            \centering 
            99\% Read-only
            \includegraphics[width=1\linewidth]{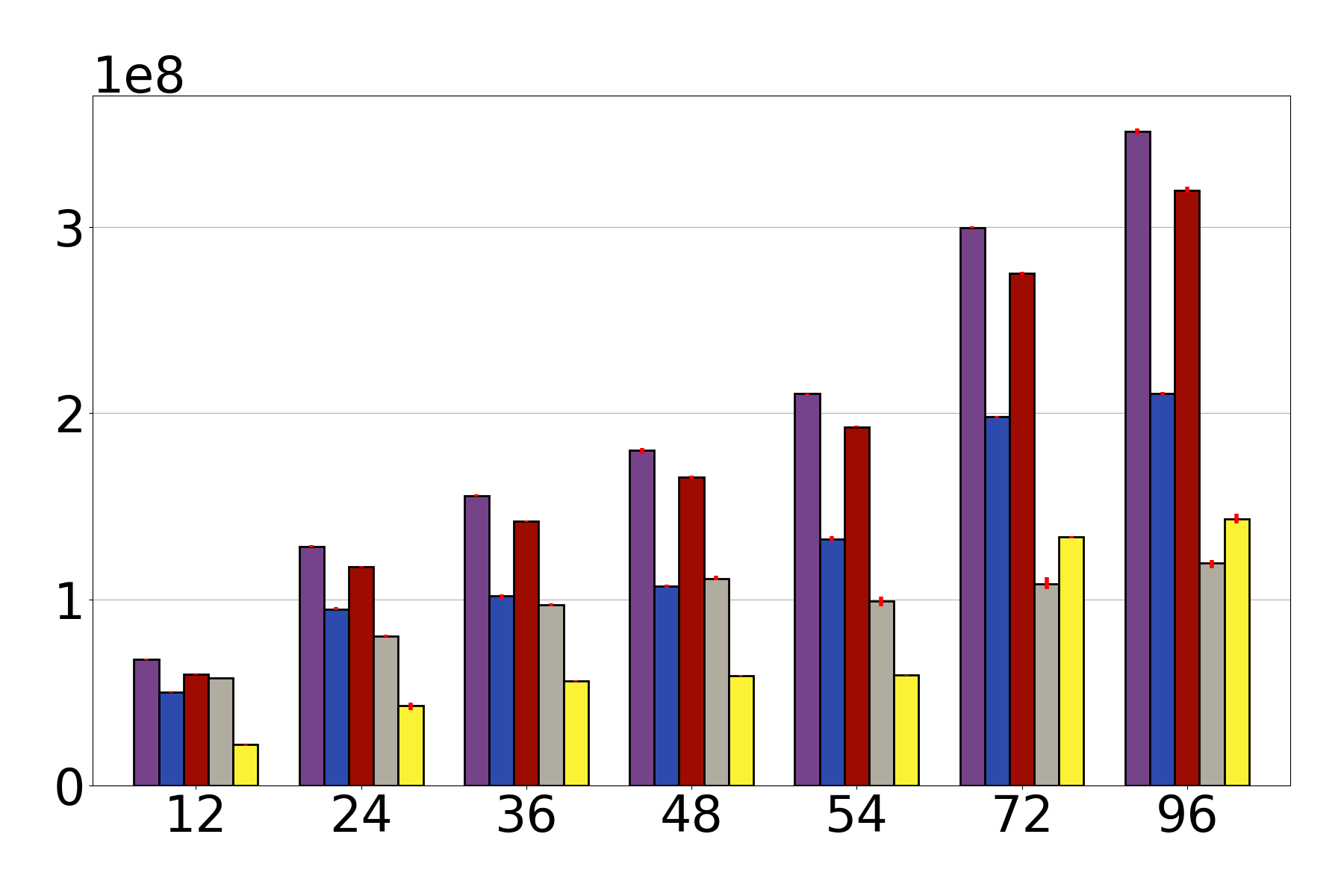} 
        \end{subfigure}     
    \end{subfigure}
    \begin{subfigure}{0.02\linewidth}
        \raisebox{-.5\height}{\rotatebox{90}{\edit{5 million keys}}}
    \end{subfigure}
    \begin{subfigure}{0.97\linewidth}
        \begin{subfigure}{0.24\linewidth}
            \centering
            \raisebox{-.5\height}{\includegraphics[width=1\linewidth]{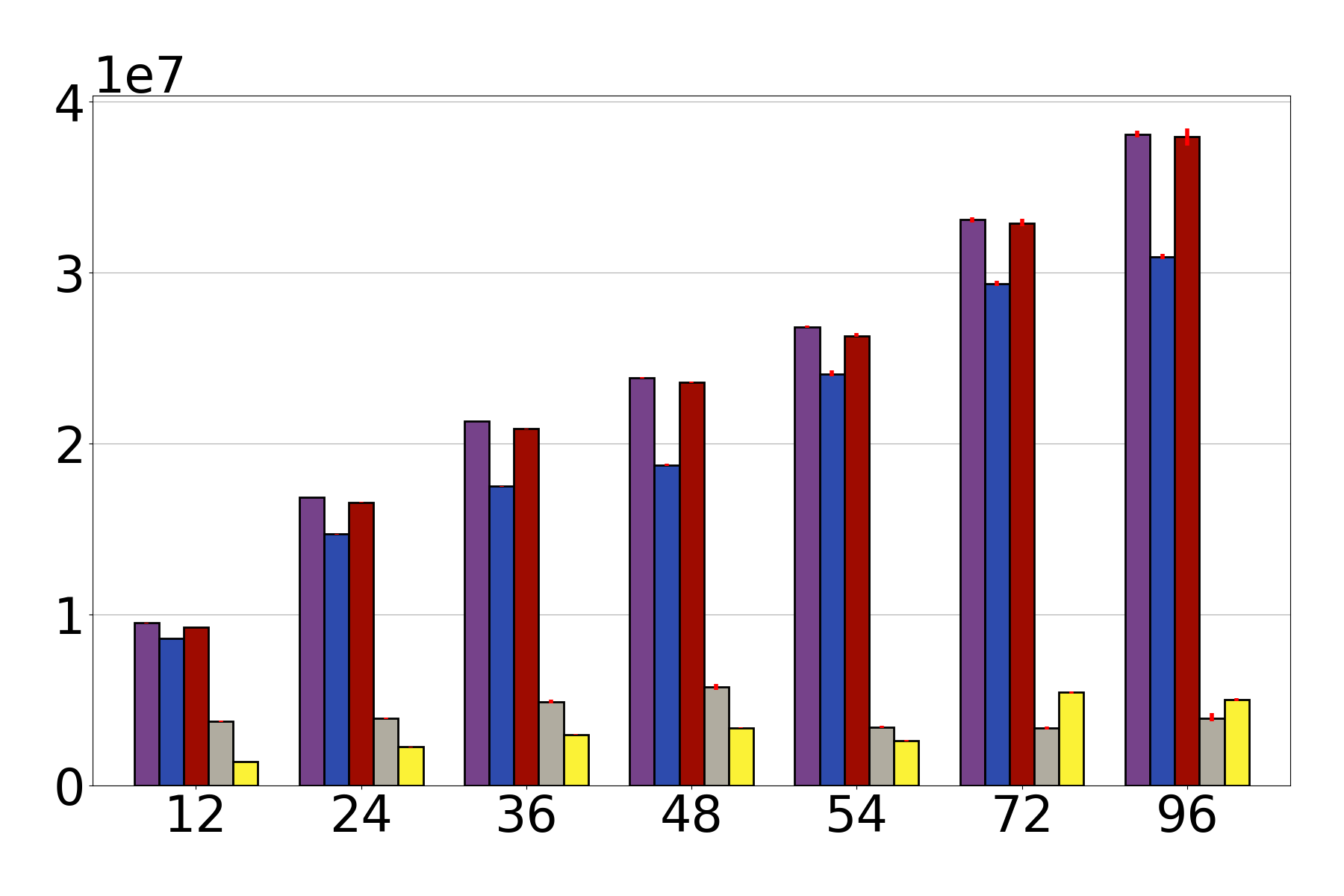}}
        \end{subfigure}
        \begin{subfigure}{0.24\linewidth}
            \centering        
            \raisebox{-.5\height}{\includegraphics[width=1\linewidth]{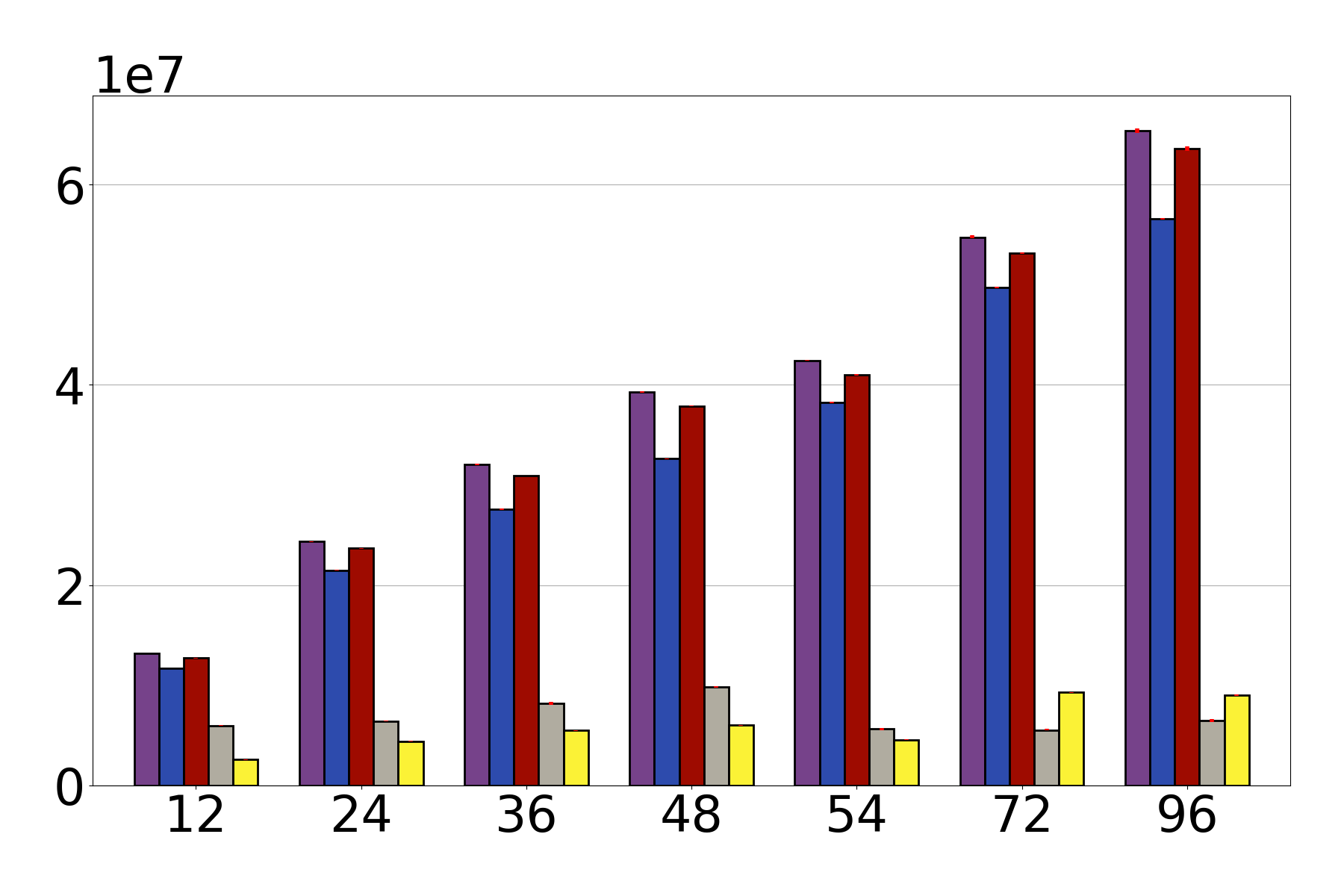}}
        \end{subfigure}
        \begin{subfigure}{0.24\linewidth}
            \centering    
            \raisebox{-.5\height}{\includegraphics[width=1\linewidth]{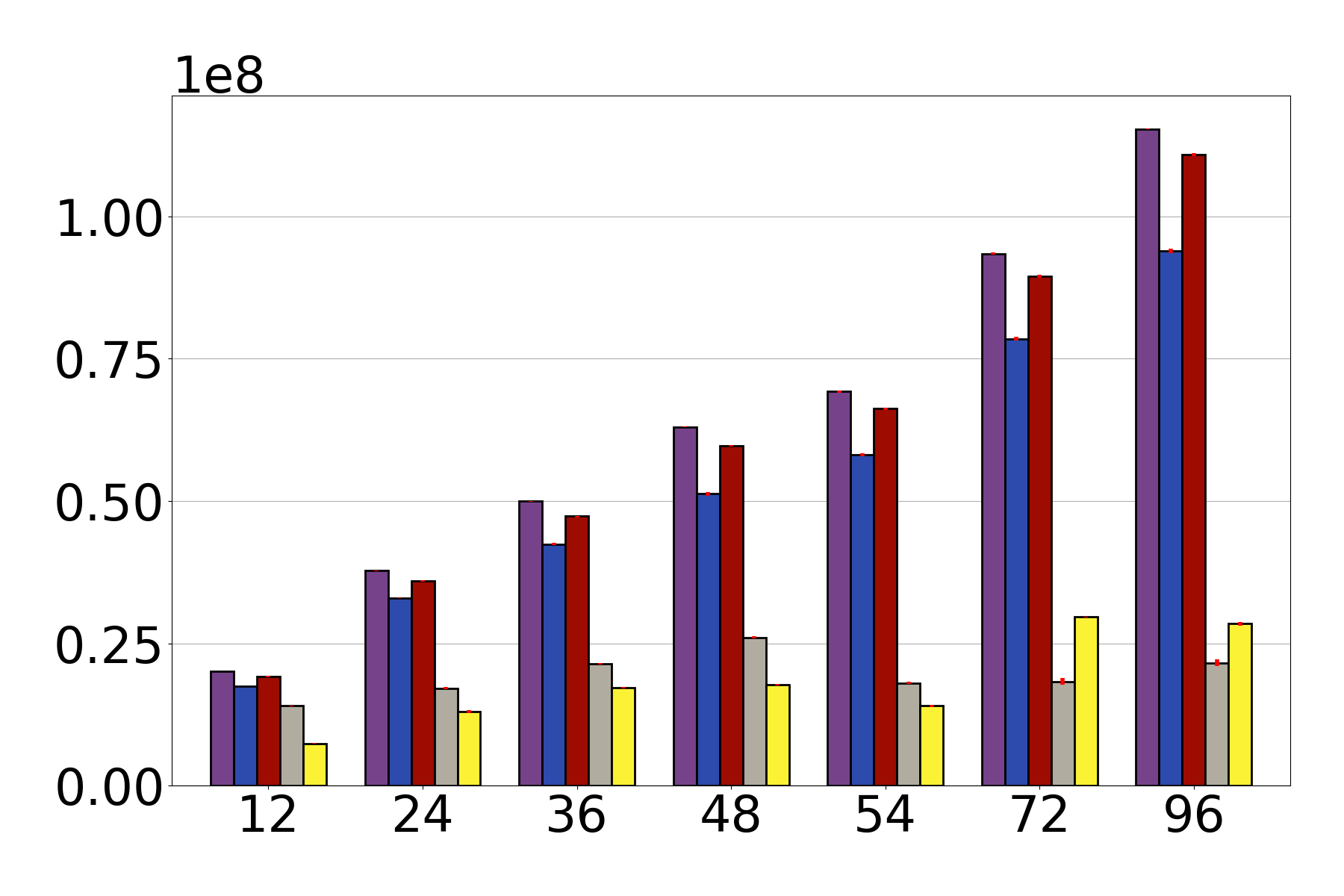}}
        \end{subfigure}        
        \begin{subfigure}{0.24\linewidth}
            \centering     
            \raisebox{-.5\height}{\includegraphics[width=1\linewidth]{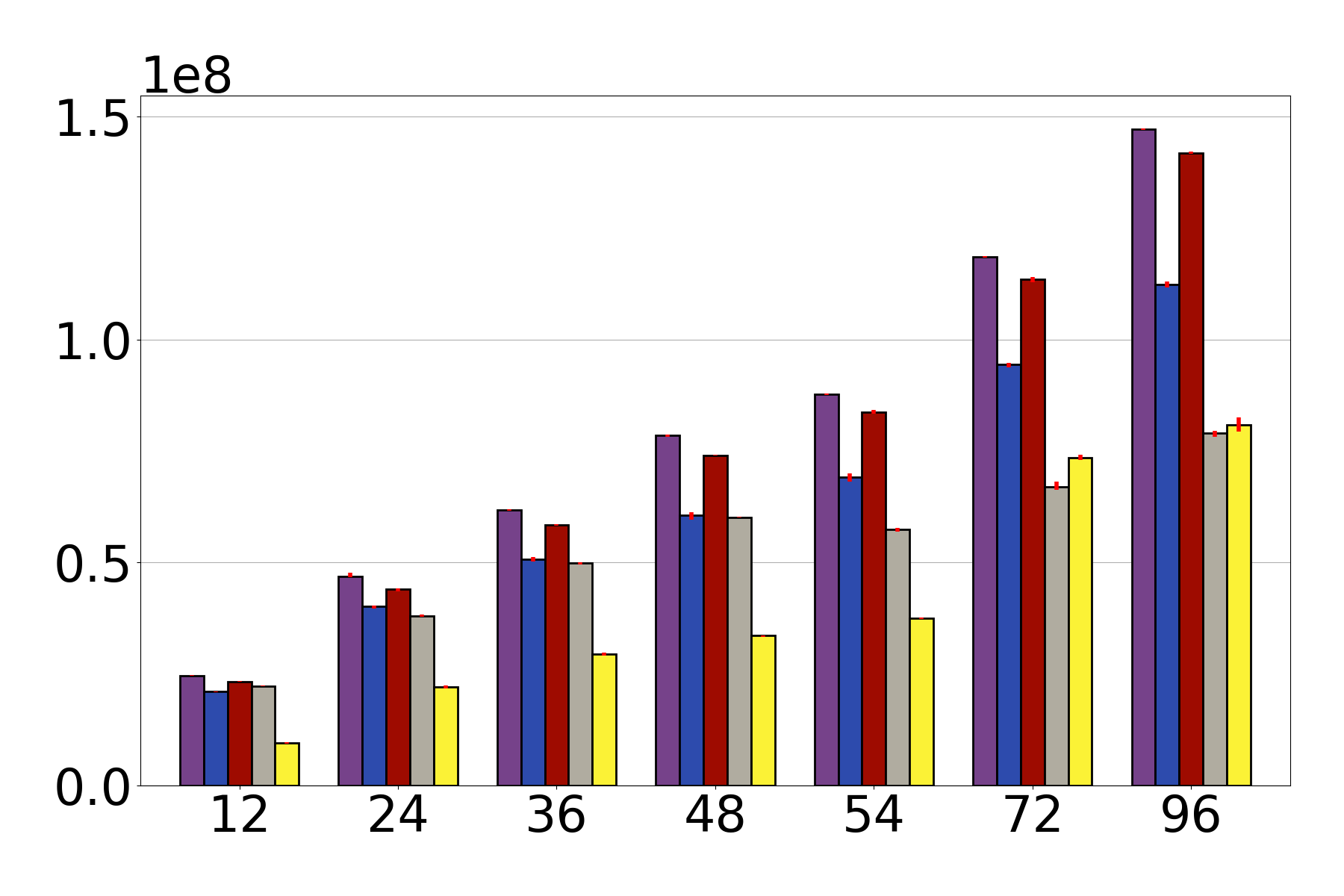}}
        \end{subfigure} 
    \end{subfigure}
    \begin{subfigure}{1.0\linewidth}
        \centering
        \includegraphics[width=0.6\linewidth]{plots/all_legend.png}
    \end{subfigure}     
    \vspace{-8mm}
    \caption{\centering Throughput for hashmap. Y-axis is ops/sec. X-axis is number of threads. Keys are accessed according to a uniform distribution.
    }
    \Description{Throughput for hashmap. Y-axis is ops/sec. X-axis is number of threads. Keys are accessed according to a uniform distribution}
    \label{fig:throughput-hashmap}
\end{figure*}

\begin{figure*}[t!]
    \begin{subfigure}{0.02\linewidth}        
        \raisebox{0.3\height}{\rotatebox{90}{Exponent=0.1}}
    \end{subfigure}
    \begin{subfigure}{0.97\linewidth}
        \begin{subfigure}{0.24\linewidth}
            \centering
            0\% Read-only        
            \includegraphics[width=1\linewidth]{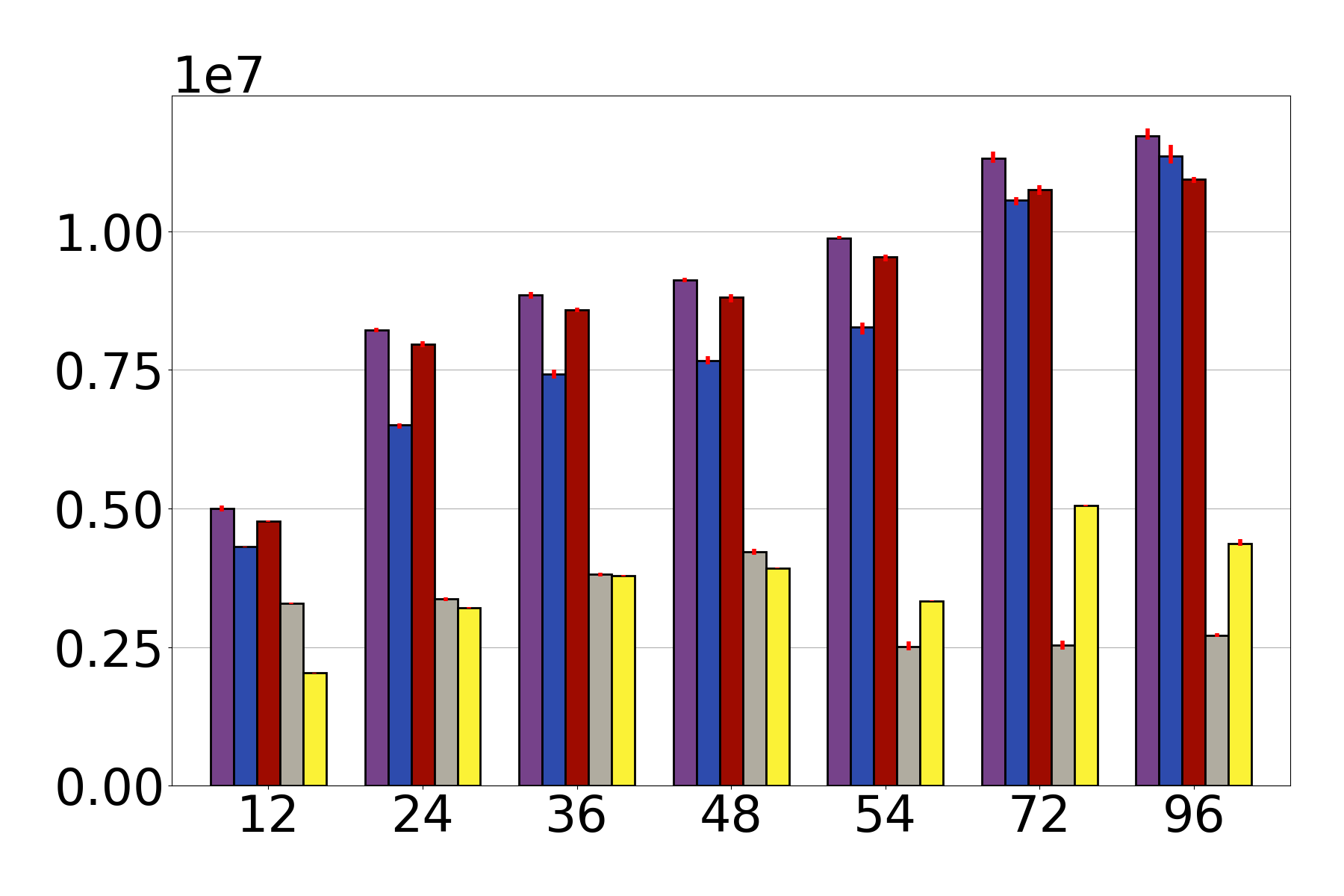}
        \end{subfigure} 
        % \vspace{-4mm}
        \begin{subfigure}{0.24\linewidth}
            \centering        
            50\% Read-only
            \includegraphics[width=1\linewidth]{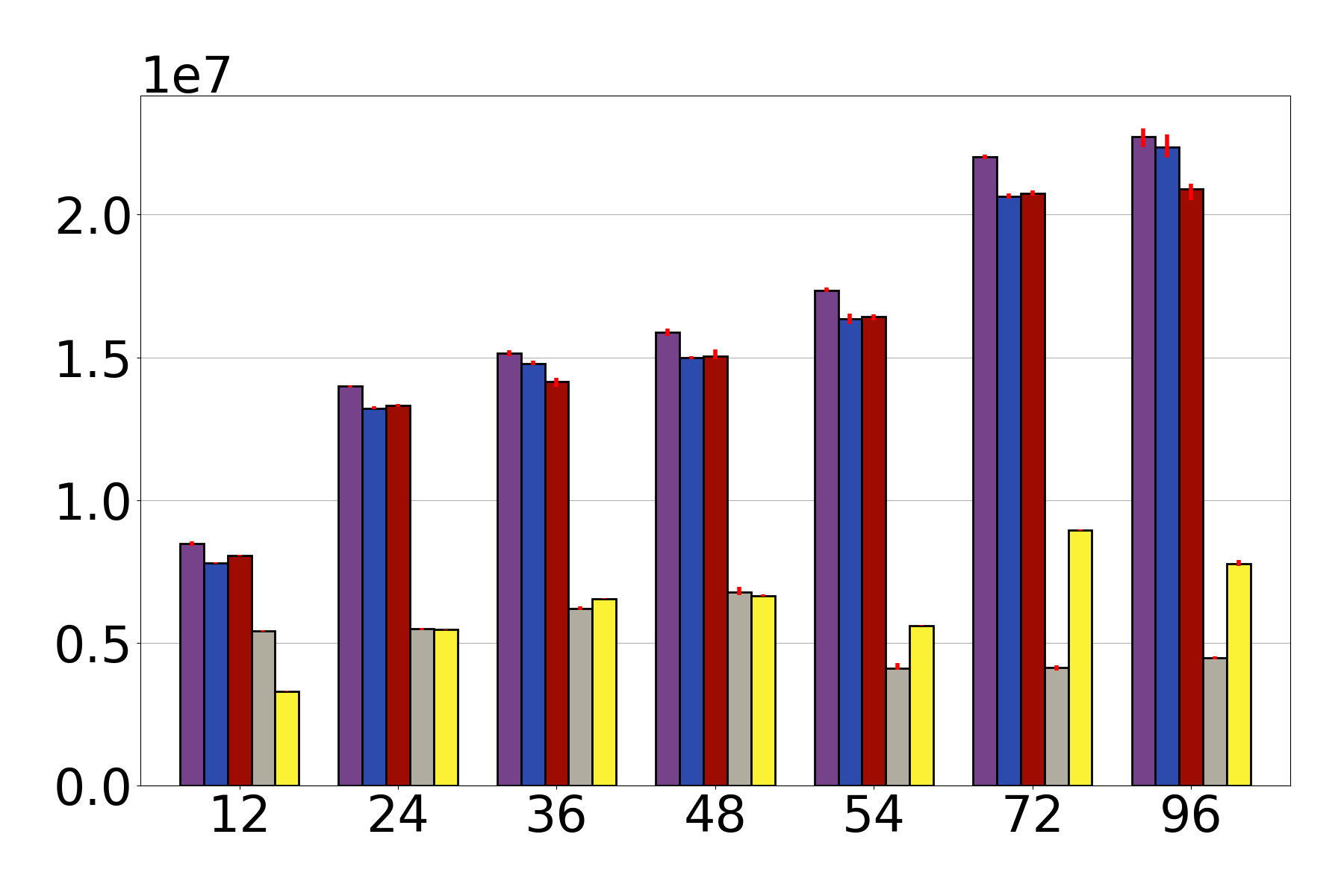}
        \end{subfigure}
        \begin{subfigure}{0.24\linewidth}
            \centering    
            90\% Read-only
            \includegraphics[width=1\linewidth]{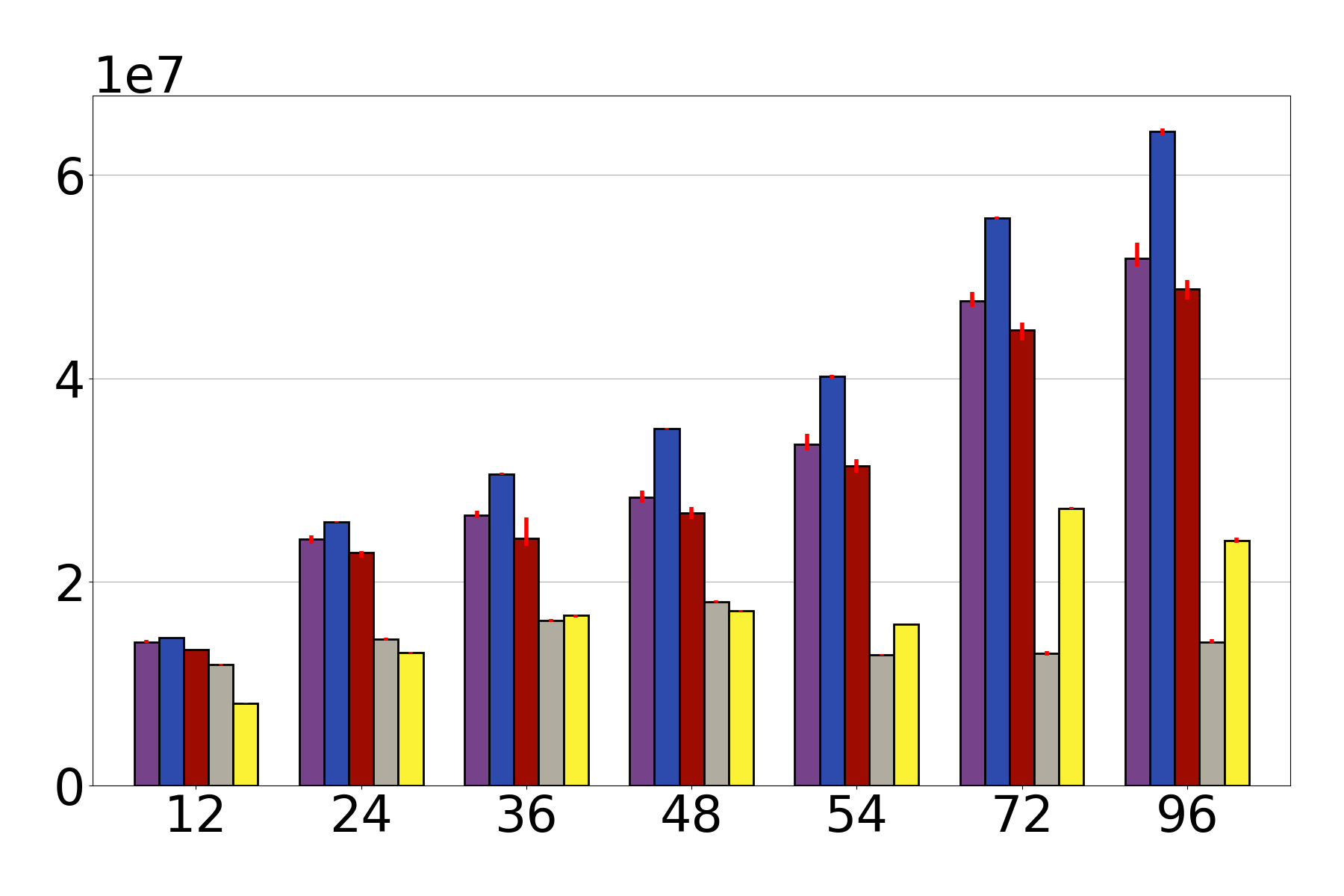}
        \end{subfigure}        
        \begin{subfigure}{0.24\linewidth}
            \centering 
            99\% Read-only
            \includegraphics[width=1\linewidth]{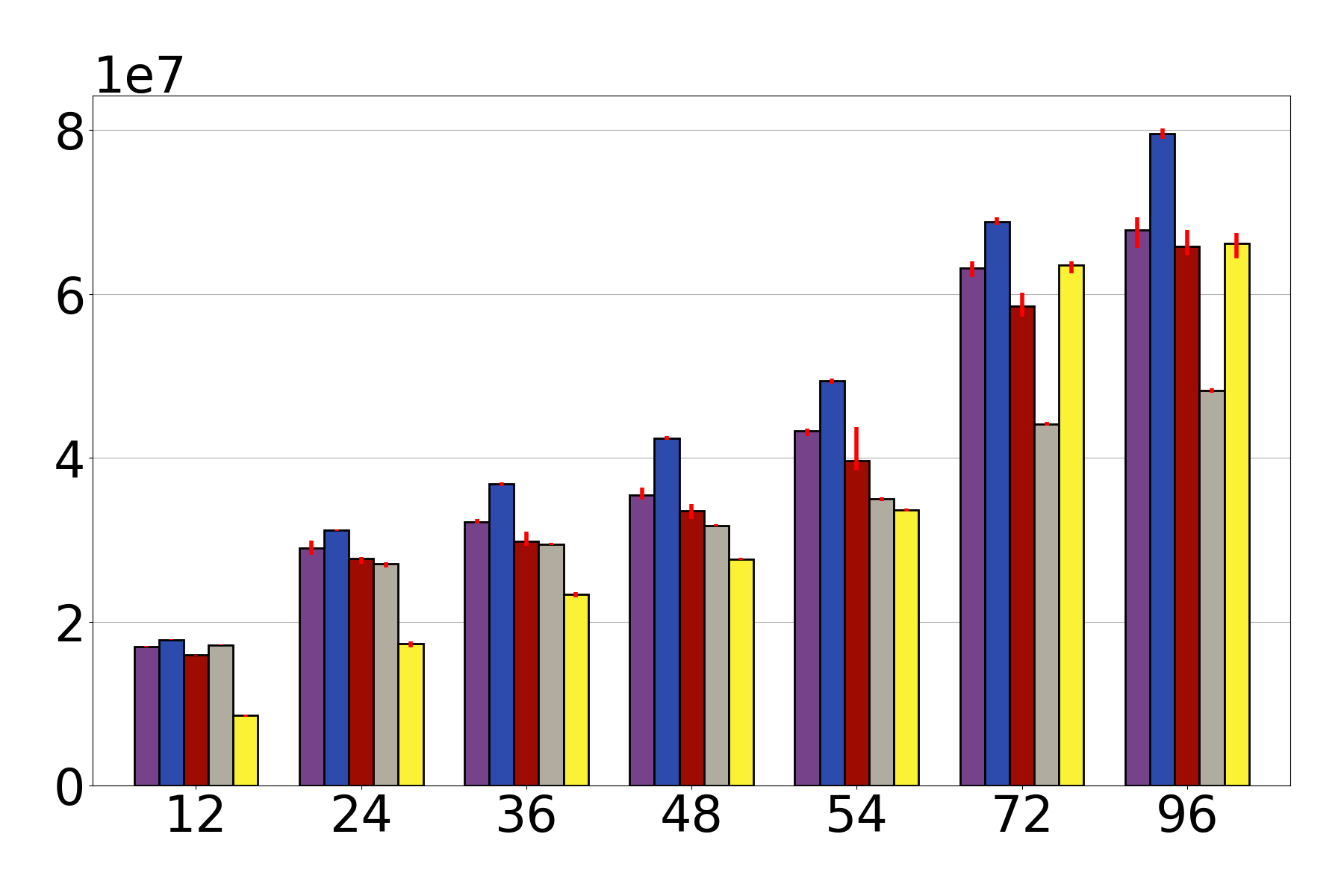} 
        \end{subfigure}     
    \end{subfigure}
    \begin{subfigure}{0.02\linewidth}
        \raisebox{-.5\height}{\rotatebox{90}{Exponent=0.9}}
    \end{subfigure}
    \begin{subfigure}{0.97\linewidth}
        \begin{subfigure}{0.24\linewidth}
            \centering
            \raisebox{-.5\height}{\includegraphics[width=1\linewidth]{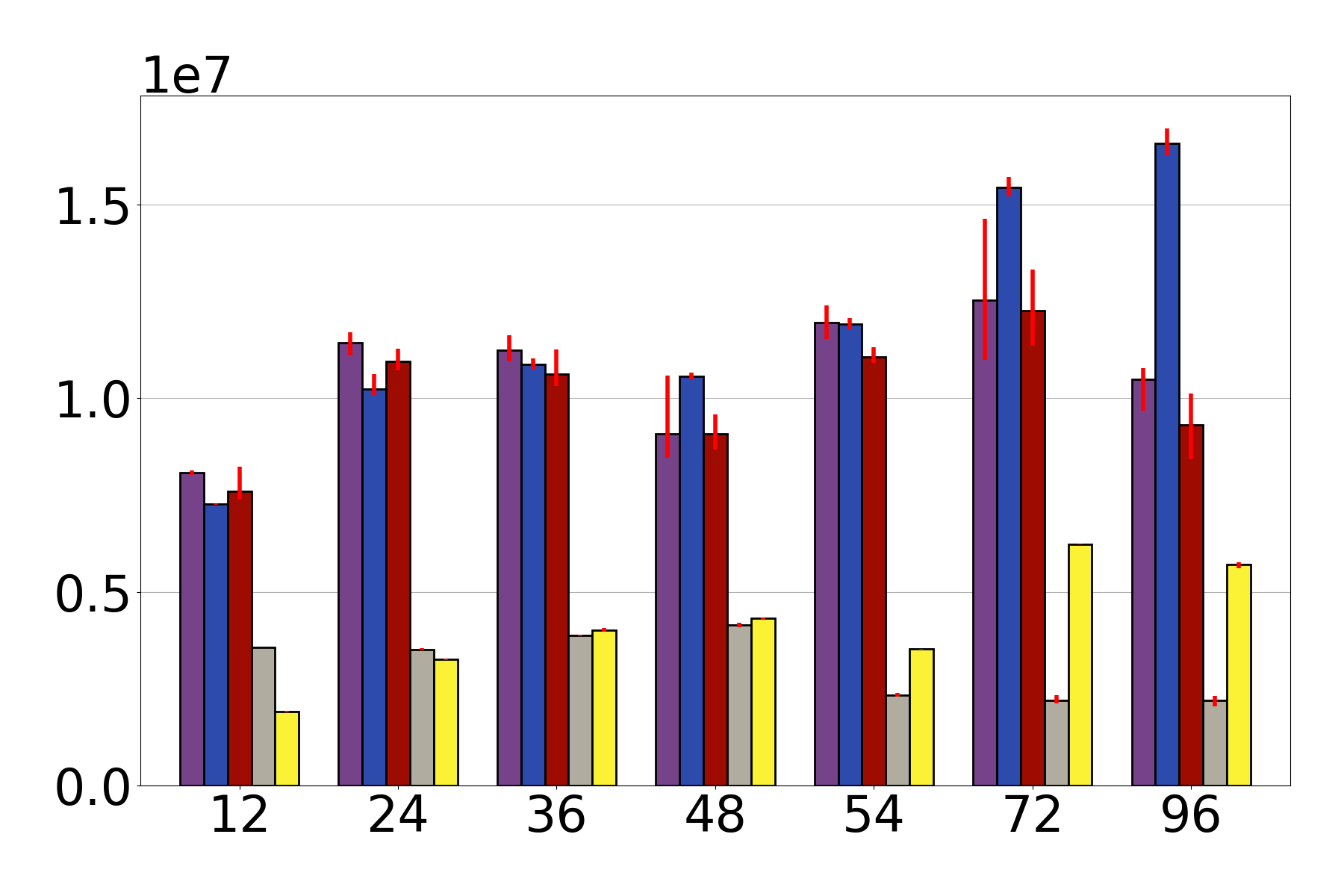}}
        \end{subfigure}
        \begin{subfigure}{0.24\linewidth}
            \centering        
            \raisebox{-.5\height}{\includegraphics[width=1\linewidth]{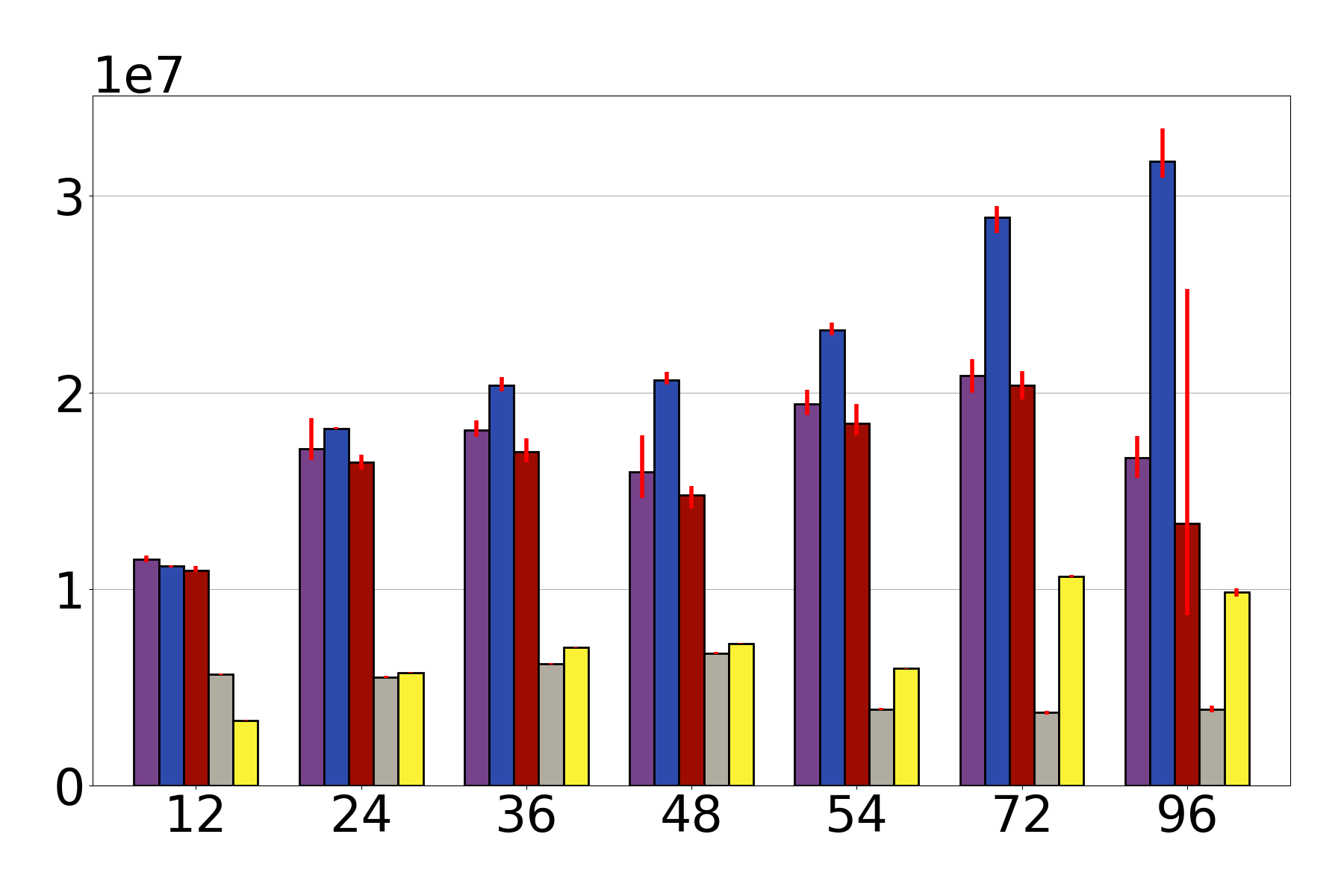}}
        \end{subfigure}
        \begin{subfigure}{0.24\linewidth}
            \centering    
            \raisebox{-.5\height}{\includegraphics[width=1\linewidth]{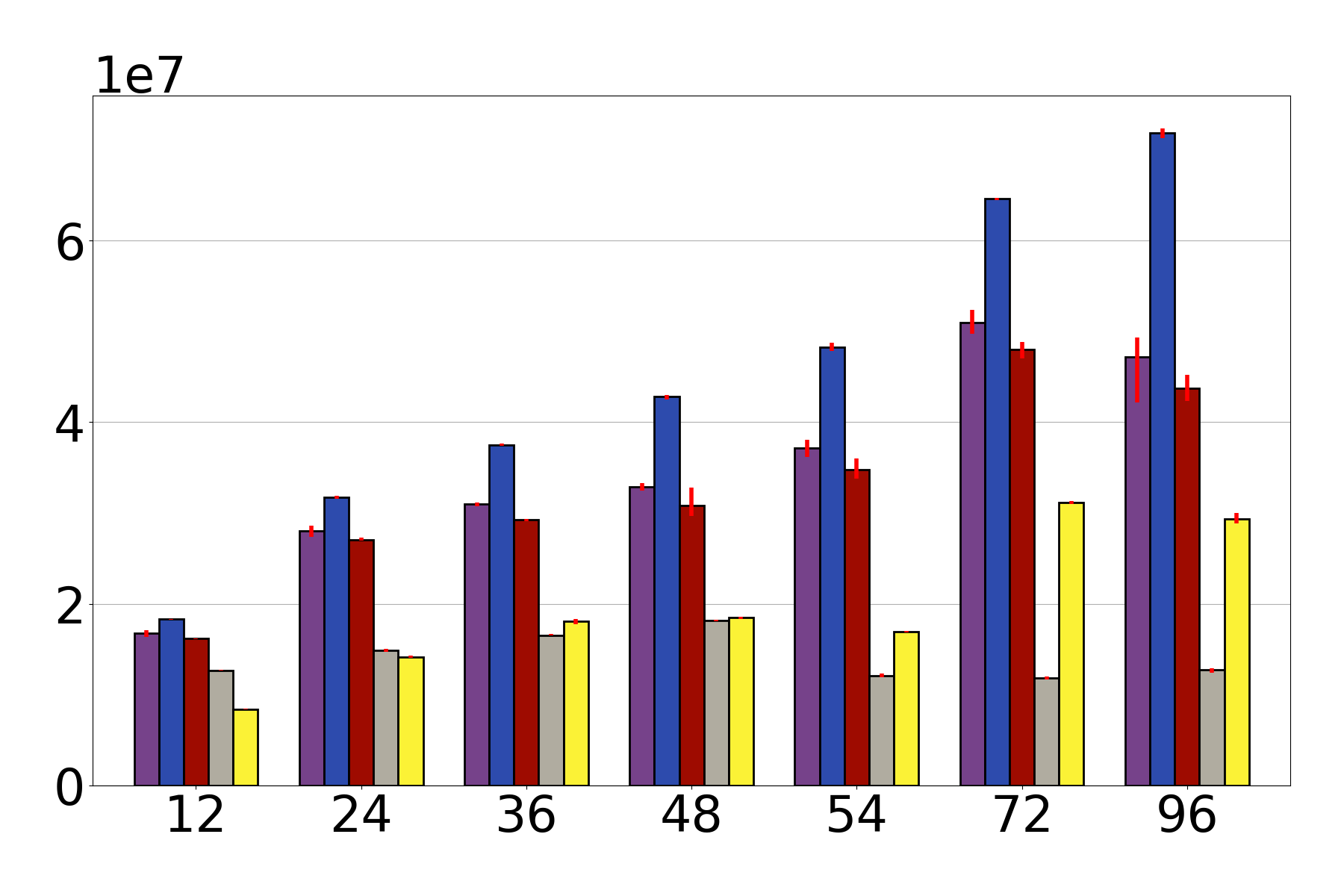}}
        \end{subfigure}        
        \begin{subfigure}{0.24\linewidth}
            \centering     
            \raisebox{-.5\height}{\includegraphics[width=1\linewidth]{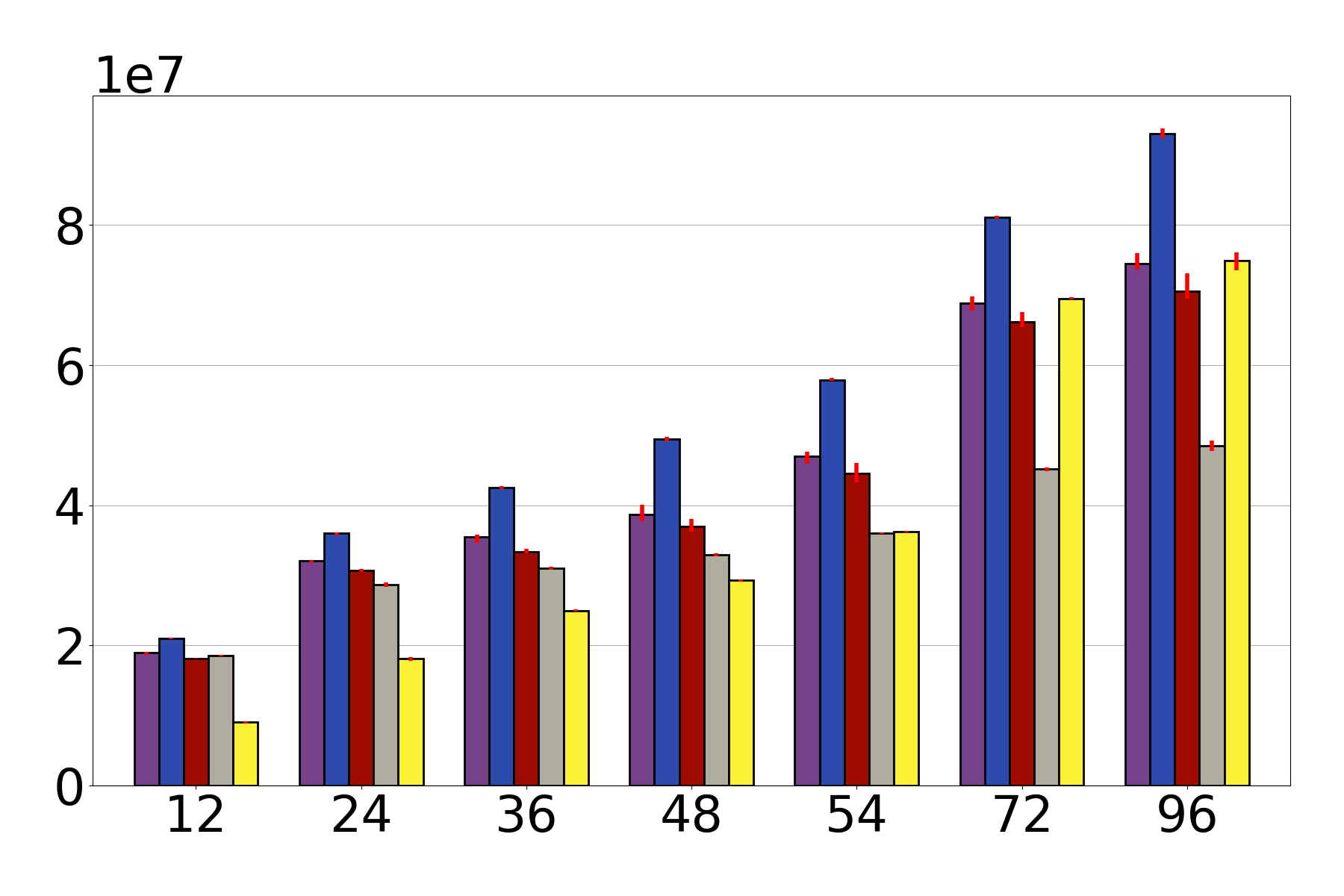}}
        \end{subfigure} 
    \end{subfigure}
    \begin{subfigure}{1.0\linewidth}
        \centering
        \includegraphics[width=0.6\linewidth]{plots/all_legend.png}
    \end{subfigure}     
    \vspace{-8mm}
    \caption{\centering \edit{Throughput for (a,b)-tree with max 1 million keys. Y-axis is ops/sec. X-axis is number of threads. Keys are accessed according to a zipfian distribution with the exponent 0.1 (first row) and 0.9 (second row).}
    }
    \Description{Throughput for (a,b)-tree with 1 million keys. Y-axis is ops/sec. X-axis is number of threads. Keys are accessed according to a zipfian distribution with the exponent listed on the left of each row.}
    \label{fig:throughput-zipf}
\end{figure*}

We compare the performance of \tmname~ against the state of the art PSTM, TrinityVR-TL2, which is Trinity combined with TL2 (we refer to this as Trinity), and the state of the art persistent HyTM, SPHT.
We test the performance in terms of throughput, in operations per second, across various workloads.
We utilize the publicly available source code for both Trinity and SPHT.
We used the same benchmark as \cite{brown2020non} and \cite{coccimiglio2022prep} for evaluating the TMs.
We ran all experiments on a NUMA system with 2 Intel Xeon Gold 5220R 2.20GHz processors, each of which has 24 cores and 48 hardware threads. 
The system has a 36608K L3 cache, 1024K L2 cache, 64K L1 cache and 1.5TB of NVRAM. 
The NVRAM modules installed on the system are Intel Optane DCPMMs \cite{optane} configured in app-direct mode.
% These NVRAM modules are local to socket 2, as such all experiments pin threads to socket 2 until it is at capacity before pinning to socket 1.
We implemented all versions of \tmname~ in C++.
The code was compiled with GCC 10.3.0 with an optimization level of -O2.
We plan to make the code public after publication.
In each trial we prefill the data structure to 50\% capacity before beginning performance measurements. 
The measurement period of each trial is 20 seconds. 
We report the average of 5 trials.

\subsection{Comparing to State of the Art PSTM}
One may wonder why we compare our persistent HyTM against the state of the art for PSTM, Trinity.
It has been shown to perform well for various workloads and we have already discussed the added overhead related to persisting hardware transactions.
Comparing against the state of the art PSTM provides a better perspective on whether or not our design is effective despite these overheads.

\paragraph{(a,b)-Tree}
\Cref{fig:throughput-abtree} shows the throughput of \tmname~ and Trinity for an (a,b)-tree with a=4 and b=16 where keys are accessed according to a uniform distribution.
\tmname~ significantly outperforms Trinity for workloads dominated by update operations.
For workloads dominated by read-only transactions, Trinity achieves throughput more comparable to \tmname.
Updates to the (a,b)-tree involve expensive rebalancing operations.
As a result, hardware transactions are more likely to experience conflicts causing aborts forcing more transactions onto the software path.
In this experiment \tmname~ performs best when using using colocated locks.
Another interesting observation is that our O(1)-abortable strongly progressive version performed well compared to the O(1)-abortable progressive alternatives.
We also show the results for this (a,b)-tree when keys are accessed according to a zipfian distribution in \Cref{fig:throughput-zipf}.
The results are similar to the experiment with a uniform key access pattern.
Even when the key access pattern is skewed, \tmname~ consistently outperforms the existing state of the art, especially for the update-heavy workloads.

\paragraph{Hashmap}
\Cref{fig:throughput-hashmap} shows the throughput of \tmname~ and Trinity for a hashmap with a fixed number of buckets corresponding to the number of keys.
In this case \tmname~ outperforms Trinity for all workloads across all thread counts.
With this configuration, hardware path transactions are unlikely to experience conflicts.
In this case, due to the layout of user data, the use of colocated locks negatively effects throughput.
As with the (a,b)-tree, we also tested the hashmap when keys are accessed according to a zipfian distribution.
We omit plots for the hashmap with zipfian access pattern due to space constraints.
The results were similar to the uniform access experiment, with \tmname~ performing better in all workloads.

\begin{figure*}[t!]
    % row 1
    \begin{subfigure}{0.24\linewidth}
        \centering      
        99\% Read-only
        \includegraphics[width=1\linewidth]{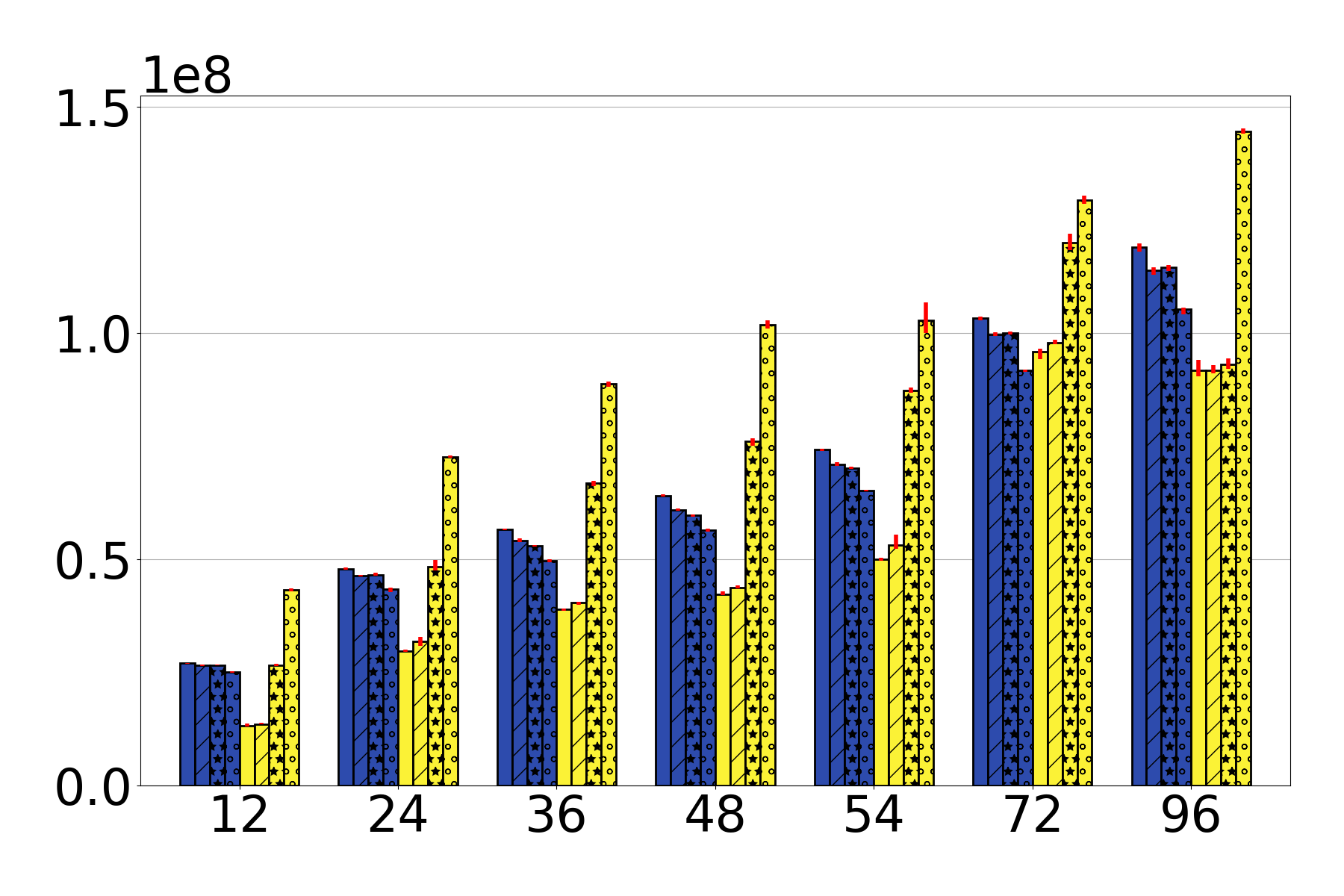} 
    \end{subfigure}
     \begin{subfigure}{0.24\linewidth}
        \centering    
        90\% Read-only
        \includegraphics[width=1\linewidth]{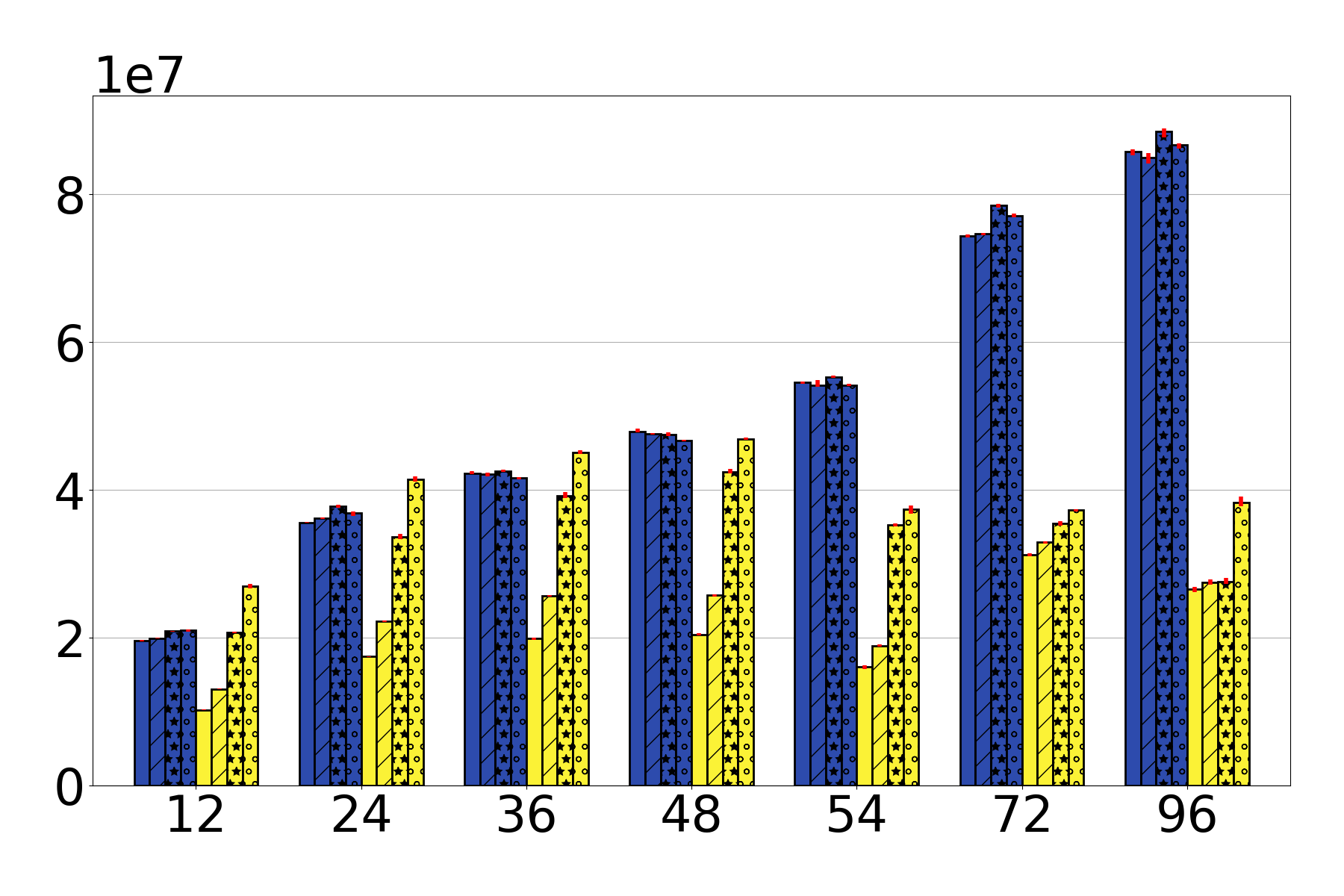} 
    \end{subfigure}    
    \begin{subfigure}{0.24\linewidth}
        \centering        
        50\% Read-only
        \includegraphics[width=1\linewidth]{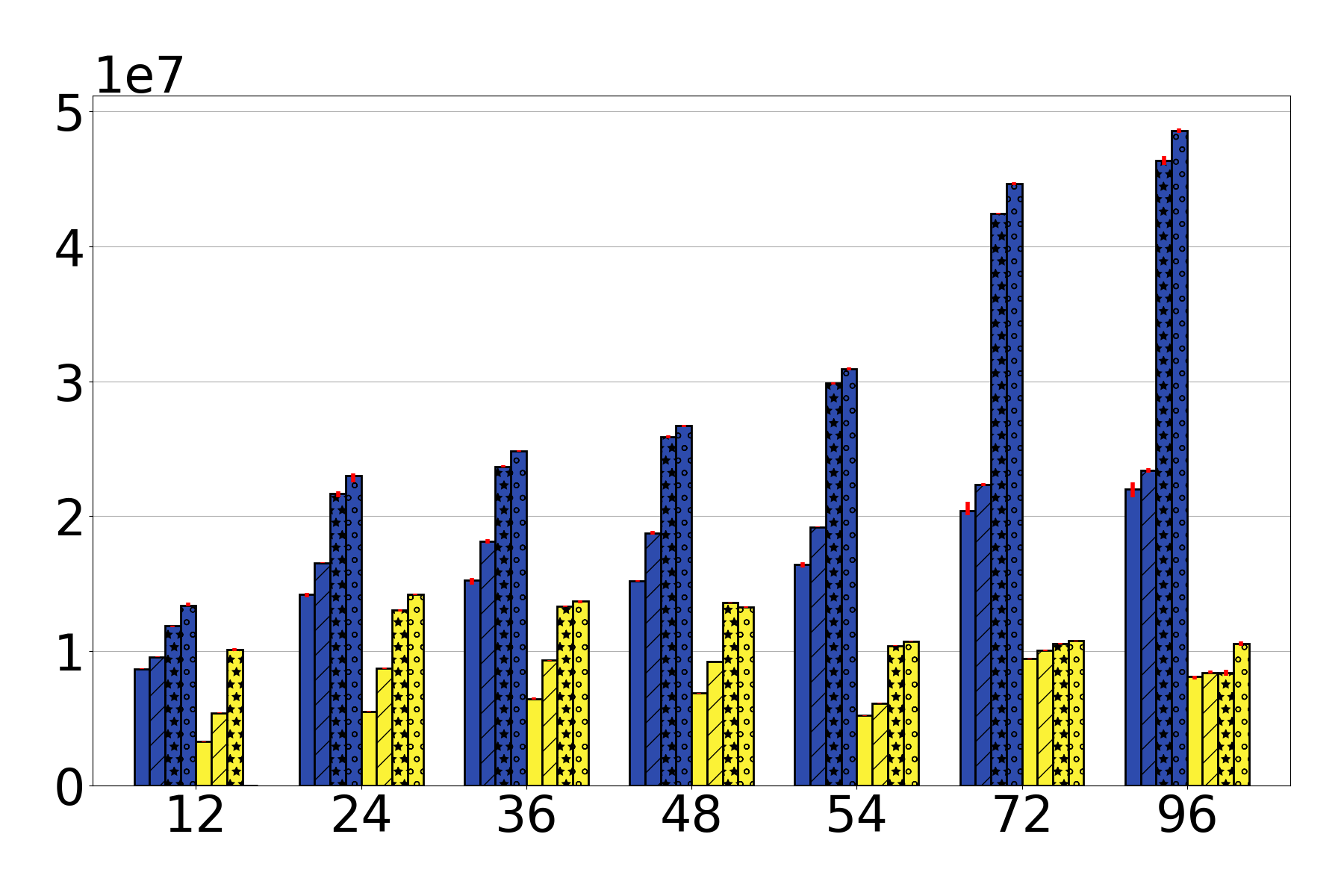} 
    \end{subfigure}
     \begin{subfigure}{0.24\linewidth}
        \centering
        0\% Read-only
        \includegraphics[width=1\linewidth]{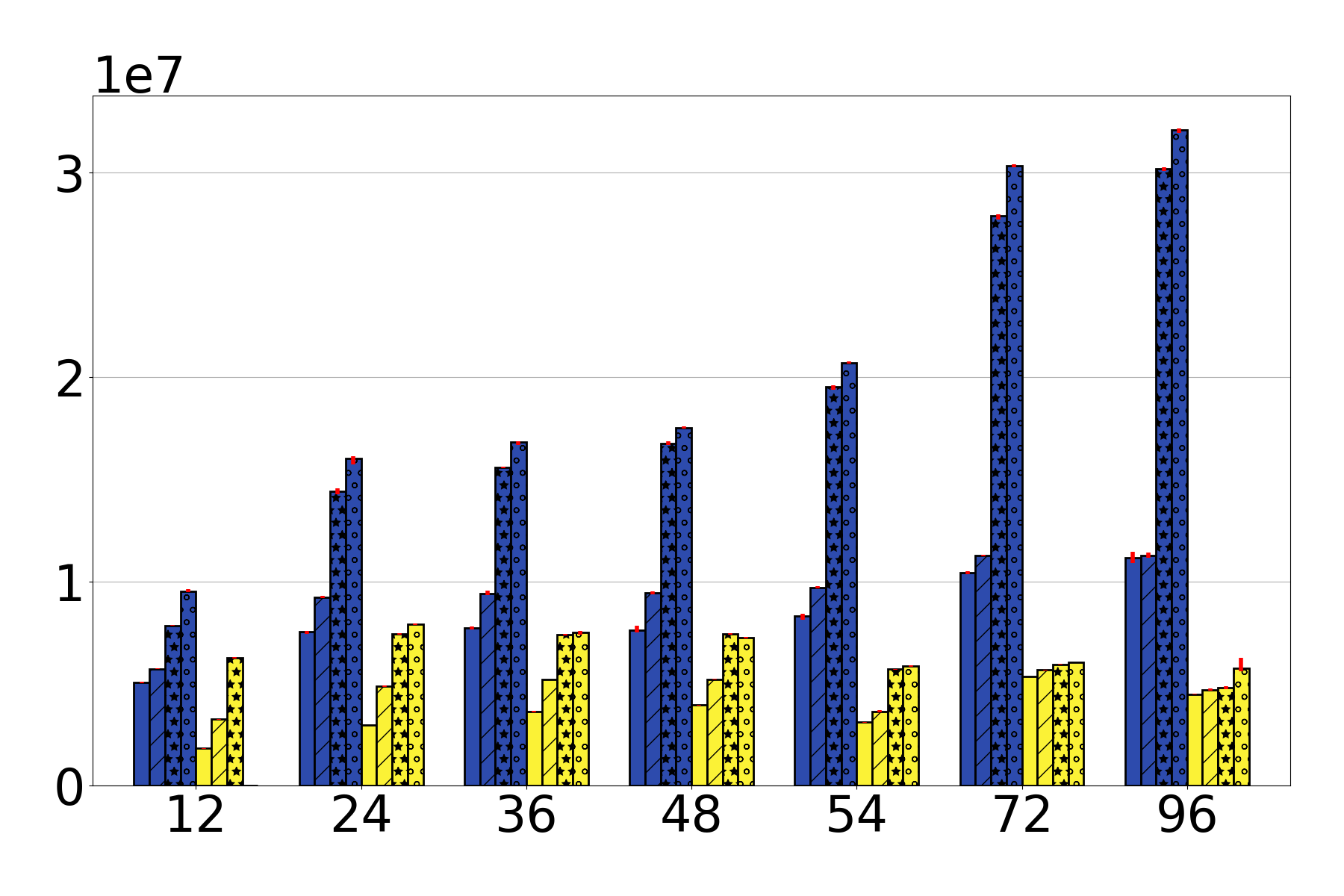} 
    \end{subfigure}  
    \begin{subfigure}{1.0\linewidth}
        \centering
        \includegraphics[width=0.23\linewidth]{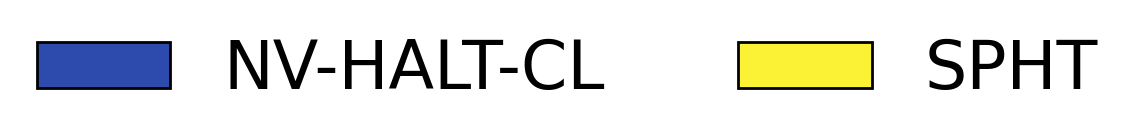}
    \end{subfigure}  
    \begin{subfigure}{1.0\linewidth}
        \centering
        \includegraphics[width=0.4\linewidth]{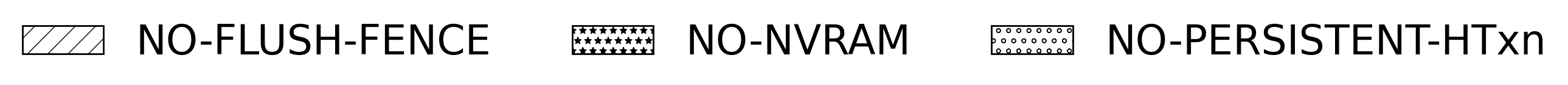}
    \end{subfigure}  
    \vspace{-8mm}
    \caption{\centering Ablation study of \tmname~-CL and SPHT comparing their base implementation to implementations with progressively fewer enabled features for the same (a,b)-tree as row 1 of \Cref{fig:throughput-abtree}. \smalledit{Y-axis is ops/sec. X-axis is number of threads. Keys are accessed according to a uniform distribution.}}
    \Description{\centering Ablation study of \tmname~-CL and SPHT comparing their base implementation to implementations with progressively fewer enabled features for the same (a,b)-tree as row 1 of \Cref{fig:throughput-abtree}. Y-axis is ops/sec. X-axis is number of threads. Keys are accessed according to a uniform distribution.}
    \label{fig:ablation}
\end{figure*}

\subsection{Comparing to State of the Art Persistent HyTM}
\label{sec:eval-spht}
SPHT is the state of the art persistent HyTM.
We compare against SPHT with its forward linking optimization enabled.
The timestamp based synchronization used by SPHT is very much unlike our hardware assisted locking approach.
% Comparing against SPHT demonstrates how, despite the awkwardness, using hardware transactions to read and acquire fine-grained locks performs well compared to existing alternatives.
% This comparison demonstrates that hardware assisted locking performs well compared to existing alternatives.
% 
We integrate the author's publicly available implementation of SPHT into our benchmark.
SPHT utilizes a very simplified custom memory allocator to allocate memory during transactions.
This allocator has artificially low overhead as it does not implement freeing.
A thread in SPHT allocates memory from a fixed sized per-thread pool simply by incrementing a pointer.
In practice this would make SPHT unusable.
% We have significant performance concerns with this allocator.
Unfortunately, the log replay mechanism utilized by SPHT directly relies on this custom allocator, so we cannot modify the allocator or utilize a different allocator without impacting log replay.
% Our concern with SPHT's allocator is the fact that it has artificially low overhead.
% The main reason for this is that the allocator does not implement freeing.
% In practice this would make SPHT unusable since SPHT would eventually run out of memory even for workloads with a balanced ratio of adding or removing data.
% This would not be a problem for Trinity or \tmname~.
% Moreover, without freeing, allocating memory becomes trivial.
% A thread in SPHT allocates memory from a fixed sized per-thread pool simply by incrementing a pointer.

The log replay mechanism of SPHT must be configured since even in a crash-free execution, the log will eventually need to be replayed.
% thus it is unreasonable to disable log replay.
In \cite{castro2021spht}, the author's show that the replay mechanism does not scale well.
As SPHT is configured out of the box, it does not replay the log until the TM is shutdown which occurs only after the regular benchmark operations have completed.
Recall that, system crashes model power failures which occur infrequently.
When crashes are infrequent, it is not realistic to delay log replay until the TM is shutdown.
Consider a long-running server application.
In this setting, the logs would need to be replayed at runtime or else they will become full and cause threads to block.
We enable SPHT's concurrent replay setting to allow a single background thread to replay the logs concurrently with regular benchmark operations.
We attempted to utilize more than 1 concurrent log replay thread but this resulted in SPHT producing errors.
We are engaged in private communication with the author, but the cause of the error is not yet known.
For our microbenchmarks, we configure SPHT to utilize a size of 512MB for each per-thread log (48GB in total at the max thread count of 96).
This log size is likely much larger compared to what one would want to use in practice.
A larger log reduces the time that worker threads can be blocked by the replay mechanism.
By default, SPHT sets the log size to 2GB per thread, however, we believe that this is an unrealistic configuration.
At 2GB per thread, SPHT can typically avoid ever being blocked by the the log replay in our microbenchmarks.
Moreover, this would require 192GB of memory just for the log for a single data structure in our benchmarks at max thread count on our machine.
This is significantly worse for larger machines.
For example to utilize all threads in an 8 socket system with 8x Intel 56 core Xeon 8480+ processors would require 1.75TB of memory just for the log in a single program that uses SPHT.
Our target log size was actually 2MB per thread but this resulted in SPHT exceeding the 4 minute timeout used in our benchmarks. 
2MB is equivalent to a typical operating system huge page.
It is also a common page size and default limit for per-thread buffers in many allocators.
For example the default size for all thread caches in tcmalloc is 2MB.
512MB per thread is a generous size that reveals a small fraction of the cost of SPHT's replay mechanism, whereas 2GB completely obscures this cost.
% Following \cite{castro2021spht} we utilize 16 replay threads in all tests as this was the thread count the authors reported to perform best, and we replay the log only after the regular benchmark operations have completed.
% We report the throughput of SPHT including log replay.
% In \Cref{appendix:spht-replay} we show the throughout of SPHT when log replay is disabled.

\paragraph{(a,b)-Tree}
\tmname~ outperforms SPHT for all workloads in the (a,b)-tree microbenchmark.
SPHT's performance is best when the workload is read-dominant.
This is not surprising.
For the 99\% read-only workload there are fewer hardware aborts since conflicts are less likely.
This means that SPHT's trivial software path is rarely utilized.
In this workload SPHT also benefits from having uninstrumented hardware path reads.
However, when we have even a moderate amount of update operations, \tmname~ significantly outperforms SPHT, especially at higher thread counts.
There are two reasons for SPHT's lower throughput.
First, in these workloads, aborts of hardware transactions are more common.
In this case, the software fallback path utilized by SPHT can become a bottleneck since it immediately claims a global lock.
Second, these workloads showcase the expensive cost of the synchronization mechanism utilized by the commit phase of SPHT as well as the cost of replaying the persistent logs. 
Both of these represent overhead that does not exist in \tmname~.
Moreover, \tmname~ has non-trivial overhead related to memory management which SPHT artificially avoids.

\begin{figure*}[t!]    
    \begin{subfigure}{0.02\linewidth}        
        \raisebox{0.3\height}{\rotatebox{90}{ Time (sec)}}
    \end{subfigure}
    \begin{subfigure}{0.97\linewidth}
        \begin{subfigure}{0.24\linewidth}
        \centering      
        Kmeans-high
        \includegraphics[width=1\linewidth]{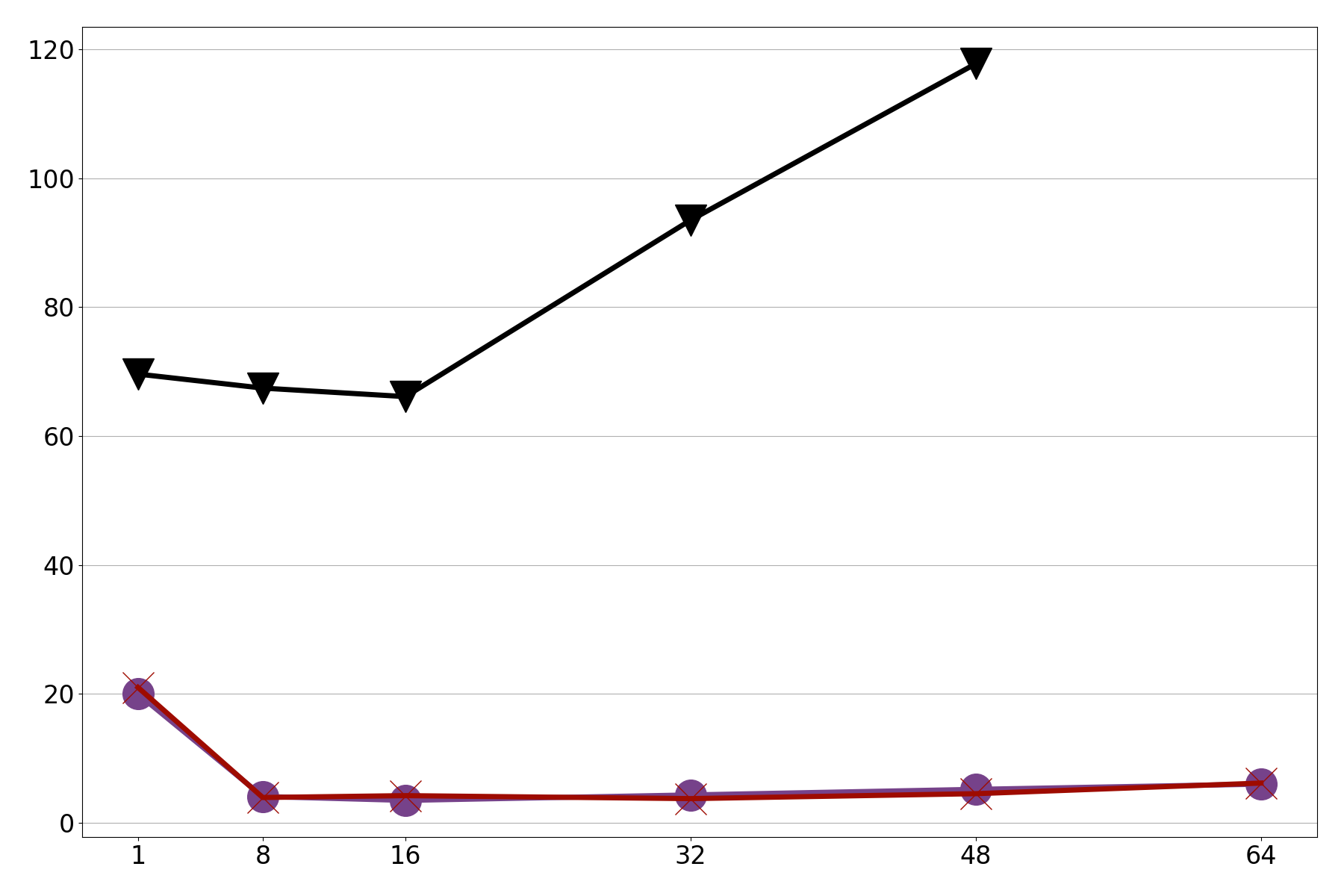} 
        \end{subfigure}
         \begin{subfigure}{0.24\linewidth}
            \centering    
            Intruder
            \includegraphics[width=1\linewidth]{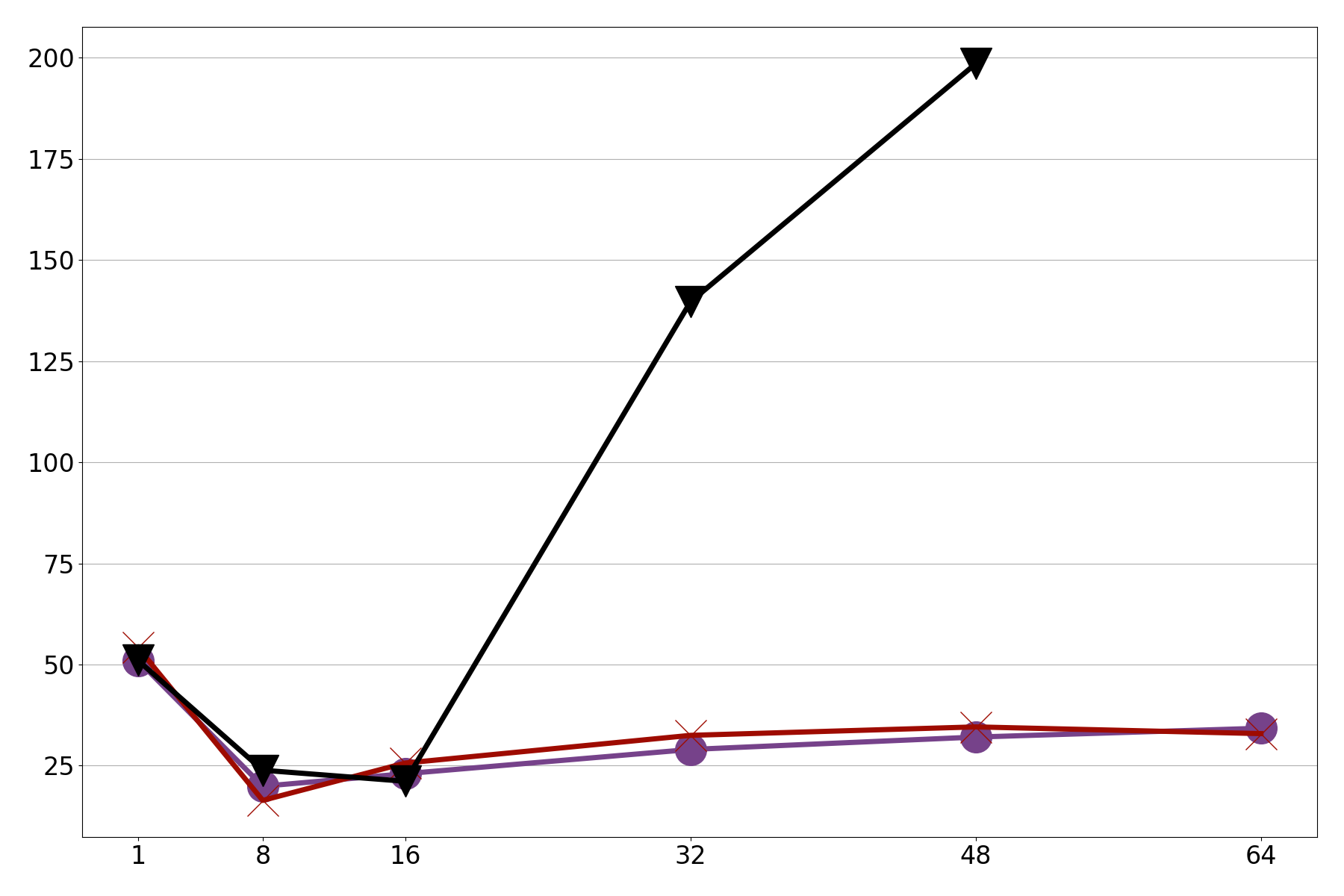} 
        \end{subfigure}        
         \begin{subfigure}{0.24\linewidth}
            \centering
            SSCA2
            \includegraphics[width=1\linewidth]{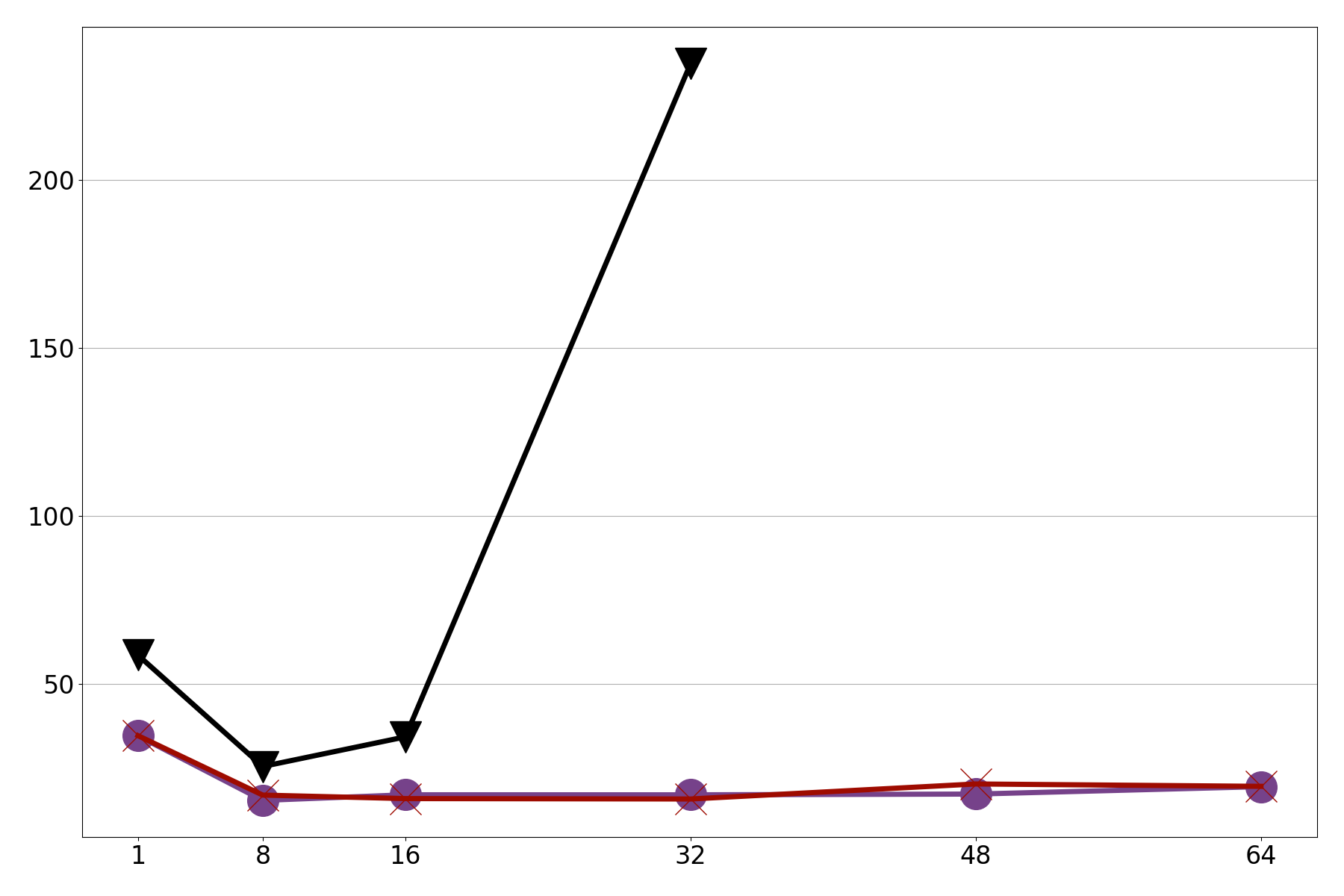} 
        \end{subfigure}      
        \begin{subfigure}{0.24\linewidth}
            \centering        
            Labyrinth
            \includegraphics[width=1\linewidth]{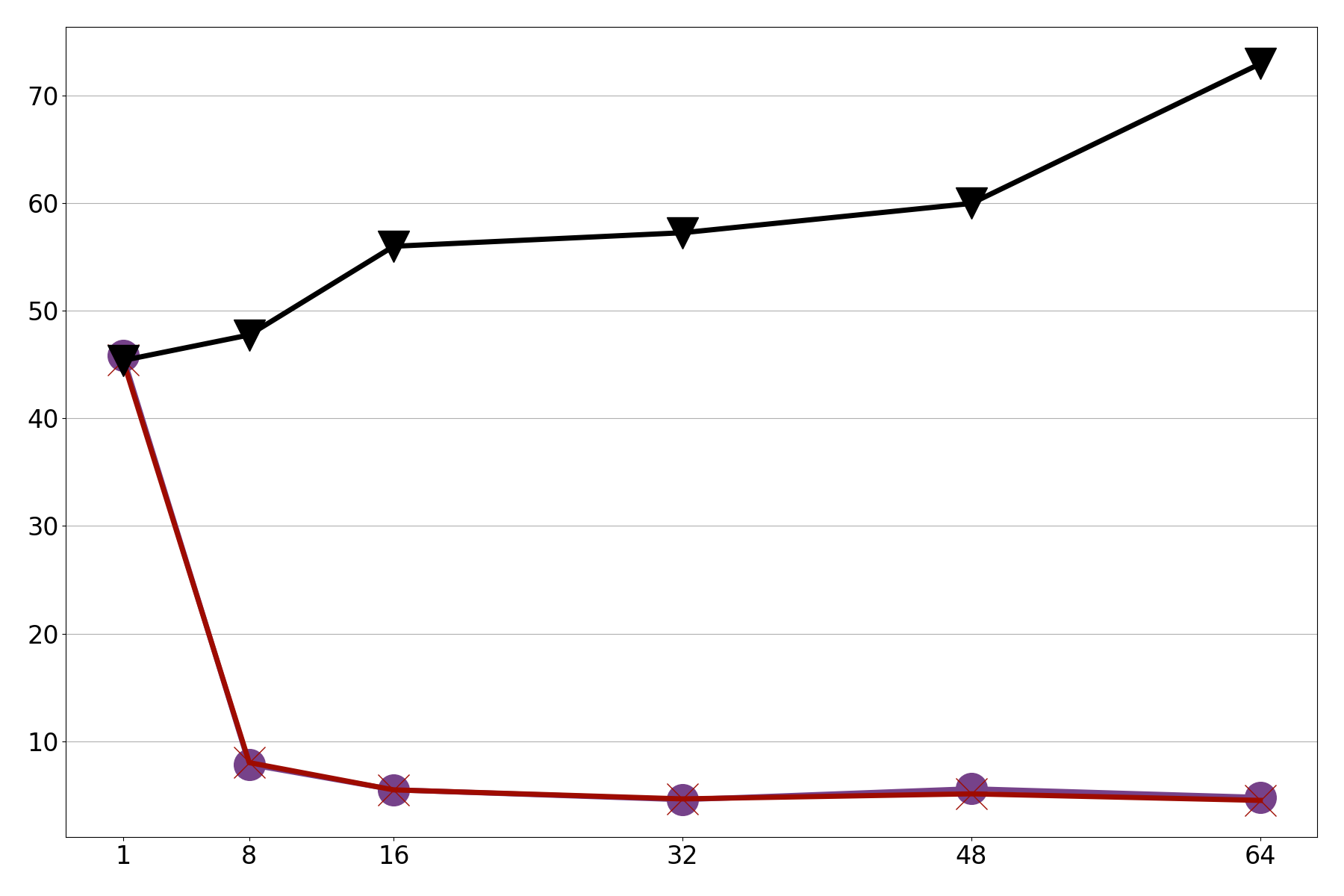} 
        \end{subfigure} 
    \end{subfigure}
    \begin{subfigure}{1.0\linewidth}
        \centering
        \includegraphics[width=0.3\linewidth]{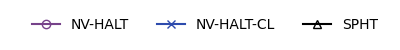}
    \end{subfigure}    
    \vspace{-8mm}
    \caption{\centering Comparing \tmname~ to SPHT using 4 of the STAMP benchmarks. All benchmarks are using ++ workloads. X-axis is number of threads. Y-axis is time in seconds, lower is better. Missing data points indicate exceeding the 4 minute timeout.}
    \Description{Comparing \tmname~ to SPHT using some of the STAMP benchmarks using ++ workloads. X-axis is number of threads. Y-axis is time in second (lower is better).}
    \label{fig:stamp}
\end{figure*}

\begin{figure*}[t!]
    % left header
    \begin{subfigure}{0.49\linewidth}
        \centering
        energy-pkg
    \end{subfigure}
    % right header
    \begin{subfigure}{0.49\linewidth}
        \centering
        energy-ram
    \end{subfigure}
    \vspace{0.2mm}
    % ow 2
    \begin{subfigure}{0.24\linewidth}
        \centering      
        90\% Read-only
        \includegraphics[width=1\linewidth]{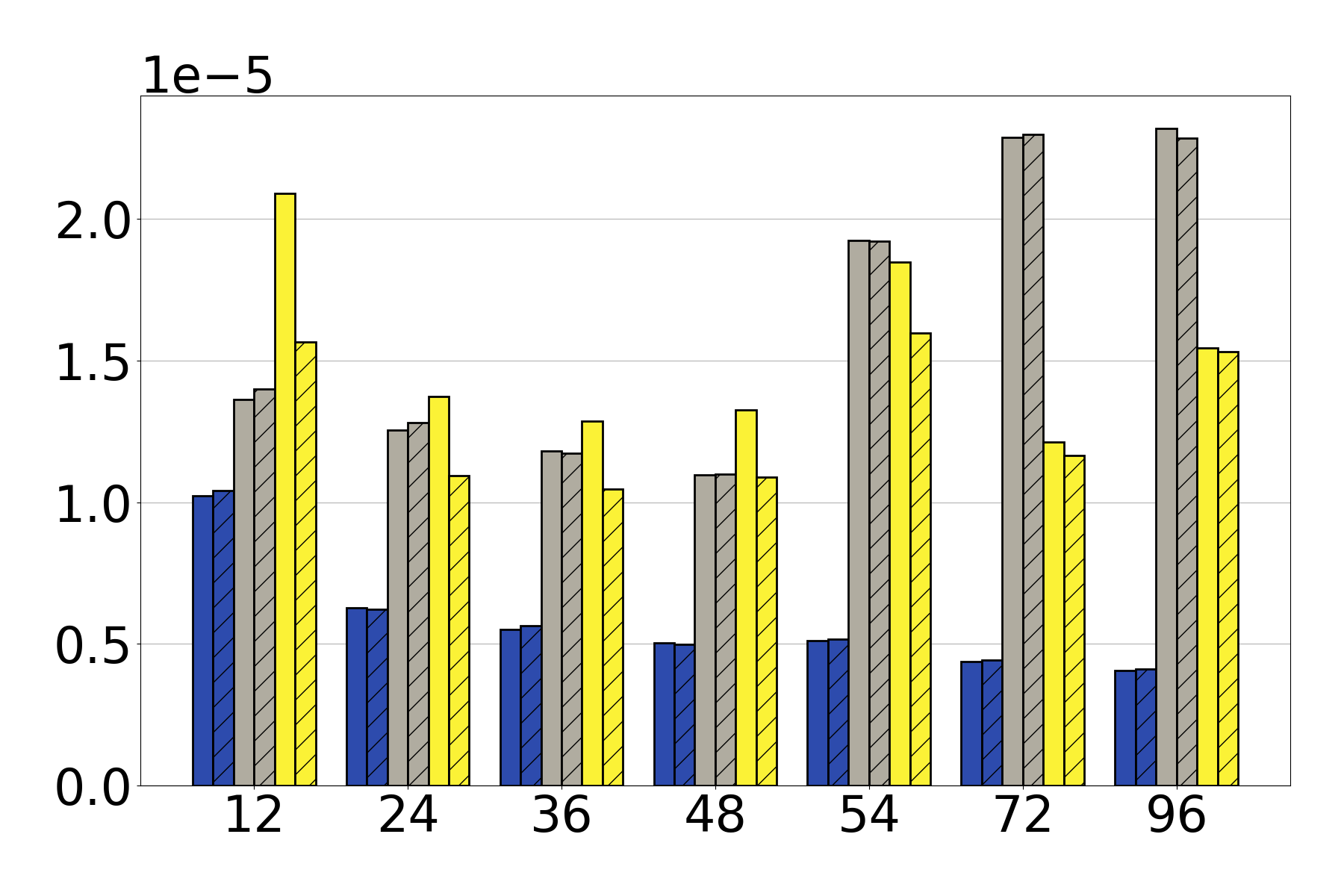} 
    \end{subfigure}
     \begin{subfigure}{0.24\linewidth}
        \centering    
        50\% Read-only
        \includegraphics[width=1\linewidth]{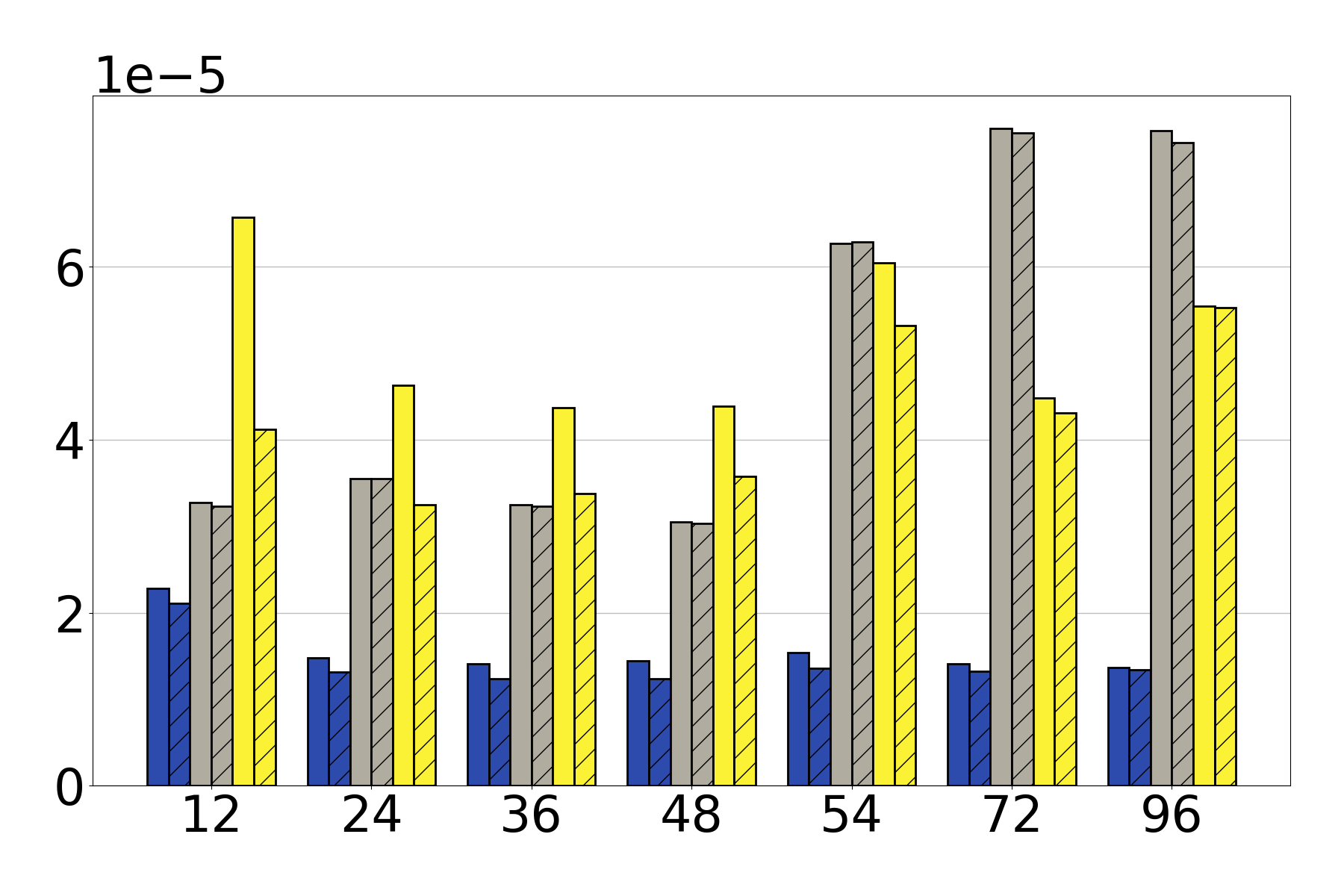}
    \end{subfigure}    
    \begin{subfigure}{0.24\linewidth}
        \centering        
        90\% Read-only
        \includegraphics[width=1\linewidth]{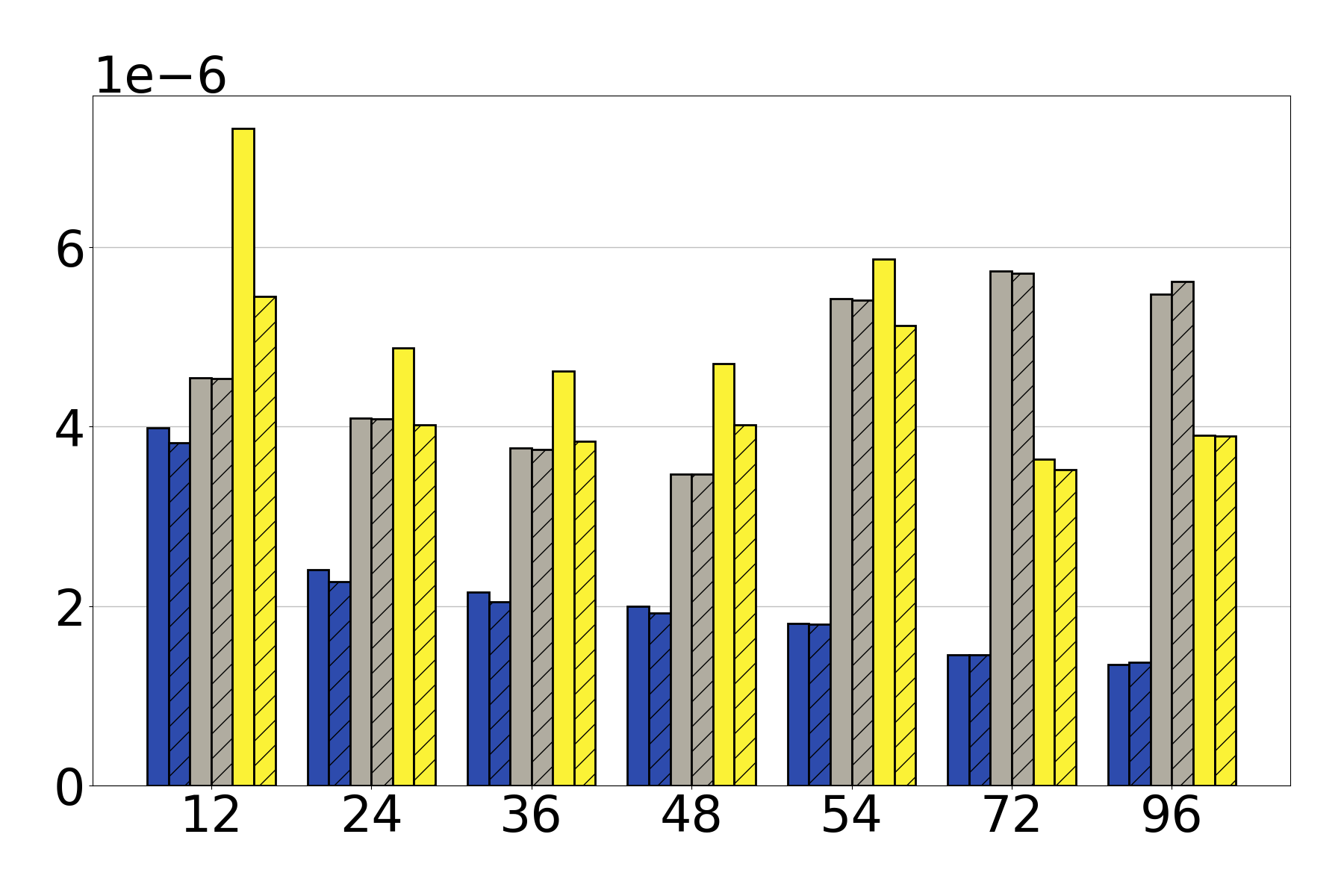} 
    \end{subfigure}
     \begin{subfigure}{0.24\linewidth}
        \centering
        50\% Read-only
        \includegraphics[width=1\linewidth]{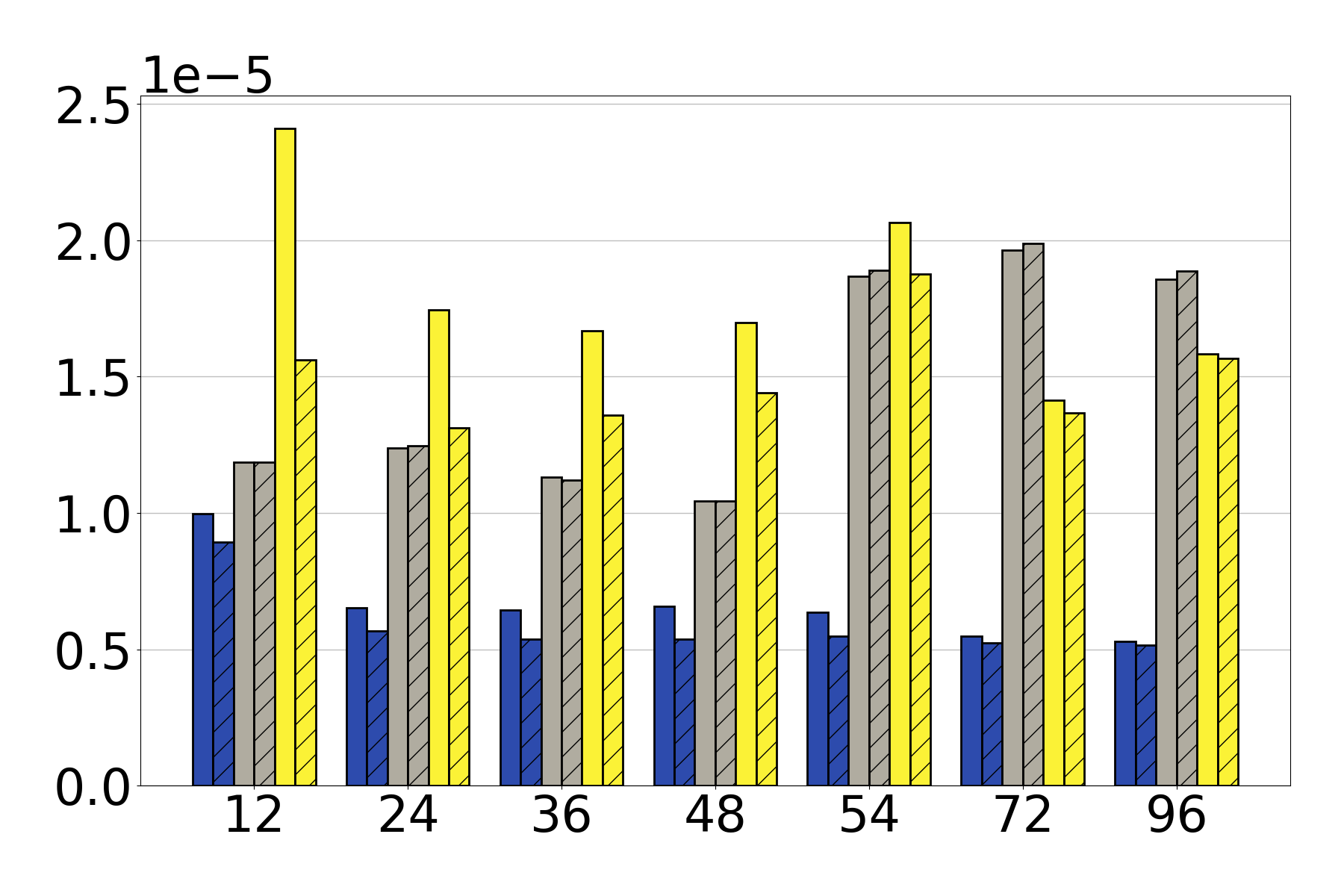} 
    \end{subfigure}  
    \begin{subfigure}{1.0\linewidth}
        \centering
        \includegraphics[width=0.23\linewidth]{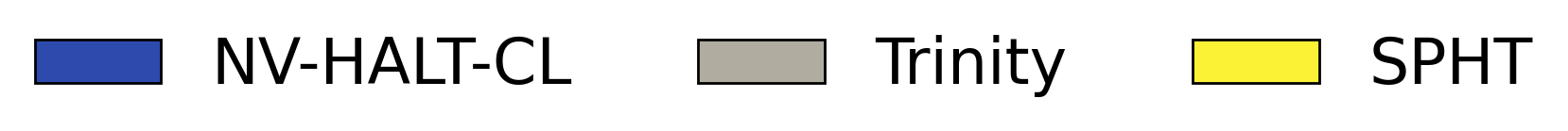}
    \end{subfigure}  
    \begin{subfigure}{1.0\linewidth}
        \centering
        \includegraphics[width=0.15\linewidth]{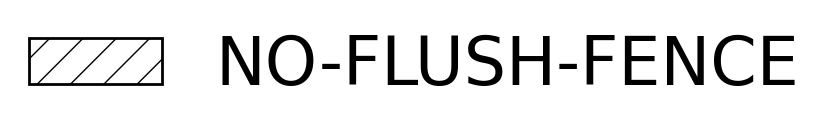}
    \end{subfigure}  
    \vspace{-8mm}
    \caption{\centering \edit{Comparison of energy consumption of persistent TMs with and without flushes/fences via \texttt{perf}. First two columns show the energy consumption of the CPU package and the latter two show the energy consumption of RAM. The workload and key access pattern is the same as row 1 of \Cref{fig:throughput-abtree}. Y-axis is average Joules per commit. X-axis is number of threads.}}
    \Description{\centering Comparison of energy consumption using \texttt{perf} for persistent TMs with and without flushes/fences. First two columns capture the energy-pkg event (energy consumption of the CPU package). Latter columns capture the energy-ram event (energy consumption of RAM). The workload is the same (a,b)-tree as in row 1 of \Cref{fig:throughput-abtree}. Y-axis is average joules per commit. X-axis is number of threads. Keys are accessed according to a uniform distribution.}
    \label{fig:energy}
\end{figure*}

\paragraph{Hashmap}
\tmname~ outperforms SPHT for all workloads and all thread counts for the hashmap microbenchmark especially for workloads with more update operations.
% A single transaction in the hashmap will typically perform fewer reads and writes compared to the (a,b)-tree.
The read and write set size of transactions in the hashmap is typically small which leads to fewer capacity related aborts.
% This reduces the size of the read and write set of hardware path transactions leading to fewer capacity related aborts.
Remove operations in this hashmap mark nodes as empty rather than freeing them.
This is somewhat of a more fair comparison in terms of memory management since neither algorithm will be performing memory reclaimation (memory allocation is still trivial in SPHT and non-trivial in \tmname~).

\subsubsection{Ablation Study}
\label{sec:ablation}
We can categorize overhead of a persistent HyTM into 3 classes:
1) the overhead of flush and fence instructions.
This represents the cost necessary to guarantee writes in volatile cache memory are written back to NVM.
Note that a system eADR would not require these instructions.
2) the overhead of reading and writing to NVRAM.
This represents the cost of persisting transactions once the transaction is guaranteed to commit.
In other words, this is the cost of the persistence mechanism without considering the cost of synchronization.
3) and finally the overhead of additional volatile synchronization.
This is the cost of enabling the TM to persist hardware path transactions.
For the PTMs that we evaluate there is no additional synchronization needed to persist software path transactions.
% It is likely that this would be true for most persistent HyTMs as it is significantly more challenging to persistent hardware path transactions.

To understand these costs for \tmname~ and SPHT we compare the throughput using implementations where we progressively remove each overhead.
The results are shown in \Cref{fig:ablation}.
For each TM the left most bar is the base implementation with all features enabled. 
Each successive bar removes a feature. 
NO-FLUSH-FENCE removes overhead class 1 by using no-ops for flush and fences.
NO-NVRAM removes overhead class 1 and 2 by extending the previous implementation to also mmap all memory utilized by the TM in DRAM instead of NVRAM.
NO-PERSISTENT-HTxn removes all three overhead classes by extending the previous implementation to also remove any synchronization specifically necessary to allow persisting hardware path transactions.

For both TMs, in the 99\% read-only workload, there is little difference in throughput between any of the implementations.
This is not surprising, when we perform fewer writes we engage with the mechanisms for persisting transactions less so we do not pay their cost.
For the 90\% read-only workload we can see some difference in throughput as we disable features particularly at larger thread counts. 
This result is also true for the other update-heavy workloads.
For these workloads, \tmname~ has a larger increase in throughput when we disable the use of NVRAM.
Neither TM gains much throughput when the third overhead class is also removed.

A naive interpretation of this result would be to assume that the persistence mechanism of \tmname~ is inferior, however, \tmname~ achieves higher throughput compared to SPHT for these workloads.
This demonstrates the importance of the underlying volatile synchronization of the TM.
In these workloads hardware aborts more common.
SPHT's synchronization becomes a major bottleneck.
Under high contention, SPHT can spend upwards of half of the entire measurement period in the fallback path where concurrency is disabled.
If there are many concurrent hardware path transactions that write and commit, SPHT will block some of these transactions (even though their write sets are disjoint).
Thus, these results demonstrate that when there is data to persist (10\% or more updates) SPHT is close to the volatile only throughput but this maximum is still lower than throughput of the persistent \tmname~.
This shows that \tmname~ will benefit more from advancements in NVRAM technology and it reinforces our suggestion that hardware assisted locking is preferable compared to approaches like SPHT.

\subsubsection{STAMP}
We compare \tmname~ to SPHT using some of the STAMP \cite{minh2008stamp} benchmarks.
Specifically, we show the kmeans-high, intruder, labyrinth and ssca2.
We omit the other STAMP benchmarks where SPHT experienced crashes.
The STAMP benchmarks are most interesting from the perspective of volatile synchronization.
STAMP was not designed with persistence in mind.
This manifests in the benchmarks in various ways that can effect the performance and correctness of a persistent TM.
For example, there are various places in the STAMP benchmarks where stack allocated memory or memory allocated outside of a transaction via the system allocator is later accessed within a transaction.
None of the PTMs that we know allow for recovering data that was stack allocated.
Likewise, PTMs including \tmname~, SPHT and Trinity cannot persist data allocated by the system allocator.
In this case, we can simply replace the calls to \texttt{malloc} and \texttt{free} with calls to the TM allocator.
As with our microbenchmarks, there is still a significant difference between \tmname~ and SPHT with regards to memory management.
All allocations by SPHT utilize their trivial allocator (which just involves a simple pointer increment) and all frees are no-ops whereas \tmname~ performs a non-trivial amount of work for allocating and freeing.

The benchmarks we tested differ in terms of data set size and contention.
Transactions in the labyrinth benchmark have the largest data set sizes while transactions in the kmeans and ssca2 benchmarks have the smallest.
The data set size of transactions in the intruder benchmark is lower than that of labyrinth and higher than that of kmeans and ssca2. 
Labyrinth and intruder are high contention benchmarks while ssca2 and kmeans are low contention. 
Though we use the higher contention version of kmeans, it is still low compared to labyrinth and intruder.
When executing the STAMP benchmarks we utilize a timeout of 4 minutes for all algorithms, which is multiple minutes beyond the worst case runtime for \tmname~ in any workload.
With our previous microbenchmarks we allow SPHT to obscure the log replay by using a large per-thread log size. 
For the STAMP benchmarks, we attempt to get a better portrayal of the cost of SPHT's log replay mechanism by using a lower per-thread log size, specifically 8MB per thread.

\Cref{fig:stamp} shows the results of running the STAMP benchmarks.
We compare TMs in terms of average processing time (lower is better) for different thread counts.
\tmname~ outperforms SPHT in these benchmarks, requiring less processing time in all cases.
Both algorithms do not scale well, however, this is not surprising since it is known that the STAMP workloads do not scale well \cite{dalessandro2010norec, christie2010evaluation}.
SPHT experiences negative scaling past 16 threads in all cases.
In many of the benchmarks, SPHT exceeds the 4 minute timeout for all trials at some of the higher thread counts.
This is primarily due to blocking caused by the log replay mechanism.
This is a more accurate representation of the behavior one should expect from SPHT in practice where we cannot delay replaying the log until some arbitrary time in the future and we cannot afford to dedicate hundreds of gigabytes of memory for the log in a single program.
These results further reinforce the effectiveness of \tmname~.

\subsection{\edit{Power Consumption}}
One may wonder about the energy consumption of persistent TMs and the energy implications related to forcing write backs from the cache.
To address this, we utilized \texttt{perf} to measure the energy consumption of \tmname~, Trinity and SPHT.
We specifically track the \texttt{perf} events \texttt{energy-pkg} and \texttt{energy-ram}.
These events correspond to the total system-wide energy consumption of the CPU packages (all cores) and RAM respectively.
Note that it is not possible to measure the energy consumption of a specific program or core \cite{khan2018rapl, raffin2024dissecting}.
We measure both the standard implementations as well as the implementations where flushes and fences are replaced with no-ops.
The results of these experiments are shown in \Cref{fig:energy}.
Note that the thermal design power (TDP) of each processor installed in our machine is 150W. 
The TDP represents the average power, in watts, the processor dissipates when operating at base frequency with all cores active under an Intel-defined, high-complexity workload \cite{chip_spec}.

The energy consumption per commit of each algorithm reflects the differences in their throughput.
\tmname~ consumes significantly less energy per commit, achieving up to 3.5x improved energy efficiency compared to Trinity and SPHT.
SPHT specifically suffers from its use of the WBINVD instruction to flush the entire cache during log replay \cite{wbinvd}.
In \Cref{sec:ablation} we discussed how replacing flushes and fences with no-ops does not significantly effect throughput for SPHT and \tmname~, especially for workloads dominated by read-only operations.
The difference in energy consumption per commit with and without flushes and fences follows this same trend.
In most cases replacing flushes and fences with no-ops does not noticeably effect the power consumption per commit of the TMs.
SPHT specifically shows a more noticable difference.
This also aligns with the results of ablation study shown in \Cref{fig:ablation}.

\section{Conclusion}
In this work we presented \tmname~, a family of persistent HyTMs utilizing a new technique for persisting hardware path transactions.
We implemented several versions of our TM with different liveness guarantees.
We demonstrate that, despite the difficulties caused by the inability to persist data within hardware transactions, our persistent HyTM achieves improved performance compared to the existing state of the art.

% 

%%
%% Bibliography
%%

\bibliographystyle{ACM-Reference-Format}
\bibliography{refs}

\clearpage
\appendix

\end{document}